\documentclass[twocolumn]{aastex62}

\shortauthors{Blancato et al.}

\usepackage[T1]{fontenc}
\usepackage{ae,aecompl}
\usepackage{lipsum}
\usepackage[]{algorithm2e}

\usepackage{graphicx}
\usepackage{amsmath}
\usepackage{amssymb}
\usepackage{enumitem}
\usepackage{rotating}

\newcommand{\RN}[1]{%
\textup{\uppercase\expandafter{\romannumeral#1}}
}

\usepackage{hyperref}

\definecolor{mycolor}{HTML}{045a8d}

\hypersetup{
colorlinks,
citecolor=mycolor,
filecolor=mycolor,
linkcolor=mycolor,
urlcolor=mycolor
}

\hypersetup{urlcolor=mycolor}

\usepackage{ulem}
\shorttitle{}

\begin{document}

\title{\uppercase{Data-driven derivation of stellar properties from photometric time series data using convolutional neural networks}}
\correspondingauthor{Kirsten Blancato}
\email{knb2128@columbia.edu}

\author{Kirsten Blancato}
\affil{\textit{Department of Astronomy, Columbia University, 550 West 120th Street, New York, NY 10027, USA}}

\author{Melissa Ness}
\affiliation{\textit{Department of Astronomy, Columbia University, 550 West 120th Street, New York, NY 10027, USA}}
\affiliation{\textit{Center for Computational Astrophysics, Flatiron Institute, 162 Fifth Avenue, New York, NY 10010, USA}}

\author{Daniel Huber}
\affiliation{\textit{Institute for Astronomy, University of Hawai`i, 2680 Woodlawn Drive, Honolulu, HI 96822, USA}}

\author{Yuxi(Lucy) Lu}
\affiliation{\textit{Department of Astronomy, Columbia University, 550 West 120th Street, New York, NY 10027, USA}}
\affiliation{\textit{American Museum of Natural History, Central Park West, New York, NY 10024, USA}}

\author{Ruth Angus}
\affiliation{\textit{American Museum of Natural History, Central Park West, New York, NY 10024, USA}}
\affiliation{\textit{Center for Computational Astrophysics, Flatiron Institute, 162 Fifth Avenue, New York, NY 10010, USA}}
\affiliation{\textit{Department of Astronomy, Columbia University, 550 West 120th Street, New York, NY 10027, USA}}

%%%%%%%%%%%%%%%%%%%%%%%%%%%%%%%%%%% ABSTRACT %%%%%%%%%%%%%%%%%%%%%%%%%%%%%%%%%%
\vspace{10px}
\begin{abstract}
Stellar variability is driven by a multitude of internal physical processes that depend on fundamental stellar properties. These properties are our bridge to reconciling stellar observations with stellar physics, and for understanding the distribution of stellar populations within the context of galaxy formation. Numerous ongoing and upcoming missions are charting brightness fluctuations of stars over time, which encode information about physical processes such as rotation period, evolutionary state (such as effective temperature and surface gravity), and mass (via asteroseismic parameters). Here, we explore how well we can predict these stellar properties, across different evolutionary states, using only photometric time series data. To do this, we implement a convolutional neural network, and with data-driven modeling we predict stellar properties from light curves of various baselines and cadences. Based on a single quarter of \textit{Kepler} data, we recover stellar properties, including surface gravity for red giant stars (with an uncertainty of $\lesssim$ 0.06 dex), and rotation period for main sequence stars (with an uncertainty of $\lesssim$ 5.2 days, and unbiased from $\approx$5 to 40 days). Shortening the \textit{Kepler} data to a 27-day TESS-like baseline, we recover stellar properties with a small decrease in precision, $\sim$0.07 dex for log $g$ and $\sim$5.5 days for $P_{\rm rot}$, unbiased from $\approx$5 to 35 days. Our flexible data-driven approach leverages the full information content of the data, requires minimal feature engineering, and can be generalized to other surveys and datasets. This has the potential to provide stellar property estimates for many millions of stars in current and future surveys. 
\end{abstract} 

%%%%%%%%%%%%%%%%%%%%%%%%%%%%%%%%%%% INTRODUCTION %%%%%%%%%%%%%%%%%%%%%%%%%%%%%%%%%%
\section{Introduction}
\label{sec:introduction}
In the coming years the number of stars with photometric time series observations is projected to increase by several orders of magnitude. The ongoing TESS mission \citep{ricker14} will deliver light curves for the order of 10$^{5}$ stars, while LSST \citep{lsst} is planned to deliver light curves for an unprecedented number of $\sim$10$^{8}$ stars. The large stellar samples covered by these space- and ground-based surveys will enable further probing of known, and possibly reveal new, empirical connections between time domain variability and stellar physics. In combination with \textit{Gaia} parallaxes \citep{gaia16, gaia18}, these observations have the additional potential to markedly extend the characterization of stellar properties and populations throughout the Milky Way. However, while high quality light curves have been used to infer a number of stellar properties, fast and automated methods that can be employed on shorter baseline and sparser cadence observations will be crucial to maximize insights from the forthcoming volume of time domain data. 

Brightness variability in the time domain encodes information about stellar properties through physical processes including oscillations, convection, and rotation. With high-cadence time domain data from the \textit{Kepler} \citep{borucki08} and CoRoT \citep{baglin06} missions, solar-like oscillations have been detected in a large ensemble of stars. These oscillations are a result of turbulent convection near the stellar surface, which induces acoustic standing waves in the interiors of both main sequence and evolved stars, generating stellar fluctuations across a range of timescales \citep[e.g.][]{aerts10}. Solar-like oscillations are typically parameterized through two average parameters, $\nu_{\mathrm{max}}$ and $\Delta\nu$, which can be precisely measured in power spectra computed from high-cadence time series data \citep[e.g.][]{hekker09, deridder09, gilliland10, bedding10, mosser10, stello13, yu18}. The frequency of maximum power, $\nu_{\mathrm{max}}$, is dependent on the temperature and surface gravity of a star \citep{brown91, belkacem11}, while the large frequency separation between consecutive overtones, $\Delta\nu$, is dependent on the stellar density \citep{ulrich86}. The combination of $\nu_{\mathrm{max}}$ and $\Delta\nu$ thus allows a direct measurement of stellar masses ($M_{*}$) and radii ($R_{*}$) \citep[e.g.][]{kjeldsen95, stello09a, stello09b, kallinger10, huber11}.

In stars with convective envelopes, stellar granulation is also imprinted in a star's photometric variability. The circulation of convective cells produces brightness fluctuations at the stellar surface, where brighter regions correspond to hotter, rising material (granules) and darker regions correspond to cooler, sinking material (intergranule lanes). Because the size of the granules is dependent on the pressure scale height \citep{freytag97, huber09, kjeldsen11}, the variability timescale of granulation has been demonstrated to scale with the surface gravity (log $g$) of a star \citep{mathur11, kallinger16, pande18}. The relationship between granulation timescale and surface gravity has led to the development of the ``Flicker method", in which brightness variations on timescales less than 8 hours are used to estimate log $g$ \citep{bastien13, bastien16, cranmer14}. With this estimate of log $g$, and a probe of effective temperature ($T_{\mathrm{eff}}$) (e.g. from spectroscopy, broad-band photometry), a stars relative position on the Hertzsprung-Russell (HR) diagram, and thus evolutionary state, can be determined. 

In addition to oscillations and granulation, stellar rotation also contributes to variability in the observed brightness of a star. Star spots on the surface of magnetically active stars quasi-periodically cross the observable stellar face, imprinting semi-regular patterns in the photometric time series \citep{strassmeier09, garcia10}. Based on these modulations, stellar rotation periods, ($P_{\rm rot}$), have been estimated by examining light curves from ground-based surveys \citep[e.g.][]{irwin09}, the \textit{Kepler} and CoRoT missions \citep[e.g.][]{mosser09, do12, reinhold13, nielsen13, garcia14, santos19}, as well as more recently the \textit{K2} and TESS missions \citep[e.g.][]{curtis19, reinhold20}. Since rotation at the surface is linked to processes occurring in the stellar interior (e.g. dynamos, turbulence) \citep[e.g.][]{zahn92, mathis04, browning06, decressin09, wright11}, there is a prospect of using rotation period measurements to probe fundamental stellar properties, as well as the magnetic and dynamical evolutionary history of stars.

Of particular value is the connection between rotation period and stellar age. As main sequence stars evolve, stellar winds transport angular momentum away from the star, slowing the rate at which it rotates \citep{weber67, kawaler88, bouvier97}. The empirical relationship between stellar age and rotation period was first realized by \cite{skumanich72}, which prompted the development of gyrochronology \citep{barnes03} as a tentative tool for estimating stellar ages from rotation and color alone. Recent theoretical work has focused on deriving the gyrochronology relations from stellar physics \citep[e.g.][]{matt12, reiners12, gallet13}, open clusters, and other stellar samples, for which precise and independent measurements of both stellar age and rotation period can be made have been used to calibrate these relationships \citep[e.g.][]{kawaler89, barnes03, barnes07, cardini07, meibom09, mamajek08, agueros18, douglas16, douglas19}. 

However, as illustrated in \cite{angus15}, a robust empirical calibration of gyrochronology has proven to be challenging. Based on \textit{Kepler} stars with asteroseismic age estimates, it is found that multiple age-period-color relationships are necessary to describe the properties of the stellar sample, which suggests that the gyrochronology relationship is under-specified. Furthermore, in {\color{mycolor} Angus et al., accepted for publication in AJ}, it is demonstrated that empirically calibrated gyrochronology models are not able to sufficiently reproduce the ages of rotating stars, particularly of late K- and early M-type dwarfs, which suggests that the simple gyrochronology relation as proposed by \cite{skumanich72} is unable to capture the full complexity of stellar spin-down. Adding additional physics appears necessary, and in the semi-empirical modeling of rotational evolution pursued in \cite{spada20}, it is found that including a mass and age-dependent core-envelope coupling timescale is needed to reproduce the rotation periods of stars in old open clusters \citep[e.g.][]{curtis19}.

The uncertainty of gyrochronology relations have also been revealed from a theoretical perspective. For instance, \cite{vansaders16} find that weakened magnetic breaking limits the predictive capability of gyrochronology, specifically for stars in the second half of their main sequence lifetimes. In the context of theoretical stellar rotation models, \cite{claytor19} determine the biases associated with the inference of stellar age from rotation period for lower main sequence stars based on current theoretical models of stellar angular momentum spin-down. Furthermore, combining theoretical models of stellar rotation with expected observational biases, \cite{vansaders19} use forward modeling to probe rotation periods across a population of stars, finding that current models of magnetic braking fail at longer rotation periods, and that particular care is necessary to correctly interpret stellar ages from rotation period distributions. 

As a result of the physical processes described above, high cadence stellar photometry contains rich information at multiple timescales about fundamental stellar properties including mass, radius, and age. Careful analysis of light curves from missions like \textit{Kepler} and CoRoT have revealed the potential of this data and enabled the determination of stellar properties for thousands of stars. However, the imminent volume of time domain data that will be delivered by surveys like TESS and LSST necessitates the development of new methods for estimating stellar properties from shorter baseline and sparser cadence data. Automated pipelines to measure the asteroseismology parameters $\nu_{\mathrm{max}}$ and $\Delta\nu$ have been developed and applied to large samples of \textit{Kepler} stars \citep{huber09}, and Bayesian methods for inferring these parameters have been tested on small (\textless~100 stars) samples \citep{davies16, lund17}, and on $\sim$13,000 \textit{K2} Campaign 1 stars \citep{zinn19}. 

Automated methods for extracting rotation periods from \textit{Kepler} photometry have also been put forth. Producing the largest catalog of homogeneously derived rotation periods to date, \cite{mcquillan14} derive rotation periods for $\sim$30,000 main sequence stars with a peak identification procedure in the autocorrelation function (ACF) domain based on a minimum baseline of $\sim$2 years of observational coverage (see \cite{mcquillan13a}). Instead, taking a probabilistic approach, \cite{angus17} infer posterior PDFs (probability distribution functions) for the rotation periods of $\sim$1,000 stars based on a Gaussian process model. This method has the benefit of not assuming strictly sinusoidal periodicities, and compared to traditional methods it provides more robust credible intervals on the inferred rotation periods. However, \cite{angus17} find that the posteriors still underestimate the true uncertainties, and the method relies on computationally expensive posterior sampling. Most recently, {\color{mycolor} Lu et al. in prep.} implement a random forest model to predict rotation periods from light curves and \textit{Gaia} data, with a particular focus on deriving the long periods of M-dwarfs from TESS data.

Data-driven techniques have shown promise in their capability to efficiently identify red giant branch (RGB) stars with solar-like oscillations, and to estimate fundamental stellar properties like $T_{\mathrm{eff}}$ and log $g$ from time domain data. Learning a generative model for RGB stars, \cite{ness18} use \textit{The Cannon} \citep{ness15} to model the ACF amplitude at each lag as a polynomial function of stellar properties ($T_{\mathrm{eff}}$, log $g$, $\nu_{\mathrm{max}}$, $\Delta\nu$). Trained on $\sim$4-year baseline data, \cite{ness18} find the variance of their log $g$ estimator to be $\textless$ 0.1 dex and the variance of their $T_{\mathrm{eff}}$ estimator to be $\textless$ 100 K, with the information required to learn these properties being contained in ACF lags up to 35 days and 370 days, respectively, for log $g$ and $T_{\mathrm{eff}}$. Taking a similar approach, {\color{mycolor} Sayeed et al. in prep.} learn a local linear regression model between the power density at each frequency of smoothed \textit{Kepler} power spectra and stellar properties. For upper main sequence and RGB stars that do not exhibit rotation, {\color{mycolor} Sayeed et al. in prep.} learn a log $g$ estimator with a variance $\textless$ 0.07 dex based on the 10 nearest neighbors in the frequency domain of the training set. Neural networks have also been implemented for RGB asteroseismology. Training a convolutional neural network (CNN) on an image representation of \textit{Kepler} power spectra, \cite{hon17} classify RGB stars versus core helium burning stars to an accuracy of 99\%, and in \cite{hon18} their approach predicts $\nu_{\rm max}$ to an uncertainty of about 5\%. In \cite{hon18-2} it is found that based on power spectra images derived from 4-year, 356-day, 82-day, and 27-day data, that the classification accuracy decreases from $\sim$98\% based on 4-year data to $\sim$93\% based on 27-day data.

In this work, we pursue systematically and consistently estimating a set of stellar properties directly from photometric time series data. We do this by fitting a flexible 1-dimensional (1D) CNN to the data, which is able to capture the structure of the data in the time domain on multiple scales and requires minimal feature engineering. Using a single quarter of \textit{Kepler} data and asteroseismology-quality stellar measurements as our training set, we build models to classify stellar evolutionary state and a set of stellar properties across the RGB (including red giants and red clump stars) and main sequence from light curves of various baselines and cadences, and compare these results to models based on the ACF and frequency domain transformations of the data. The CNN classification model distinguishes RGB stars from main sequence and sub-giant stars to an accuracy of $\sim$90\%, and for RGB stars we demonstrate that the CNN regression model trained on 27-day \textit{Kepler} light curves is able to predict log $g$ to an rms precision of $\sim$0.07 dex, $\Delta\nu$ to an rms precision of $\sim$1.1 $\mu$Hz, $\nu_{\rm max}$ to an rms precision of $\sim$17 $\mu$Hz, and $T_{\rm eff}$ to an rms precision of $\sim$300 K. For main sequence stars, we predict rotation periods up to $P_{\rm rot}\sim$35 days based on 27-day and even 14-day data, with an rms precision of $\sim$6 days. We also find that for observations spaced 1 day apart (over 97 days), we can recover $P_{\rm rot}$ from $\approx$5 to 40 days with an rms precision of $\sim$6.2 days. Our approach, which leverages the full information content of the data, serves as a proof of concept in the pursuit of estimating stellar properties for many millions of stars from variable quality time domain data.

%%%%%%%%%%%%%%%%%%%%%%%%%%%%%%%%%%% TRAINING DATA %%%%%%%%%%%%%%%%%%%%%%%%%%%%%%%%%%
\section{Training Data}
\label{sec:data}

\subsection{The Kepler data}
\label{sec:data_kepler}
To predict stellar and asteroseismology parameters from time domain variability, we build models trained on long-cadence (29.4-minute sampling) \textit{Kepler} data. We download all available Q9 light curves from the \textit{Kepler} mission archive\footnote{https://archive.stsci.edu/pub/kepler/lightcurves}, which supplies $\sim$97 days of time domain observations for 166,899 stars. To minimize the amount of data-processing, we train our models based on this single quarter of observations. 

Light curves are often transformed to different representations, in the frequency and time-lag (i.e. ACF) domains. This is commonly done in order to extract signals that concentrate in these forms: $\nu_{\mathrm{max}}$ and $\Delta\nu$ from the frequency spectrum, and rotation period from the peak in the ACF. In this work, we do not collapse the data to a few measurable signatures. We leverage the entire set of flux observations to predict the stellar properties. It is therefore unclear as to whether a particular choice of data representation will better than another. We examine how well we can derive stellar properties using (i) time, (ii) frequency and (iii) time-lag representations of the data. We report the differences between the approaches in Section \ref{sec:regression}. We note however that any preferential representation, in terms of prediction performance, may simply reflect the compatibility of the data representation, given our modeling choice.  

\subsection{Light curve processing}
\label{sec:data_space}
\subsubsection{The time domain}
\label{sec:time_space}
The flux measurements we use are the Pre-search Data Conditioning Simple Aperture Photometry (PCDSAP) flux values, which have been corrected for systematic errors and anomalies caused by the spacecraft and instrument \citep{twicken10, jenkins10}. Before transforming the time series data to different domains, we apply two additional processing steps to each light curve. First, we remove observations with a \texttt{SAP\_QUALITY} flag greater than 0. We then apply a local sigma clipping algorithm to each light curve, removing observations with flux values more than three standard deviations away from the mean flux computed in a sliding window of 50 consecutive observations. Finally, we transform the light curves to be in units of relative flux, $\Delta f/f$. These three pre-processing steps are applied to the light curves before any further processing and transformation into other data domains.

For the models we build based on the data in the original time domain, we apply the following additional processing steps. First we enforce the light curves to be on a common time grid, so that the structure of the input data is standardized. Since the cadence of the \textit{Kepler} data is mostly regular with flux measurements at every 29.4 minutes, we set any missing flux values in the time grid to zero. In Section \ref{sec:discussion} we discuss alternative imputation choices that can be explored, however our model is successful taking the simplest zero-imputation approach. We then normalize the relative flux values of each light curve by subtracting the mean ($\mu$) and dividing by the standard deviation ($\sigma$) of the relative flux values, so that $f_{scaled}$ = ($f - \mu)/\sigma$. Because the standard deviation of each individual light curve provides useful information in comparing across the collection of light curves, we supply this as an additional feature to the model as discussed in Section \ref{sec:architecture}. 

\begin{figure*}[htp!]
\centering
\includegraphics[width=1.9\columnwidth]{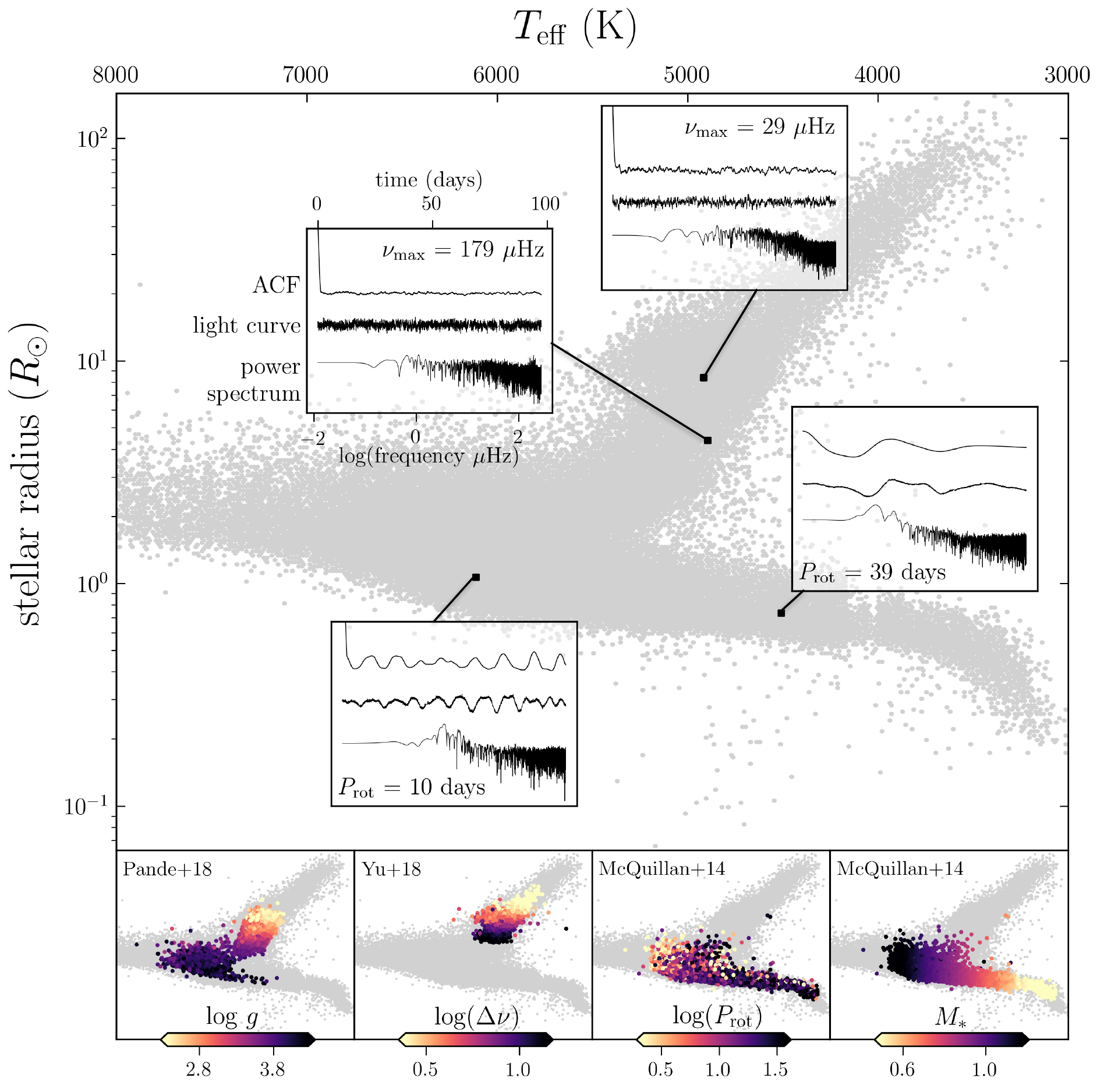}
\caption{Demonstration of how the \textit{Kepler} light curves and stellar properties jointly vary across different regions of the HR diagram, illustrating the potential to learn the fundamental properties of a star from its light curve alone. The HR diagram (shaded in grey) is from the \cite{berger18} catalog of derived stellar radii and $T_{\rm eff}$ for $\sim$150,000 stars. In the \textit{\textbf{main panel}}, the insets show the original light curve, as well as the ACF and power spectrum computed from the light curve, at various positions along the RGB and main sequence. The \textit{\textbf{lower panels}} indicate how various stellar properties (log $g$, $\Delta\nu$, $P_{\rm rot}$, and $M_{*}$) from different catalogs also vary across the HR diagram. \bigskip}
\label{fig:data}
\end{figure*}

In Figure \ref{fig:data} we illustrate how the time domain data varies for stars across the HR diagram. In the main panel of this figure, we show (in grey) the distribution of stars in the stellar radius against effective temperature plane ($R_{*}$-$T_{\rm eff}$) for the set of $\sim$150,000 stars from \cite{berger18} that have \textit{Kepler} Q9 light curves available. The main sequence is distributed across log($R_{*}$) $\lessapprox$ 0.5 $R_{\odot}$ and 6500 K $\lessapprox$ $T_{\rm eff}$ $\lessapprox$ 3000 K, and the RGB is located at log($R_{*}$) $\gtrapprox$ 0.5 $R_{\odot}$ and 5500 K $\lessapprox$ $T_{\rm eff}$ $\lessapprox$ 3000 K with the red clump resolved at log($R_{*}$) $\sim$ 1 $R_{\odot}$ and $T_{\rm eff}$ $\sim$ 4800 K. Stars with $T_{\rm eff}$ $\gtrapprox$ 6500 K have thin convective or entirely radiative envelopes, and thus include ``classical" pulsators such as delta Scuti or gamma Doradus stars. To demonstrate how the shape of the light curves vary across the regions of the HR diagram, the middle signal in the inset panels shows the time domain data for two stars with different property values in the main sequence, as well as for two stars in the RGB/red clump with different property values. For the main sequence stars (at about the same stellar radius) their light curves clearly indicate a stellar rotation signal with the hotter, upper main sequence star having a shorter rotation period of 10 days and the cooler, lower main sequence stars having a longer rotation period of 39 days. For the two RGB stars with $T_{\rm eff}$ $\sim$ 5000 K, we see that unlike the main sequence stars the light curves don't exhibit a rotation signal within the 97-day baseline that is shown, but that the amplitudes of the short timescale variations differs between the stars at different $R_{*}$. From these example light curves in the main sequence and RGB we see that the time domain data varies across the HR diagram, where stars with different properties exhibit distinctive light curves characteristics. What this suggests is that from the light curves alone we can place stars (to some degree of precision) on the HR diagram and predict other fundamental stellar properties. 

\subsubsection{The ACF}
\label{sec:acf_space}
In addition to working with data in the original light curve space, we also test building models based on the ACF of the time series. The ACF describes the strength of periodic signals present in time series data by measuring the similarity of the time series with itself at different lags. The ACF has been shown to be an effective domain for measuring the surface gravity of stars \citep{kallinger16}, the rotation periods of main sequence stars \citep{mcquillan14}, as well as the temperatures, surface gravity's, and asteroseismology observables of RGB stars \citep{ness18}. 

For observations evenly space in time, $t_{k}$ = $(k-1)\Delta t$, the ACF at each lag $k$ is, 

\begin{equation}
    \mathrm{ACF}_{k} = \frac{\sum^{N-k}_{i=1}[(x_{i} - \overline{x})(x_{i+k} - \overline{x})]}{\sum^{N}_{i=1}(x_{i} - \overline{x})^{2}},
\smallskip
\label{eq:acf}
\end{equation}

\noindent where the numerator is the co-variance between the time series and itself at lag $k$, and the denominator is the variance of the time series, which normalizes the ACF to be 1 at lag $k=0$ and defined over the range [$-$1, 1] \citep[e.g. see][Chapter 10]{ivezic}. To compute the ACF according to Equation \ref{eq:acf}, we first linearly interpolate the flux of each light curve to a common, evenly spaced time grid defined from 0 to 97.4 days with a $\Delta t$ = 0.0204 days (i.e. the long-cadence sampling). 

The main panel of Figure \ref{fig:data} shows example ACFs for stars in the main sequence and RGB. For the two stars in the main sequence, we see that the second peak of the ACF corresponds to the rotation period of the star, with the peaks at later lags being integer multiples of the period. However, for the two RGB stars, ACF shows less visible structure. For these stars that don't exhibit strong rotation over the baseline of the data, the information contained in the ACF is more subtle. For example, granulation, as a stochastic process, is much less coherent than rotation, which results in a less structured imprint of this signal in the ACF. 

\subsubsection{The frequency domain}
\label{sec:frequency_space}
Another representation of stellar time series data is in the frequency domain. The power spectrum of a star's light curve quantifies the strength of the flux signal across a range of timescales ($T$), represented as a the spectral density ($P$) as a function of frequency ($f$ = 1/$T$). The primary asteroseismology observables, $\nu_{\rm max}$ and $\Delta\nu$, are defined and identified in the power spectrum representation of stellar light curves \citep[e.g.][]{bedding10, yu18}. For discretely sampled data the fast Fourier transform (FFT) algorithm, which represents the light curves as a summation of sinusoidal functions, is typically used to compute the power spectrum of stellar time series. However, the FFT algorithm requires that the time series be regularly sampled over the entire observation window. In the case of unevenly sampled or missing data, an alternative method for generating a frequency domain representation of time series data is to compute a periodogram as an estimate of the true power spectrum. A commonly used algorithm in astronomy is the Lomb-Scargle (LS) periodogram \citep{lomb76, scargle82}, which is a least squares method for detecting sinusoidal periodic signals in time series data.

To compute the LS periodogram of the \textit{Kepler} light curve data, we use the implementation provided by the \texttt{astropy} package. Following the recommendations of \cite{vanderplas18}, we compute the periodogram on a frequency grid with a minimum frequency of $f_{\rm min}$ = 0 Hz, a maximum frequency of $f_{\rm max}$ = 1/(2$\delta t$) Hz, and a frequency spacing of $\Delta f$ = 1/($n_{o} T$) Hz, where $T$ is the baseline of the observations (e.g. 97.39 days for Q9) and $n_{o}$ is the oversampling factor, which we set to $n_{o}$ = 10. The value for the Nyquist frequency, $f_{\rm max}$, is a pseudo-windowing limit, where we take $\delta t$ to be most frequent spacing of the time series observations (0.0204 days).

In the main panel of Figure \ref{fig:data} we show example periodograms for stars at different locations in the HR diagram. Considering the two RGB stars, the frequency of maximum power is a prominent feature of the power spectra. For the star with the larger stellar radius, $\nu_{\rm max}$ is at a lower frequency of 29 $\mu$Hz while the $\nu_{\rm max}$ of the star with a smaller stellar radius is at a higher frequency of 179 $\mu$Hz. For the main sequence stars $\nu_{\rm max}$ is not visible. For these stars, the frequency of maximum power resides at frequencies greater than the range permitted by the Nyquist frequency ($\gtrapprox$ 240 $\mu$Hz). Even though $\nu_{\rm max}$ lies beyond the frequency grid of the power spectra for these stars, the overall shape and other features of the spectrum contain useful information that can potentially be indicative of the properties of the star.

\begin{figure*}[htp!]
\centering
\includegraphics[width=1.75\columnwidth]{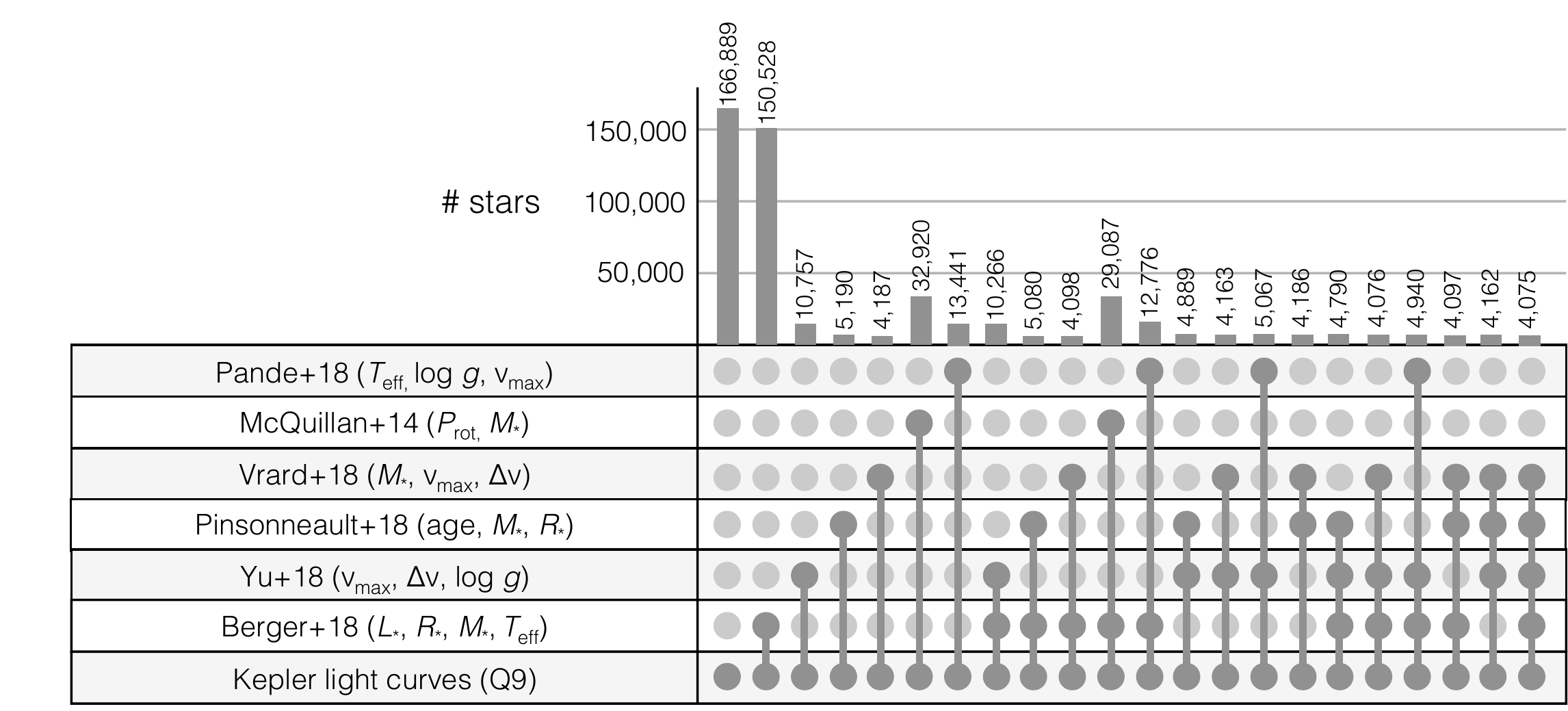}
\caption{UpSet plot \citep{upset} showing the set intersections of the \textit{Kepler} stars with Quarter 9 light curves and the various stellar property catalogs available in the literature. The histogram indicates the number of stars contained in the set defined in each column, where the shaded circles indicate which catalogs are intersected. For conciseness, only the intersections that contain a minimum of 4,000 stars is displayed. This plot demonstrates the various datasets that can be used to train a model to predict stellar properties from light curves. \bigskip}
\label{fig:catalog_coverage}
\end{figure*}

\subsection{Stellar property catalogs}
\label{sec:data_params}
There are a number catalogs in the literature providing stellar property estimates for \textit{Kepler} stars. Many of these catalogs have stars in common, but there is no joint database that exists. Here we try to systematically explore the intersect of several important and relevant catalogs for data-driven inference work. Figure \ref{fig:catalog_coverage} shows the coverage and set intersection (e.g. catalog 1 $\cap$ catalog 2) of six stellar property catalogs with the stars that have \textit{Kepler} Q9 light curves available. As seen in the figure, the \cite{berger18} catalog includes a majority of the \textit{Kepler} stars, delivering estimates of $R_{*}$ and evolutionary state across the HR diagram. The catalog that provides stellar property estimates for the next greatest number of stars is the \cite{mcquillan14} rotation period catalog for $\sim$30,000 main sequence stars, and following this the \cite{yu18} and \cite{pande18} catalogs provide $\nu_{\rm max}$, $\Delta\nu$, $M_{*}$, $R_{*}$, and log $g$ for $\sim$13,000 stars and $\nu_{\max}$, log $g$, and $T_{\rm eff}$ for $\sim$10,000 stars respectively, primarily for stars on the RGB. The remaining catalogs shown in Figure \ref{fig:catalog_coverage} provide stellar properties for fewer stars, with a minimum of $\sim$4,000 to be included in the figure. 

As an initial proof of concept of our modeling approach, we focus on the three catalogs covering the greatest number of stars \citep{mcquillan14, yu18, pande18} where the stellar properties are homogeneously derived. However, models can certainly be tested on the other catalogs, as well as on a set of stellar properties combining the estimates from multiple catalogs. The stellar property catalogs compiled in Figure \ref{fig:catalog_coverage} demonstrates the various datasets that can be constructed and used to train data-driven models of stellar properties.

We now provide a brief description of how the stellar properties we predict in Section \ref{sec:regression} are derived. For the \cite{yu18} sample, which includes RGB stars, we successfully recover the asteroseismology observables, $\Delta\nu$ and $\nu_{\rm max}$, as well as log $g$, each which are derived as follows:

\begin{itemize}
    \item $\Delta\nu$: derived from the \textit{Kepler} 29.4-minute cadence data across available quarters using the SYD pipeline described in \cite{huber09}, which consider the light curves in both the frequency and ACF domain of the data (see \cite{huber09} for details). The mean of the reported uncertainties on $\Delta\nu$ is $\sim$0.05 $\mu$Hz, and the mean fractional uncertainty is $\sim$1\%.
    
    \item $\nu_{\rm max}$: derived with the same pipeline as $\Delta\nu$ (see \cite{huber09} for details). The mean of the reported uncertainties on $\nu_{\rm max}$ is $\sim$0.9 $\mu$Hz, and the mean fractional uncertainty is $\sim$2\%.
    
    \item log $g$: derived along with mass and radius from scaling relations. The mean of the reported uncertainties on log $g$ is $\sim$0.01 dex, and the mean fractional uncertainty is $\sim$0.5\%.
\end{itemize}

\noindent For the \cite{pande18} sample, which includes RGB as well as sub-giant stars, we successfully recover $T_{\rm eff}$ and log $g$, which are derived as follows:

\begin{itemize}
    \item $T_{\rm eff}$: taken from \cite{mathur17}, which compiled temperatures from various sources including spectroscopic and photometric based measurements (see \cite{mathur17} for details). The mean of the reported uncertainties on $T_{\rm eff}$ is $\sim$140 K, and the mean fractional uncertainty is $\sim$2.5\%.
    
    \item log $g$: determined from \textit{Kepler} 29.4-minute cadence data based on an empirical relationship between log $g$, $T_{\rm eff}$ and $\nu_{\rm max}$ which has been established using the Fourier transform of the 1-minute cadence \textit{Kepler} benchmark dataset, consisting of $\sim$500 stars \citep{huber11, bastien13}. The mean of the reported uncertainties on log $g$ is $\sim$0.25 dex, and the mean fractional uncertainty is $\sim$8\%.
\end{itemize}

\noindent and finally for the \cite{mcquillan14} sample, which covers main sequence stars, we successfully recover the stellar rotation period, and weakly recover $M_{*}$, which are derived as follows:

\begin{itemize}
    \item $M_{*}$: derived from the \cite{baraffe98} isochrone models taking $T_{\rm eff}$ as input, where $T_{\rm eff}$ is either from the \textit{Kepler} Input Catalog (KIC) or \cite{dressing13}, if available.
    As reported in \cite{mcquillan14}, given a $\sim$200 K precision for the $T_{\rm eff}$ estimates the typical uncertainty on $M_{*}$ is $\sim$0.1 $M_{\odot}$. Assuming a 0.1 $M_{\odot}$ uncertainty across the entire stellar mass range, this translates to a mean fractional uncertainty of $\sim$12\%.
    
    \item $P_{\rm rot}$: derived from a minimum of 8 of the 12 \textit{Kepler} 29.4-minute cadence quarters from Q3 - Q14. The rotation period for each star is identified using an automated peak identification procedure in the ACF domain (see \cite{mcquillan13a}), excluding stars from the sample that are eclipsing binaries, KOIs, and without convective envelopes ($T_{\mathrm{eff}}$ \textgreater~6500 K). The mean of the reported uncertainties on $P_{\rm rot}$ is $\sim$0.6 days, and the mean fractional uncertainty is $\sim$3\%.
\end{itemize}

%%%%%%%%%%%%%%%%%%%%%%%%%%%%%%%%%%% METHODS %%%%%%%%%%%%%%%%%%%%%%%%%%%%%%%%%%
\section{Methods}
\label{sec:methods}
In this section we discuss our modeling approach, as well as outline our training and evaluation procedures. The modeling code is made publicly available on GitHub at \href{https://github.com/kblancato/theia-net}{https://github.com/kblancato/theia-net}.

\subsection{Modeling approach}
\label{sec:model}
As demonstrated in Figure \ref{fig:data}, the properties of stars and the traits of their light curves vary jointly across the HR diagram. Given these correlations, our goal is to predict the properties of a star based on its light curve alone. To achieve this, the model we choose should capture the time structure of the data. That is, how each flux value is related to other values in the time series. Typically, the time structure of light curves is characterized by transforming the data to either the ACF domain or the frequency domain, described in Sections \ref{sec:acf_space} and  \ref{sec:frequency_space}, respectively. After performing these data transformations, informative features in these domains are identified and used to infer stellar properties that features are known to correlate with. These data transformations require additional computational time and preconceptions of how to transform the data to produce the features of interest. Transformations of data may also result in information loss. Given these considerations, in this paper our goal is to build a model that can learn directly from the time series data itself, requiring minimal pre-processing or handcrafted engineering of the raw data.

To do this, we implement a 1D CNN to accomplish the supervised learning task of mapping light curve data to stellar properties. CNN based models have been very successfully used for many supervised learning tasks, particularly for image classification \citep[e.g.][]{alexnet, resnet, vgg, gans, unet}. They are built from a hierarchy of artificial neural networks, known as ``universal function approximators" \citep{hornik90, hornik91}, which learn increasingly abstract representations of the input data, $\vec{X}$, by non-linearly transforming the data through a series of hidden layers that relate $\vec{X}$ to an output prediction $\vec{Y}$. CNNs are a special class of neural network architecture, that differ from fully-connected neural networks, by their inclusion of only partially connected, or so-called convolutional layers, which detect the topological structure of the input data, capturing how neighboring image pixels are related spatially, or how adjacent time series measurements are related temporally. The convolution operation relates elements of the input data to each other through weight sharing. This makes the modeling more efficient and less prone to overfitting than the fully-connected counterpart, by effectively reducing the number of model parameters that need to be learned. CNN models have been successfully used for a variety of tasks in astronomy, including the classification of galaxy morphology and properties based on galaxy images \citep[e.g.][]{dieleman15, huertascompany18, sanchez18}, to predict characteristics of stellar feedback in CO2 emission maps \citep{vanoort19, xu20}, and to predict the 3D distribution of galaxies from the underlying dark matter distribution in large-volume cosmological simulations \citep{zhang19, yip19}. 

With the CNN as our model of choice, the modeling approach we take is a so called end-to-end discriminative approach. A model is learned from a set of objects, for which the input data (light curves), and labels (stellar properties) which describe it, are both defined.  The model takes the time series light curve data as an input, and through the training process learns an informative set of features from the data which optimize the stellar property predictions. This procedure requires no handcrafted transformations or feature engineering of the data as a separate procedure before model training. For the task that we tackle here of predicting stellar properties from time series data, there are a number of alternative models that can capture the time dependence of the data. We discuss an alternative method that is also suited to this problem, recurrent neural networks (RNNs), in the discussion.

\subsection{Model architecture}
\label{sec:architecture}
The CNN model architecture we implement has two convolutional layers, followed by three fully-connected layers, which together perform the stellar property prediction. Given that the size of our training sets are on the order of 10$^{4}$ examples, we define a relatively small network architecture so as to minimize the number of network parameters that need to be learned and to prevent overfitting. For comparison, AlexNet \citep{alexnet}, with 5 convolutional layers and 3 fully-connected layers, had a total of 60 million network parameters and was trained on 1.2 million images.

\begin{figure*}[tp!]
\centering
\includegraphics[width=2\columnwidth]{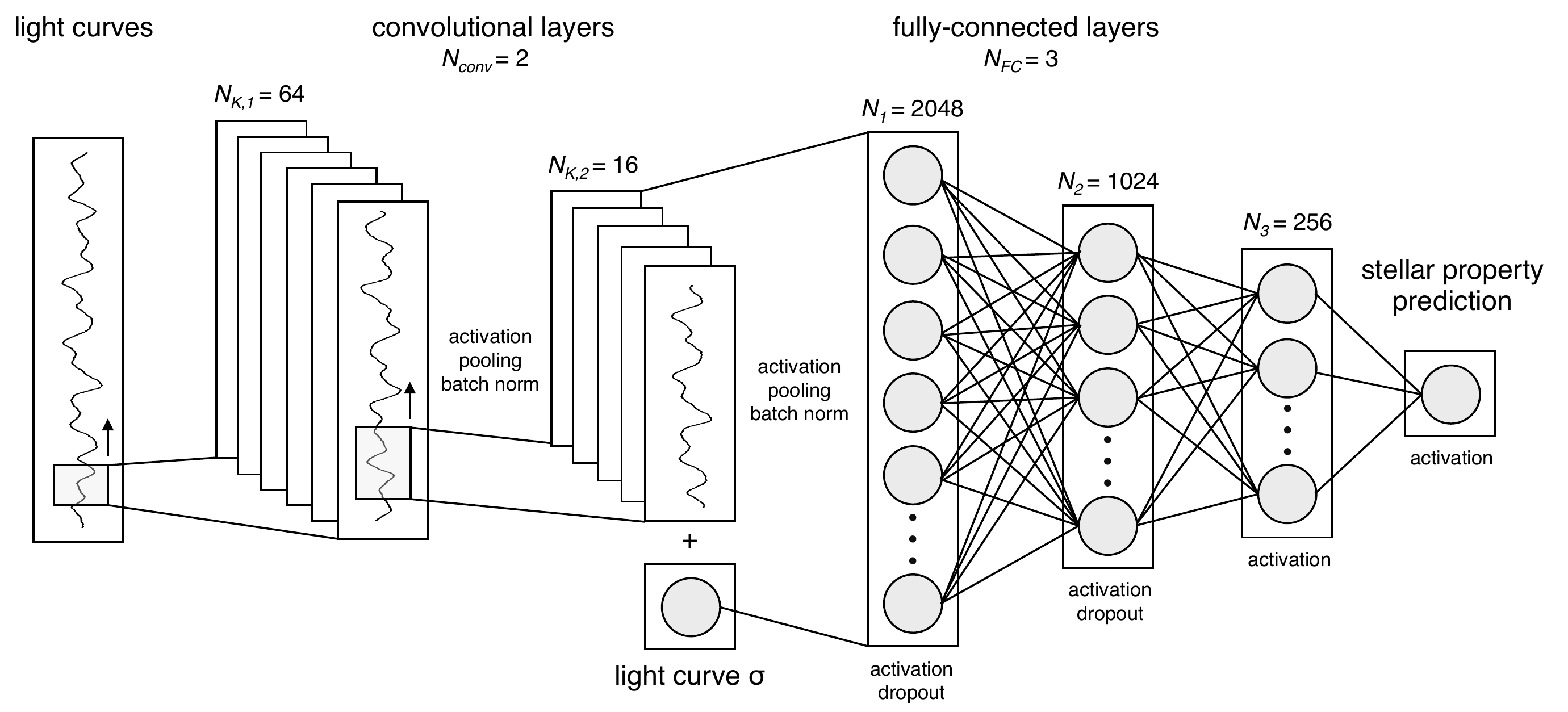}
\caption{Schematic of the CNN architecture implemented to predict stellar properties from light curves. The left-most panel represents the input time series data, the next two panels indicate the two convolutional layers with different kernel widths and output channels, the fourth and fifth layers show the two fully-connected layers where each circle represents a hidden unit, and the last layer is the stellar property prediction. The symbols describing the network architecture are defined in Table \ref{tab:nnparams}. For the classification of evolutionary state, the last layer is replaced with a prediction of the probability of the star belonging to the RGB, sub-giant branch, and the main sequence. \bigskip}
\label{fig:nn}
\end{figure*}

Figure \ref{fig:nn} is a visual representation of the model, showing the operations performed to transform the light curve data to a stellar property prediction. The left-most block represents the light curve data itself, which has been pre-processed and scaled as described in Section \ref{sec:time_space}. The first operation applied to the time series data is a 1D convolution with one input channel, i.e. the scaled flux values at each time, and a specified number of output channels, $N_{K}$, which corresponds to the number of learned kernels each having its own weight matrix and bias. This makes the number of parameters to learn for each convolutional layer [($K_{W} \times K_{H}$)+1]$\times N_{K}$, where $K_{W}$ is the kernel width and $K_{H}$ is the kernel height (in the 1D case $K_{H}=1$). The addition of one accounts for the single bias parameters learned per kernel. The convolution operation takes the input vector, $\vec{X}$, of length $n(X_{\rm in})$, and transforms it into a new vector of length $n(X_{\rm out})$, which is computed as:

\begin{equation} \label{eq:output_length}
n(\vec{X}_{\rm out}) = \Big[\frac{n(\vec{X}_{\rm in})+2\times P-D\times (K_{W}-1)-1}{S}+1\Big],
\medskip
\end{equation}

\noindent where $P$ is number of zeros padded to either side of the time series, $D$ is the dilation factor, and $S$ is the stride over which the convolution is taken. In Figure \ref{fig:nn}, the block to the immediate right of the light curve data represents the output of the first convolution layer, where each of the $N_{K}$ output channels has a length described by Equation \ref{eq:output_length}.

After each convolution, three additional operations are performed on the data before it is passed to the next layer of the model. First, an activation function is applied to introduce non-linearities into the model. This captures non-linear relationship between the data and the labels that describe it. To do this, we implement the commonly used ReLU (Rectified Linear Unit) activation function, defined as $max(0, \vec{X}_{\rm out})$. Following the activation function, a pooling operation is applied. Pooling, or ``down-sampling", reduces the dimensionality of the data vector that will be passed to the following convolutional layer and aids in the prevention of overfitting. The pooling operation slides over the data vector, and typically takes either the maximum or average of the data values within each window, resulting in an output vector of length $n(\vec{X}_{\rm out})$ = $n(\vec{X}_{\rm in})$/$K_{\rm pool}$ when the stride is set equal to the pooling kernel width, $K_{\rm pool}$. Lastly, batch normalization \citep{batchnorm} is applied to each output channel. Batch normalization solves the problem of ``internal covariate shift", in which the distribution of each hidden-layer value changes during training, as the parameters of the previous layers are updated. To enforce that the distribution of the hidden layer values is similar throughout the training process, the batch normalization operation standardizes the values of each hidden layer by subtracting and dividing by the batch mean. This operation adds two new parameters for the model to learn, that weight and shift the normalized vector, but leads to faster and more stable training and also acts to regularize the model.

After the activation function, pooling, and batch normalization operations, convolutions are performed on each of the $N_{K}$ output channels from the previous layer, which has the same properties as the first convolution, as described above. Following this second convolution operation, an activation function, pooling, and batch normalization are again applied to the data. After the two convolution layers, the output channels produced by the second convolution are flattened to a single dimension, and the data is then passed to the fully-connected part of the network. The fully-connected part of the network, represented in the last four panels of Figure \ref{fig:nn}, is a typical multilayer perceptron (MLP). Each element of the flattened data vector produced by the second convolution layer is mapped to $N_{\rm 1}$ hidden units in the first MLP layer, with each hidden unit having its own learnable weight parameter, $w_{1}$. With the addition of a bias parameter, $b_{1}$, the output of the first MLP layer is $\vec{h}(X)$ = $f\big(\sum_{i=0}^{n(\vec{X}_{\rm out})}(w_{1,i}X_{i} + b_{\rm 1}\big)$ where $f$ is the specified activation function and $n(w_{1,i}) = N_{1}$. To this layer we also pass the scaled standard deviation of the flux values (light curve $\sigma$) for each light curve, to capture how the amplitude of the light curves vary across the sample. The second and third fully-connected layers take the output of the layer immediately preceding it, and perform the same operation with each layer learning its own set of weights and biases. 

The last operation of the model architecture, shown as the right-most panel of Figure \ref{fig:nn}, is the prediction of the output stellar property $\vec{Y}$. In the case of regression, $\vec{Y}$ = $\big(\sum_{i=0}^{N_{3}}(w_{4,i}g(X)_{i} + b_{\rm 4})\big)$, where $n(w_{4,i}) = 1$. We experiment with one hyperparameter describing the fully-connected part of the model architecture, which is the dropout probability, $D_{\rm FC}$, applied to the first and second MLP layers. Dropout is a form of regularization, where, during each training iteration, the values for a number of hidden units are randomly set to zero with a probability of $p$ \citep{hinton12}. The two dropout probabilities we consider are $D_{\rm FC}$ = [0.0, 0.3].

The architecture described above is a smaller capacity, 1D regression version of the ``vanilla" end-to-end CNN architectures that are commonly used for the task of image classification. In Table \ref{tab:nnparams}, we summarize the parameters of the model architecture, and note which (hyper)parameters we experiment with varying, which we will discuss in Section \ref{sec:parameters}. In the following sections we also describe how we split the data for training, validation, and testing, and we describe our training procedure and model evaluation metrics. 

\subsection{Datasets}
\label{sec:data_split}
We split each of the data samples into three sets to form a training set (72\%), a validation set (13\%), and a test set (15\%). This split of the data was chosen to include as many stars as possible in the training sets, while having at least a thousand stars in the validation and test sets to be representative of the entire parameter range. We test a 50\%-25\%-25\% and 90\%-5\%-5\% train-validate-test split for two parameters, $\Delta\nu$ and $P_{\rm rot}$, and find only marginal differences in model performance, with variations in the $r^{2}$ score of the best models being on the order of a few percent. The full \cite{yu18} sample includes 10,755 stars with 7,769 in the training set, 1,372 in the validation set, and 1,614 in the test set. The full \cite{pande18} sample includes 13,439 stars with 9,709 in the training set, 1,714 in the validation set, and 2,016 in the test set. The full \cite{mcquillan14} sample includes 27,001 stars with 19,507 in the training set, 3,443 in the validation set, and 4051 in the test set. Figure \ref{fig:data_density}, in the Appendix, shows the distribution of the stellar properties we learn for each sample, including $\Delta\nu$, $\nu_{\rm max}$, and log $g$ for the \cite{yu18} sample, log $g$ and $T_{\rm eff}$ for the \cite{pande18} sample, and $P_{\rm rot}$ and $M_{*}$ for the \cite{mcquillan14} sample. In forming the training, validation, and test sets, we draw stars evenly from the underlying stellar property distribution.

The training sets are used to train a given model, i.e. learn the optimal weights and biases of the network. The validation sets, which don't contribute to learning the network parameters, are used to evaluate the performance of the network throughout the training process, as well as perform the hyperparameter selection. To prevent overfitting, we implement early stopping based on monitoring the loss of the validation set, which we describe further in Section \ref{sec:training}. The test sets, which are independent of learning the network parameters, and are used to evaluate the performance of the model after training has been terminated. We describe the model evaluation and selection procedure in more detail in Section \ref{sec:evaluation}

For model training we scale the distribution of each stellar property we predict to the range [0, 1] by computing $\vec{Y}_{\rm scaled}$ as, 

\begin{equation} \label{eq:minmax}
\vec{Y}_{\rm scaled} = \frac{\vec{Y} - {\rm min}(\vec{Y})}{{\rm max}(\vec{Y}) - {\rm min}(\vec{Y})},
\medskip
\end{equation}

\noindent where this operation is performed separately for the training, validation, and test sets to prevent information leakage. In this context, information leakage refers to when the distribution of one dataset is incorrectly used to inform the scaling of another dataset, making them no longer independent, which often leads to inflated model performance.

\subsection{Training procedure}
\label{sec:training}
The models are trained using NVIDIA Tesla GPUs. We implement our model architecture and training procedure in the machine learning library \texttt{PyTorch} \citep{pytorch}, which includes the \texttt{nn} module that can be used to define a variety of network architectures, as well as compute model gradients and perform tensor computations with GPU support.

For our training task, to predict continuous stellar properties, the loss function, $\mathcal{L}$, we optimize is the mean square error (MSE), which is a common choice for regression problems. The mean squared difference between true and predicted target value is computed as,

\begin{equation} \label{eq:loss}
{\rm MSE} = \frac{1}{N_{\rm batch}}\sum\limits_{i-1}^N(Y_{i} - \hat{Y}_{i})^2,
\medskip
\end{equation}

\noindent where $N_{\rm batch}$ is the number of data examples in the batch, $Y_{i}$ is the true stellar property of interest, and $\hat{Y}_{i}$ is the predicted stellar property, computed through the series of convolution and fully-connected network operations as described in Section \ref{sec:architecture}.

For each model we train, the training data is batched into sets of $N_{\rm batch}$ = 256 stars. For the \cite{yu18} sample this results in 31 training batches, for the \cite{pande18} sample this results in 38 training batches, and for the \cite{mcquillan14} sample this results in 77 training batches. During each training iteration, which includes the forward and backward pass, one batch of the training data through the network architecture and used to update the model parameters. One epoch of training has been completed once all of the training batches have been passed through the network. Batching the training data reduces the memory requirements during each training iteration, decreases the training time since the weights are updated more frequently, and acts to improve how well the model generalizes to unseen data.

The training procedure, which is typical for neural network models, can be summarized as follows: (1) forward pass of the batch through the network architecture to compute $\vec{\hat{Y}}_{\rm batch}$, (2) compute $\mathcal{L}$ according to Equation \ref{eq:loss}, (3) backpropagation of $\mathcal{L}$ through each layer of the network architecture, (4) compute $\nabla_{\vec{\theta}}\mathcal{L}$, the gradient of the loss function with respect to each model parameter $\vec{\theta}$, (5) update the value of each model parameter to minimize $\mathcal{L}$. Steps 1 through 5 are repeated for every training batch iteration, and the model is trained for $N_{\rm epochs}$ = 800 epochs or until an early stopping criterion is met. During training, we also compute the loss function for a validation set described in Section \ref{sec:data_split}. At the beginning of each training epoch, steps 1 and 2 listed above are carried out on the validation dataset and the loss is monitored as the model trains. Since the validation set is not used to update the model weights, the performance of the model on this dataset is diagnostic of how generalizable the model is to new data. To combat overfitting, we implement an early stopping criterion based on the validation loss as a function of epoch. If the validation loss increases for $N_{\rm stop}$ = 50 epochs within a tolerance of $N_{\rm tol}$ = 10$^{-2}$, then training is terminated and the model parameters before the validation loss increased is saved as the final model. 

To update the values of the model parameters during training (i.e. step 5 above), we use \texttt{PyTorch}'s implementation of the AdamW (Adaptive Moment Estimation) optimization method \citep{adam}, with ``Decoupled Weight Decay Regularization" \citep{adamw}. AdamW is an adaptive learning rate optimization method that computes individual learning rates for each model parameter based on the exponential moving average of the first and second moments of the loss function gradient, $\nabla_{\vec{\theta}}\mathcal{L}$, with two parameters $\beta_{1}$ and $\beta_{2}$, that set the exponential decay parameters for each moment. For the models we train, we fix the exponential decay parameters to their defaults in the original Adam paper of $\beta_{1}$ = 0.9 and $\beta_{2}$ = 0.999. Adam differs from traditional stochastic gradient descent, which uses a single learning rate for all parameters throughout the duration of the training process. Each model parameter, $\theta_{i}$, is updated at timestep $t$:

\begin{equation} \label{eq:adam1}
\theta^{t}_{i} = \theta^{t-1}_{i} - \alpha\frac{\hat{m}^{t-1}_{i}}{\sqrt{\hat{v}^{t-1}_{i}} + \epsilon}
\end{equation}

\noindent where $\epsilon$ = 10$^{-8}$ is typically added to promote numerical stability, $\hat{m}^{t}_{i}$ is the bias corrected exponential average of the first moment of the gradient with respect to parameter $\theta^{t}_{i}$ and $\hat{v}^{t}_{i}$ is the exponential average of the second moment of the gradient with respect to parameter $\theta^{t}_{i}$, both computed as defined in \cite{adam}.

The initial learning rate, $\alpha$, controls the step size at which the model parameters are updated. The optimal learning rate is problem specific, but is typically set in the range of [10$^{-4}$, 10$^{0}$]. Learning rates that are too low can result in training that takes many iterations to find a minimum in the loss function gradient, and without sufficient training time the parameter space may not have been explored sufficiently and a local minimum solution is returned. Learning rates that are too high can overstep the minimum in the loss function gradient and ultimately fail to converge on a desirable solution. As will be described in Section \ref{sec:parameters}, we experiment with three different initial learning rate values, $\alpha$ = [10$^{-5}$, 10$^{-4}$, 10$^{-3}$], and choose the one that leads to the best results for predicting a given stellar property. 

Lastly, to introduce regularization into the optimization routine, we add a weight decay term to the loss function described in Equation \ref{eq:loss}. The weight decay term, $\lambda$||$\theta$||$^{2}$, is a typical L2 regularization that penalizes model parameters that become too large by a factor of $\lambda$. As will be described in the following section, we test two different weight decays parameters, $\lambda$ = [10$^{-5}$, 10$^{-1}$].

\begin{table*}
\begin{centering}
\tabletypesize{\scriptsize}
\caption{Model and training parameters}
\label{tab:nnparams}
\tablewidth{0pt}

\begin{tabular}{rccl}

\hline
 & Parameter &  Value(s)/Setting(s)  &  Description \\ 
\hline
\hline
\vline & $N_{\rm conv}$ & 2 & number of convolutional layers \\
\vline & $N_{K,1}$ & 64 & number of output kernels \\
\vline & $K_{W, 1}$ & 3, 5, 6, 8, 12, 20 & kernel width \\
\vline & $P_{1}$ & 4, 5, 2, 3, 1, 5 & padding \\
\vline & $S_{1}$ & 3, 3, 2, 2, 2, 2 & convolution stride \\
\vline & $T_{pool,1}$ & average & pooling type \\
\vline & $K_{pool,1}$ & 4 & width of pooling kernel \\
\vline & $f_{\rm conv,1}$ & ReLU & convolution activation function \\
\vline & $N_{K,2}$ & 16 & same as above for second convolution \\
\vline & $K_{W, 2}$ & 5, 8, 10, 12, 16, 30 & --- \\
\vline & $P_{2}$ & 1, 0, 0, 1, 1, 1 & --- \\
\vline & $S_{2}$ & 1, 2, 2, 2, 1, 1 & --- \\
Architecture~~~~\vline & $T_{pool,2}$ & average & pooling type \\
\vline & $K_{\rm pool,2}$ & 2 & --- \\
\vline & $f_{\rm conv,2}$ & ReLU & --- \\
\vline & $N_{\rm FC}$ & 3 & number of fully-connected layers \\
\vline & $N_{1}$ & 2048 & number of hidden units in fully-connected layer \\
\vline & $f_{\rm FC,1}$ & ReLU & fully-connected activation function \\
\vline & $D_{\rm FC,1}$ & 0.0, 0.3 & dropout probability applied to fully-connected layer \\
\vline & $N_{2}$ & 1024 & same as above for second fully-connected layer \\
\vline & $f_{\rm FC,2}$ & ReLU & --- \\
\vline & $D_{\rm FC,2}$ & 0.0, 0.3 & --- \\
\vline & $N_{3}$ & 256 & same as above for third fully-connected layer \\
\vline & $f_{\rm FC,3}$ & ReLU & --- \\
\vline & $D_{\rm FC,3}$ & 0.0 & --- \\

\hline
 \vline & optimizer & AdamW & --- \\
\vline & $\alpha$ & 10$^{-5}$, 10$^{-4}$, 10$^{-3}$ & learning rate \\
Optimization~~~~\vline & $\lambda$ & 10$^{-5}$, 10$^{-1}$ & weight decay parameter \\
\vline & $\epsilon$ & 10$^{-8}$, 10$^{-2}$ & numerical stability term \\
 \vline & $\mathcal{L}$ & mean squared error (MSE) & loss function \\

 \hline
 \vline & $N_{\rm batch}$ & 256 & training batch size \\
Training~~~~\vline & $N_{\rm epochs}$ & 800 & maximum number of training epochs \\
 \vline & $N_{\rm stop}$ & 50 & number of epochs to stop training if no improvement \\
 \vline & $N_{\rm tol}$ & 10$^{-2}$ & early stopping tolerance \\

\hline
\hline
\vspace{-2ex}

%\medskip

\end{tabular}

\end{centering}
\end{table*}

\subsection{Model Hyperparameters}
\label{sec:parameters}
As is evident in Table \ref{tab:nnparams}, there are numerous parameters that must be set to define the model architecture as well as the training procedure. These ``hyperparameters" are parameters whose values are determined before training begins, and are not updated through the course of the training process. Since the dimensionality of the hyperparameter space is large, it is not feasible to evaluate all possible hyperparameter combinations and the effect each has on model performance. However, as an improvement beyond choosing ad-hoc or values selected empirically, we heuristically choose a small set of hyperparameters that are varied systematically, and perform a grid search over the combinations. We train one model with each hyperparameter combination defined in the grid, and select the preferred hyperparameter values based on how the model performs on the validation dataset. Limiting the search to just five hyperparameters, we test varying $K_{W}$, $\alpha$, $\lambda$, $\epsilon$, and $D_{\rm FC}$. We define a grid over the values of these parameters, and train a model with each combination of hyperparameters. For each run, all other architecture and training hyperparameters are set to the values listed in Table \ref{tab:nnparams}. 

As described in Section \ref{sec:training}, the optimal learning rate is problem specific and the consequences for choosing too low or high of a rate can result is poor model performance. Therefore, we experiment with three values for the initial learning rate, $\alpha$ = [10$^{-5}$, 10$^{-4}$, 10$^{-3}$]. We also experiment with two values for the weight decay parameter, $\lambda$ = [10$^{-5}$, 10$^{-1}$]. We prioritize varying this parameter because the amount of regularization in the optimization procedure directly impacts the values of the model parameters and controls how well the model generalizes to unseen data. We also experiment with two values of the numerical stability term $\epsilon$, testing values of both 10$^{-8}$ and 10$^{-2}$.

In addition to the two optimization-related hyperparameters, we also test varying one of the model architecture parameters. Motivated by our physical understanding of how information about different stellar properties are encoded at different timescales in the light curves, we decide to test varying the kernel widths of the convolution layers. Presumably, smaller kernel widths are more sensitive to information encoded on shorter timescales, while larger kernel widths will pick up information imprinted on longer timescales. We choose 8 different kernel widths to test for the first convolution layer, $K_{W, 1}$ = [3, 5, 6, 8, 12, 20], which corresponds to convolution over timescales of $t_{\rm conv,1}$ = [.061, .102, .123, .163, .245, .408] days respectively. For the second convolution layer we choose larger kernel widths, with $K_{W, 2}$ = [5, 8, 10, 12, 16, 30], where each element in $K_{W, 2}$ is paired with its corresponding element in $K_{W, 1}$. After the second kernel is applied, these kernel widths result in time series that are convoluted over timescales of $t_{\rm conv,2}$ = [.306, .817, 1.23, 1.96, 3.92, 12.25] days respectively. To ensure that each convolution and pooling operation results in an integer number of output data elements, we modify the zero-padding and stride parameters for each layer as necessary. The values for $P_{1}$, $P_{2}$, $S_{1}$, and $S_{2}$ for each element of $K_{W, 1}$ and $K_{W, 2}$ is listed in Table \ref{tab:nnparams}.

\subsection{Model evaluation and selection}
\label{sec:evaluation}
As described in Section \ref{sec:data_split}, we select the best model (over the grid of hyperparameters tested) based on the models performance on the validation set. The validation set is not used to train the model, and thus yields a more realistic report of how the model performs on unseen data. To assess the performance of a given model, we compute three evaluation metrics: the coefficient of determination ($r^{2}$), the bias ($\Delta$) and the rms. The $r^{2}$ score is computed as,

\begin{equation} \label{eq:r2}
r^{2} = 1 - \frac{1}{N\sigma^{2}}\sum_{i}\Big({Y_{i} - \hat{Y_{i}}\Big)^{2}},
\end{equation}

\noindent where $Y$ and $\hat{Y}$ are the true and model predicted values of the dependent variable, $N$ is the number of observations in the validation or test set, and $\sigma^{2}$ is the variance of $\vec{Y}$. An $r^{2}$ score closer to 1 indicates that the model predicts the variation in $Y$ well, whereas an $r^{2}$ score of 0 indicates that the model does not capture any of the variation. The bias and root mean square of the estimator are computed as,

\begin{equation} \label{eq:bias}
\Delta = \frac{1}{N}\sum_{i}\Big(\hat{Y_{i}} - Y_{i}\Big),
\end{equation}

and,

\begin{equation} \label{eq:rms}
\mathrm{rms} = \Bigg[\frac{1}{N}\sum_{i}\Big(Y_{i} - \hat{Y_{i}}\Big)^{2}\Bigg]^{\frac{1}{2}},
\end{equation}

\noindent respectively. Both of these metrics are in the units of the stellar property, $Y$, where less bias and smaller rms values both indicate better model performance.

For each stellar property that we predict for the \cite{yu18}, \cite{pande18}, and \cite{mcquillan14} samples, we train 144 models with the hyperparameter choices described in Section \ref{sec:parameters}. Each of these models is trained according to the procedure outline in Section \ref{sec:training}, with the model architecture described in Section \ref{sec:architecture}. Based on the same validation set for each stellar property, we compute the $r^{2}$, $\Delta$, and rms for each of the 144 models we train to predict the property. For each stellar property, we select the best model according to a two-step procedure. If possible, we first we eliminate all models with bias values greater than 10\% of the mean or rms values greater than 50\% of the standard deviation of the stellar property validation set distribution, described as follows:

\begin{equation} \label{eq:bias_cut}
\Delta~\leq~0.1\times\frac{\sum_{i}Y_{i}}{N}, 
\end{equation}

\begin{equation} \label{eq:rms_cut}
{\rm rms}~\leq~0.5\times\sqrt{\frac{1}{N}\sum_{i}(Y_{i}-\overline{Y})^{2}}, 
\end{equation}

\noindent after eliminating models that don't meet both of the criteria above, we then rank the models according to their $r^{2}$ scores. We then visually inspect the performance of the top 10 models trained for each property, and select the model with the highest $r^{2}$ score that doesn't exhibit structure in the true versus predicted plots for the validation set. As evident by comparing Equations \ref{eq:r2}, \ref{eq:bias}, and \ref{eq:rms}, the three evaluation metrics are closely related, so a high $r^{2}$ score is correlated with small $\Delta$ and rms values. Depending on the specific use case of the stellar property predictions, this model selection process can be easily modified to emphasize a particular or different evaluation metric. The final performance results we show in the following sections are based on the test set performance, which is data that was not used to train, validate, or select the best model.

%%%%%%%%%%%%%%%%%%%%%%%%%%%%%%%%%%% CLASSIFICATION %%%%%%%%%%%%%%%%%%%%%%%%%%%%%%%%%%
\section{Classification of evolutionary state}
\label{sec:classification}
Before attempting the regression problem described in Section \ref{sec:methods}, we start with the broader task of predicting a star's evolutionary state based on its light curve. By determining a star's general location on the HR diagram, this classification task serves as an initial probe of our modeling capabilities, before we move on to the task of predicting continuous (as opposed to categorical) stellar properties. This classification model also has the utility to be used a front-end to an automated stellar property derivation pipeline.

\subsection{Data}
To train the classification model we build a dataset based on the overlap between the stars listed in the \cite{berger18} catalog and the stars with \textit{Kepler} Q9 light curves, which includes $\sim$150,000 stars as shown in Figure \ref{fig:catalog_coverage}. Stars in the \cite{berger18} catalog are classified into three evolutionary states; main sequence, sub-giant, or RGB, based on fitting solar-metallicity evolutionary tracks to the transition between the end of the main sequence and start of the RGB in the temperature-stellar radius plane as shown in Figure 5 of \cite{berger18}. Of the total catalog, 67\% of stars are classified as main sequence stars, 21\% as sub-giant stars, and 12\% as RGB stars. We randomly sample the same number of stars from the three classes to ensure a balanced classification problem, with the dataset including 13,355 stars from each the main sequence, sub-giant branch, and RGB (which includes red clump stars), totaling 40,065 stars. We split this dataset into three parts as described in Section \ref{sec:data_split}, which results in 28,945 stars in the training set, 5,109 stars in the validation set, and 6,010 stars in the test set.

\subsection{Methods}
For the classification problem we make two main modifications to the model, one to the model architecture described in Section \ref{sec:architecture} and one to the training procedure described in Section \ref{sec:training}. First, instead of the output of the final fully-connected layer of the model being a single value (as shown in Figure \ref{fig:nn}), for the classification problem the output of the model is equal to the number of distinct classes $C$ (in this case, $C$=3). To convert the model output to a prediction probability over classes we apply the softmax function, $\sigma(y)_{i}$ = $e^{y_{i}}$/$\sum_{j=1}^{C} e^{y_{i}}$, and assign each star to the class with the highest probability. The second change we make is to the loss function. Instead of computing the mean square error described by Equation \ref{eq:loss}, we instead compute the cross entropy loss which is appropriate for training multi-class classification problems. The cross entropy over $C$ classes is computed as,

\begin{equation} \label{eq:ce}
CE = \sum^{C}_{j = 0} \Big[- Y_{c}+ \mathrm{log}\big(\sum^{C}_{j=0} \mathrm{exp}~{\hat{Y}_{j}} \big)\Big],
\end{equation}

\noindent where the first term, $Y_{c}$, is the indicator variable of the star's true class membership, and the second term is the log of the sum of the un-normalized class probabilities $\hat{Y}_{j}$ over the $C$ classes output from the model. In addition to the above changes to the model architecture and loss function, we also evaluate model performance with metrics that are relevant for classification models, which are different than the metrics used in Section \ref{sec:evaluation}. We focus on three performance metrics; the accuracy, average precision, and the area under the receiver operator curve. The multi-class accuracy, which quantifies the number of correct predictions averaged across $C$ classes, is computed as, 

\begin{equation} \label{eq:accuracy}
{\rm accuracy} = \frac{1}{N}\sum^{C}_{j=0} (TP_{j} + TN_{j}),
\end{equation}

\noindent where $N$ is the total number of stars in the validation or test set, $TP_{j}$ is the number of stars correctly identified as belonging to class $j$ (i.e. true positives), $TN_{j}$ is the number of stars correctly identified as not belonging to class $j$ (i.e. true negatives). An accuracy closer to unity indicates better model performance. 

\begin{figure*}[ht]
\centering
\includegraphics[width=2\columnwidth]{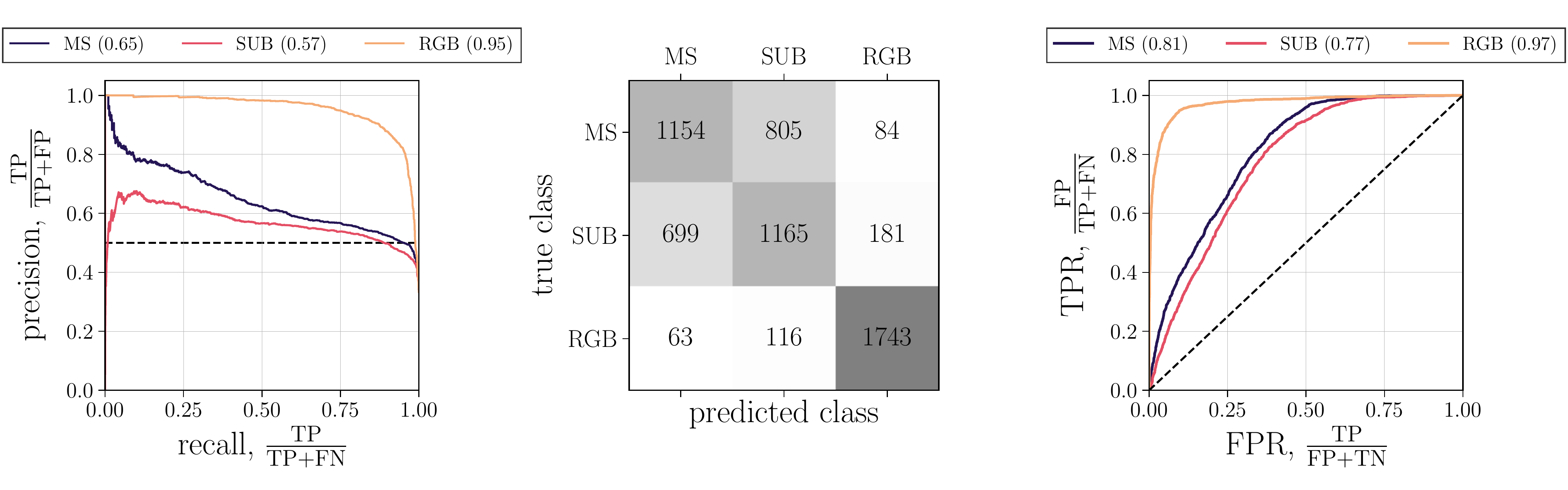}
\caption{Evolutionary state classification performance of the CNN model for a test set of stars. \textit{\textbf{Left panel:}} the precision-recall curve showing the one versus rest classification of stars as belonging to the main sequence (MS), sub-giant branch (SUB), or RGB. \textit{\textbf{Middle panel:}} confusion matrix showing the number of false positives and false negatives for each class on the off-diagonal entries. \textit{\textbf{Right panel:}} one versus rest receiver operating characteristic curve for each evolutionary state. \bigskip}
\label{fig:classification}
\end{figure*}

We also compute the average precision across classes (in a one-versus-rest manner), which summarizes the precision-recall curve. Given the class probabilities output by the model as described above, different probability thresholds can be placed to define the boundary between the classes. Precision, defined as $P$ = $TP/(TP+FP)$, where $FP$ is the number of false positives, measures how many correct predictions are made for stars belonging to a certain class at a given threshold. Recall, $R$ = $TP/(TP+FN)$, where $FN$ is the number of false negatives, measures how many stars belonging to a classes are recovered from the total population of that class. The precision-recall curve describes the trade-off between precision and recall at different class threshold boundaries, with the best threshold being one that produces both a high precision and high recall. The average precision, computed as,

\begin{equation} \label{eq:accuracy}
\mathrm{AP} = \sum_{n} P_{n}(R_{n} - R_{n-1}),
\end{equation}

\noindent where $P_{n}$, $R_{n}$, and $R_{n-1}$ are the precision and recall values at the $n$th and $n$th-1 probability thresholds, is the weighted mean of precisions at each recall threshold, with AP closer to unity indicating better model performance. In addition to accuracy and average precision, we also measure model performance by computing the area under the receiver operator characteristic (AUROC) curve. At different probability thresholds, the ROC curve shows the true positive rate, $TPR = TP/(TP+FP)$ (i.e. the recall), as a function of the false positive rate, $FPR = FP/(FP+TN)$, which describes the number of stars incorrectly classified as belonging to a class relative to the total number of stars that do not belong to the class. Models with low FPRs and higher TPRs indicate good performance, which corresponds to an area under the ROC curve closer to unity.

While the various classification metrics described above are related, they each emphasize different aspects of the model performance. The accuracy is the most general, measuring the fraction of total correct predictions. While this is a good overall metric of model performance, for more specific use cases of the predictions it is often not detailed enough. The precision captures how often the model is correct when the model predicts a specific class instance, which is relevant when the consequences of a false positive prediction are high. On the other hand the recall, which captures the fraction of a class that is correctly identified, is relevant when the consequences of a false negative prediction are high. Depending on the specific application of the classification model, it can be important to consider these different metrics together, and not only the accuracy alone.  For example, if the classifier is used to select targets for follow up spectroscopy of one class, a classifier with high precision, but with low recall, would lead to an inefficient observing program.

For the classification problem, we perform the same hyperparameter grid search as described in Section \ref{sec:parameters}, training a total of 144 models. In the following section we report all of the metrics described above, however, since our goal here is to demonstrate the general performance of the classification model, we select the best model based on which hyperparameter combination results in the best overall accuracy on the validation set. The model performance reported in the next section is on the independent test set. 

\subsection{Results}
Figure \ref{fig:classification} shows the performance of the best model we train, evaluated using the metrics described above, to classify stars as main sequence, sub-giant, or RBG based on their light curves. The middle panel of the figure shows the confusion matrix, with the true class labels along the $y$-axis and the predicted class labels along the $x$-axis. Stars that fall into the diagonal bins are correctly classified by the model, while the stars that fall into the off-diagonal bins are incorrectly classified. As evident by the confusion matrix, the model performs the best at distinguishing RGB stars from the other evolutionary states, at an accuracy of 91\%. For the remaining true RGB stars, 3\% are misclassified as main sequence stars and 6\% are misclassified as sub-giant stars. Examining the predictions for main sequence stars, 56\% are correctly classified, while 40\% of main sequence stars are misclassified as sub-giants and 4\% as RGB. For the sub-giant stars, only 57\% are classified correctly, while 34\% are incorrectly classified as main sequence stars and 9\% are misclassified as RGB stars.

The precision-recall and ROC curves in Figure \ref{fig:classification} show how varying the class discrimination threshold based on the prediction probabilities results in classifiers with different performance properties. The precision-recall curve shows that the RGB stars are clearly separable from the main sequence and sub-giant stars, with high precision values maintained at most recall thresholds resulting in a average precision of AP = 0.95. The main sequence and sub-giant stars exhibit worse performance, with average precisions of AP = 0.65 and AP = 0.57, respectively. The ROC curve shows similar behavior with regards to the classification performance. The TPR for the RGB stars is high across nearly the entire range of FPR thresholds, with an AUROC = 0.97. For the main sequence and sub-giant branch stars, high TPRs are only achieved along with higher FPRs. The TPR of main sequence stars reaches $\sim$0.95 at FPRs greater than 0.5, with an AUROC = 0.81, and the TPR of sub-giant stars reaches $\sim$0.95 at FPRs greater than 0.6, with an AUROC = 0.77. This is still better than the performance of a random classifier, characterized by an AUROC = 0.5. 

To summarize, we find that the model does well at distinguishing between main sequence and RGB stars, but mixes up the identification of a significant portion of main sequence and sub-giant branch stars. The performance of the classification model we train likely reflects that the light curves of main sequence and RGB stars vary enough to be informative as to these evolutionary states, but that light curves vary across large regions of the HR diagram in a continuous (rather than discrete) manner. Part of the reason for the poorer results may also be related to the quality of classifications. Due to the lack of spectroscopy, \cite{berger18} used solar-metallicity isochrones to separate evolutionary stages, which will introduce significant noise since the exact border between main-sequence and sub-giant stars is sensitive to metallicity. In contrast, the sub-giants and red giants are clearly separated by luminosity with relatively small dependence on metallicity, thus yielding more accurate classifications.

%%%%%%%%%%%%%%%%%%%%%%%%%%%%%%%%%%% REGRESSION %%%%%%%%%%%%%%%%%%%%%%%%%%%%%%%%%%
\section{Predicting stellar properties}
\label{sec:regression}
\subsection{Results of CNN stellar property recovery}
\label{sec:cnn_results}
In the previous section, we demonstrated the potential of using a 1D CNN model in the time domain to classify a star's evolutionary state. We now turn our attention to the main goal of this paper, which is to predict stellar properties from light curve data. As shown in Figure \ref{fig:catalog_coverage}, there are many possible training sets that can be constructed to predict a variety of stellar properties given the catalogs that are available in the literature. Here, we focus on the three catalogs with the large numbers of stars available: the \cite{yu18} catalog which includes 10,757 stars, the \cite{pande18} catalog with includes 13,441 stars, and the \cite{mcquillan14} catalog which includes 32,920 stars. We split each of these stellar samples into a training set, a validation set, and a test set as described in Section \ref{sec:data_split}, and train individual models to predict each sample and stellar property combination according to the procedure described in Section \ref{sec:training}. We perform the hyperparameter search as described in Section \ref{sec:parameters}, and for each parameter we present the predictions resulting from the best of the 144 models trained, selected as described in Section \ref{sec:evaluation}.

\begin{figure*}[!htb]
\centering
\includegraphics[width=1.8\columnwidth]{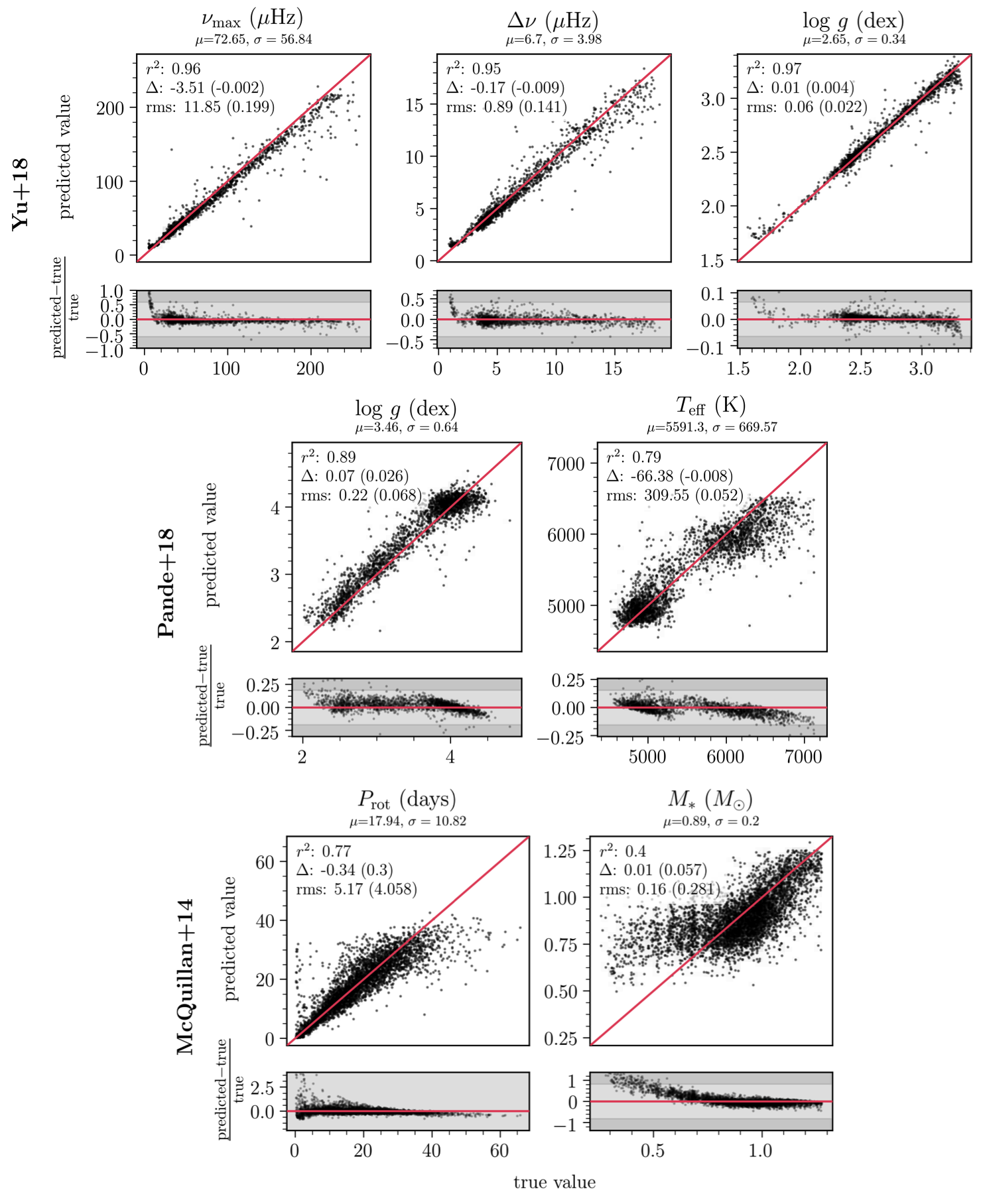}
\caption{Performance of the best CNN model (as selected in Section \ref{sec:evaluation}) in predicting $\nu_{\rm max}$, $\Delta\nu$ and log $g$ for the test set of \cite{yu18} RGB stars, log $g$ and $T_{\rm eff}$ for the test set of \cite{pande18} stars, and $P_{\rm rot}$ and $M_{*}$ for the test set of \cite{mcquillan14} stars. For each predicted stellar property, we show both the predicted values (top panels) and the fractional difference between the predicted and true values (bottom panels) as a function of the true stellar property. The $r^{2}$, $\Delta$, and rms of the predictions are indicated in each panel, as well as the bias and rms of the fractional differences shown in parenthesis. We also report the mean ($\mu$) and standard deviation ($\sigma$) of the distribution of the true property values. The red line indicates a perfect prediction and the shaded regions in the bottom panels indicate the standard deviations of the fractional difference (3$\sigma$ light grey, 5$\sigma$ dark grey). For $P_{\rm rot}$, the scale of the $y$-axis is from -0.5$\sigma$ to 1$\sigma$ to show the prediction quality across the entire range of true values. We note that the fractional metrics for the $P_{\rm rot}$ predictions is greatly inflated by the over-prediction of short period stars. If we remove these stars with fractional differences \textgreater~2.5, the fractional bias and fractional rms become 0.016 and 0.30 respectively. \bigskip \bigskip \bigskip}
\label{fig:regression_results}
\end{figure*}

Figure \ref{fig:regression_results} shows the stellar property predictions for each sample's test set derived from the best trained models. For each stellar property, we show in the top panel the true stellar property value versus the model predicted stellar property value, where the one-to-one line indicates a perfect prediction. In the bottom panels, we show the fractional difference between the model predicted and the true stellar property values, as a function of the true values. The bottom panel therefore more clearly highlights the parameter space where the model is biased. In this panel we also indicate the regions of 3 and 5 standard deviations from a perfect prediction (expect for the $P_{\rm rot}$ panel which shows -0.5$\sigma$ to 1$\sigma$), when the fractional difference is equal to zero. For each stellar property the mean ($\mu$) and standard deviation ($\sigma$) of the true test set values are also indicated, as well as the model evaluation metrics, $r^{2}$, $\Delta$, and rms, and the fractional bias and rms in parenthesis. 

First, we examine the predicted stellar properties based on the \cite{yu18} RGB stellar sample, showing the $\nu_{\rm max}$, $\Delta\nu$, and log $g$ predictions in the top row of Figure \ref{fig:regression_results}. As seen in the figure, we find that we recover all three of these stellar properties well, with $r^{2}$ scores greater than 0.95. Demonstrating the importance of the hyperparameter search, the worst performing models for these three parameters result in $r^{2}$ values of $\sim$0.8 - 0.85. Examining the best $\nu_{\rm max}$ model, the overall bias of $\Delta$ = -3.5 $\mu$Hz is $\sim$5\% of the mean of the true test set values, while the rms of the predictions is 11.85 $\mu$Hz, over the range of $\nu_{\rm max}$ values from 5 to 250 $\mu$Hz. Considering the prediction quality as a function of $\nu_{\rm max}$, we see that the predictions of $\nu_{\rm max}$ values less than $\sim$10 $\mu$Hz and greater than $\sim$150 $\mu$Hz are more biased. This is seen most clearly in the bottom panel of Figure \ref{fig:regression_results} which shows the fractional difference, with some predictions falling in the 5$\sigma$ range (and 8 examples comprising 0.5\% of the test set fall outside the plot limits). As seen in Figure \ref{fig:data_density} in the Appendix, the \cite{yu18} sample includes far fewer stars with these smaller and larger $\nu_{\rm max}$ values. This means that there are fewer examples for the model to learn this region of the parameter space well during training.

Similar to the $\nu_{\rm max}$ prediction, $\Delta\nu$ for the \cite{yu18} sample is also recovered well. The overall bias of $\Delta$ = -0.17 $\mu$Hz is $\sim$2.5\% of the mean of the true test set values. The rms of the predictions is 0.89 $\mu$Hz over the range of $\nu_{\rm max}$ values from 0.9 to 18.8 $\mu$Hz. Similarly to $\nu_{\rm max}$, $\Delta\nu$ is biased for the smallest and largest values. For $\Delta\nu$ values less than $\sim$2 $\mu$Hz and greater than $\sim$12 $\mu$Hz, the predictions are more biased. This is seen most clearly in fractional difference plot, with a few predictions falling in the 5$\sigma$ range (and 8 examples comprising 0.5\% of the test set fall outside the plot limits). Again, as with the $\nu_{\rm max}$, the \cite{yu18} sample includes far fewer stars with these smaller and larger $\Delta\nu$ values, which means that there are fewer examples for the model to learn this region of the parameter space well at training time.

The final stellar property we predict for the \cite{yu18} stellar sample is log $g$. This property is also recovered well by the best trained model, with an overall bias of $\Delta$ = 0.01 dex and an rms of the predictions of 0.06 dex, over the range of log $g$ values from 1.6 to 3.3 dex. Considering the prediction quality across the range of log $g$ values, we find again find that the predictions are more biased in the parameter space regions with fewer representative stars in the training set, as shown in Figure \ref{fig:data_density} of the Appendix. As evident in the bottom panel of Figure \ref{fig:regression_results}, there is more bias in the predictions for stars with log $g$ values less than 2 dex and also greater than 3.2 dex (and 11 examples comprising 0.7\% of the test set fall outside the plot limits).

We now discuss the property recovery for the \cite{pande18} stellar sample, which, as shown in Figure \ref{fig:data}, includes stars from the RBG as well as the sub-giant branch and the upper main sequence. The first property we consider is log $g$. As seen in Figure \ref{fig:regression_results}, the best CNN model recovers log $g$ with an $r^{2}$ score of 0.89, an overall bias of $\Delta$ = 0.07 dex, which is $\sim$2\% of the mean of the true test set values, and an rms of 0.22 dex over the range of log $g$ values from 2 to 4.8 dex. In the fractional difference plot (which excludes 6 examples comprising 0.3\% of the test set given the axis limits), we see that for log $g$ values less than $\sim$3.75 dex, there is a systematic positive bias. This is perhaps cause by the model trying to correctly predict the larger number of less evolved stars (log $g$ $\sim$ 4) at the expense of biasing the log $g$ predictions for the red giants. Compared to the recovery of log $g$ for the \cite{yu18} sample of RGB stars, log $g$ is recovered less precisely for the \cite{pande18} sample, as evident by both the difference in $r^{2}$ scores between the two models, 0.89 for the \cite{pande18} versus 0.97 for \cite{yu18}, as well as the higher rms of the \cite{pande18} model at an rms of 0.22 dex, compared to 0.06 dex for the \cite{yu18} sample. One reason for the difference in log $g$ prediction quality between these two stellar samples is the precision of the stellar properties used to train the models. As mentioned in Section \ref{sec:data_params}, \cite{yu18} use asteroseismology with an uncertainty of 0.01 dex for the derived log $g$ values, while the reported uncertainty on the \cite{pande18} log $g$ values based on granulation is much higher, at $\sim$0.25 dex.

The other property we successfully predict for the \cite{pande18} stellar sample is $T_{\rm eff}$. With an $r^{2}$ score of 0.79, the bias of the best $T_{\rm eff}$ model is $\Delta$ = -66.4 K, which is $\sim$1\% of the mean of the true test set values, and the rms is 310 K over a range of temperature values from 4520 K to 7123 K. As seen in the fractional difference plot of Figure \ref{fig:regression_results} (which excludes 2 examples comprising 0.1\% of the test set), the prediction quality varies across the range of values for both modes of the $T_{\rm eff}$ distribution. For the cluster of stars with $T_{\rm eff}$ $\sim$ 5000 K, the bias is larger at both cooler and hotter temperatures, and similarly for the cluster of stars with $T_{\rm eff}$ \textgreater~5500 K. Of the properties we've discussed so far, including both the \cite{yu18} and \cite{pande18}, the prediction of $T_{\rm eff}$ is the least precise, achieving and $r^{2}$ score of $\sim$0.8 compared to $r^{2}$ scores greater than 0.9 achieved for log $g$, $\nu_{\rm max}$ and $\Delta\nu$. This is expected due to the more indirect relation of $T_{\rm eff}$ to the physical processes causing brightness variations. Granulation and oscillation amplitudes are predominantly determined by evolutionary state (such as log $g$, radius and luminosity), which are only indirectly traced by the effective temperature of a star. This is particularly the case for main sequence and sub-giant stars, which can have a wide range of temperatures for a given log $g$. This is also consistent with larger spread towards hotter $T_{\rm eff}$ in Figure \ref{fig:regression_results}.

Finally, the last stellar sample we make predictions for is the \cite{mcquillan14} sample, which as shown in Figure \ref{fig:data} includes stars from across the main sequence with temperatures ranging from $T_{\rm eff}$ = 3500 - 7000 K. The first stellar property we consider for this sample is rotation period, which, as discussed in Section \ref{sec:introduction}, is of particular interest for its potential use as a probe of stellar age. With an $r^{2}$ score of 0.77, the bias of the best $P_{\rm rot}$ model is $\Delta$ = -0.34 days, which is $\sim$2\% of the mean of the true test set values, and the rms is $\sim$5 days over the range of periods from 0.2 to 66 days. In the bottom panel of Figure \ref{fig:regression_results} we show the fractional difference of the predictions as a function of $P_{\rm rot}$ spanning -0.5$\sigma$ to 1$\sigma$ from the line of perfect prediction, which excludes 35 stars comprising 0.9\% of the test set. These excluded stars are fast rotators, for which we see that the prediction quality for stars with $P_{\rm rot}$ $\lessapprox$~5 days is the most biased. The large reported fractional metrics are inflated by the short rotation period stars that the model greatly over-predicts. Examining the sample of stars with the highest fractional differences, we find that 49 stars have fractional differences larger than 2.5, all of which have true rotation periods \textless~6.2 days. These 49 stars comprise $\sim$8\% of the stars in the test set with $P_{\rm rot}$ \textless~6.2 days. If we remove these stars from the fractional bias and rms calculations, these metrics become 0.016 and 0.30 respectively. We suspect that most of the short period stars that the model over-predicts could be binary systems whose rotation periods, as measured in their light curves, does not reflect the true rotation periods of the stars.

In Figure \ref{fig:regression_results} we also see that the predictions of $P_{\rm rot}$ values greater than $\sim$35 days become increasingly more biased. As with the predicted stellar properties for the \cite{yu18} and \cite{pande18} catalogs, a potential reason for this behavior of the model is that there are simply fewer examples of stars in the \cite{mcquillan14} catalog with these longer rotation periods, and therefore examples for the model to learn from and be able to sufficiently learn this region of the parameter space. Since deriving stellar rotation periods is of special interest in light of upcoming photometric surveys, in Section \ref{sec:rotation} we investigate the ability to recover rotation periods from both shorter baseline and longer cadence time series data.

The other property we predict for the \cite{mcquillan14} sample is $M_{*}$. As seen in Figure \ref{fig:regression_results}, the model predicts stellar mass well only at the upper mass range, $M_{*}$ $>$ 0.8 $M_{\odot}$, resulting in an $r^{2}$ of 0.7. While the bias of the model is only $\Delta$ = 0.01 $M_{\odot}$, which is $\sim$1\% of the mean of the true test set values, the rms of 0.16 $M_{\odot}$ is large compared to the range of masses covered, from 0.26 to 1.28 $M_{\odot}$. The fractional difference plot excludes 18 of the low stellar mass stars, comprising 0.4\% of the test set, where the fractional bias of the predictions is large. We note that where the model does poorly, at $M_{*}$ \textless~0.7 $M_{\odot}$, the density distribution of this property is underrepresented in the training objects, as seen in Figure \ref{fig:data_density} of the Appendix. Of the properties we present, our recovery of $M_{*}$ is the least successful. One reason for this could be the high fractional uncertainties associated with the $M_{*}$ values ($\sim$12\%), which were derived without the use of \textit{Gaia} parallaxes. Another factor could be that the light curve data alone is not sufficient to predict this property, and perhaps adding additional information to the model, like \textit{Gaia} distances, may improve the recovery.

\subsection{Comparison to modeling in the ACF and frequency domains}
\label{sec:comparison}
\begin{figure*}[tp!]
\centering
\includegraphics[width=1.8\columnwidth]{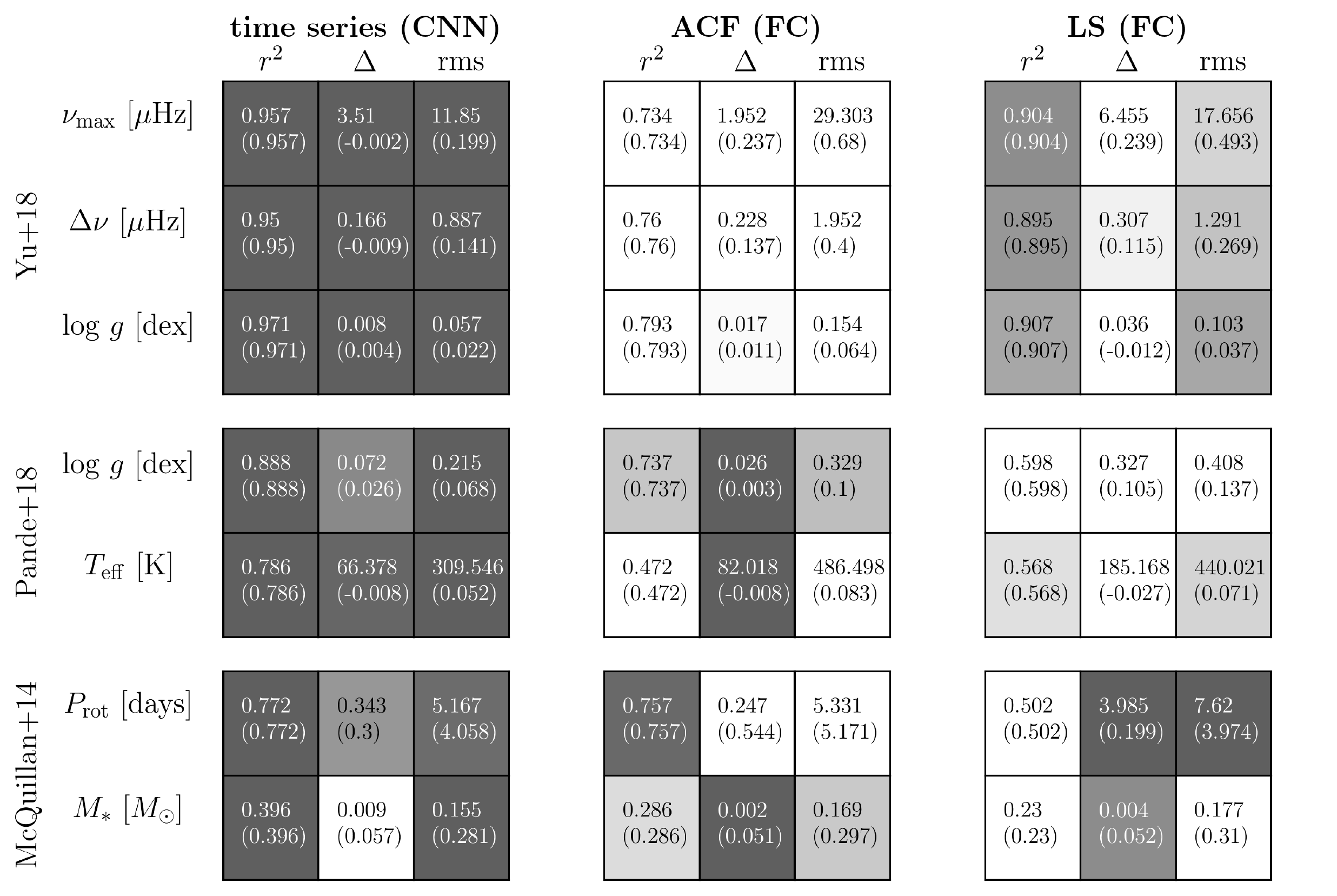}
\caption{Model performance comparison based on different transformations of the light curve data, considering the time series, ACF, and LS periodogram from left to right. For each stellar sample, the predicted stellar properties are shown along the rows, while the performance metrics ($r^{2}$, $\Delta$, rms) are shown along the columns. The fractional bias and fractional rms are indicated in parenthesis. Entries shaded in dark grey indicate better model performance, according to the $r^{2}$ score, fractional bias, and fractional rms. \bigskip}
\label{fig:model_comparison}
\end{figure*}

As discussed in Section \ref{sec:model}, when deriving stellar properties from photometric time series data, the light curves are often first transformed to an alternate representation of the original data. Two common representation are the ACF as described in Section \ref{sec:acf_space} and the power spectrum as described in Section \ref{sec:frequency_space}. Each of these representations highlights different features of the data, which are known to correlate with particular stellar properties. For example, peaks in the ACF are informative to stellar rotation periods, and the asteroseismic parameters of $\nu_{\rm max}$ and $\Delta\nu$ are defined in the frequency domain. One aim of this paper is to investigate how well a deep learning approach can learn various stellar properties from the time domain data itself, because it requires minimal feature engineering and leverages the full information content of the data. 

To test how well we learn stellar properties in the time domain compared to the ACF and frequency domains, for each of the properties we predict in Section \ref{sec:cnn_results} we also train models to predict these properties based on the ACF and the LS periodogram. We do this for the three stellar samples of (\cite{yu18}, \cite{pande18}, and \cite{mcquillan14}), as described in Section \ref{sec:acf_space} and \ref{sec:frequency_space}, respectively. Since the ACF and frequency domain already capture the time dependence of the data, we build and train fully-connected neural network models to predict stellar properties from these data representations. This means that unlike in the CNN case, which includes weight sharing to capture the time-dependence of the input data, in the fully-connected model a weight term is learned for each element of the input. Appropriately, we scale each $n^{th}$ element of the ACF and periodogram relative to the range of values exhibited by the corresponding $n^{th}$ element across all of the stars in the training set. The last four layers represented in Figure \ref{fig:nn} show the fully-connected architecture we implement, where each flux measurement of the light curve is passed to its own hidden node in the first model layer, each with its own weight term.

For these models, instead of implementing different kernel widths, as for the CNN, the architecture hyperparameter we search over is the number of hidden layers, as well as the number of hidden units in each layer. We test four different model architectures; two with three hidden layers consisting of [$N_{1}$=2048, $N_{2}$=1024, $N_{3}$=256] and [$N_{1}$=4096, $N_{2}$=1024, $N_{3}$=256] hidden units, as well as two with two hidden layers consisting of [$N_{2}$=1024, $N_{3}$=256] and [$N_{2}$=2048, $N_{3}$=512] hidden units. The other hyperparameters we optimize over are the same as those in Section \ref{sec:cnn_results}. These are the learning rate, weight decay, numerical stability term, and the dropout fraction. With the architecture choices described above, for each stellar property we train 96 models in total and select the best model as outlined in Section \ref{sec:evaluation}.

\begin{figure*}[htp!]
\centering
\includegraphics[width=2\columnwidth]{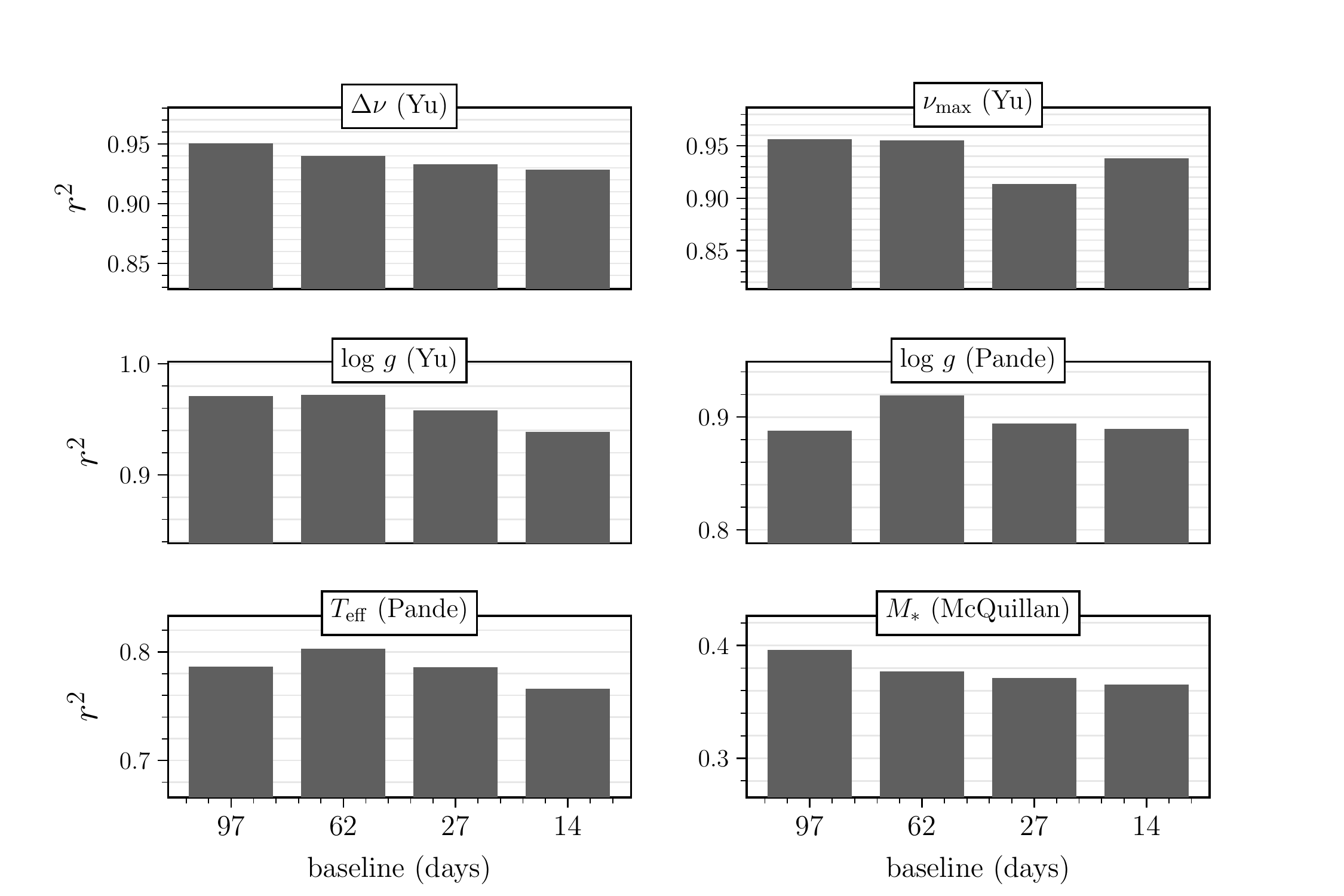}
\caption{Performance of the CNN model as a function of light curve baseline for each of the stellar properties examined in Section \ref{sec:regression} (except $P_{\rm rot}$, which is discussed in Section \ref{sec:rotation}). In each panel the bar chart shows how the $r^{2}$ score of the best model varies with baselines of 97, 62, 27, and 14 days. The bias, rms, fractional bias, and fractional rms of each of these models is reported in Table \ref{tab:shortbaseline} in the Appendix. \bigskip}
\label{fig:all_baseline}
\end{figure*}

For the stellar properties we consider in Section \ref{sec:cnn_results}, Figure \ref{fig:model_comparison} compares the test performance of the best model, $r^{2}$, $\Delta$, and rms, for the three models we train: a CNN based on the time series data, a fully-connected NN based on the ACF, and a fully-connected NN based on the LS periodogram. Compared across the three models for each property, better model performance for the evaluation metrics is indicated by darker shades of its entry in Figure \ref{fig:model_comparison}, where the fractional values of the bias and rms (in parenthesis) are used to indicate model performance. In summary, the CNN model results in the best overall performance. As detailed in Section \ref{sec:cnn_results}, taking a CNN approach, the $\nu_{\rm max}$, $\Delta\nu$, and log $g$ are recovered the most successfully with $r^{2}$ \textgreater~0.9, rotation period and $T_{\rm eff}$ are recovered to $r^{2}\sim$0.8, and stellar mass is recovered the least successfully with an $r^{2}$ = 0.4.

\subsection{Short baseline predictions}
\label{sec:all_short_baseline}

We now explore the prospect of deriving stellar properties from shorter baseline data using the 1D CNN model in the time domain. As discussed in Section \ref{sec:introduction}, ongoing and upcoming photometric missions such as TESS and LSST will observe stars with different baselines of observations, most of which will be shorter than 97 days. In particular, the TESS mission is delivering thousands of stellar light curves with a baseline of 27 days. With the application in mind of being able to estimate stellar properties from this short baseline data, in this section we train CNN models to predict stellar properties using baselines of 62, 27, and 14 days. We do this by truncating the \textit{Kepler} light curves to these shorter baselines.

For the shorter baseline models, we use the same sample of \cite{yu18}, \cite{pande18}, and \cite{mcquillan14} stars, described in Section \ref{sec:data_params}. We simply truncate the light curves to each baseline length, starting at the first observation of the Q9 \textit{Kepler} data. For each of the baselines we test, we recompute the standard deviation of the light curve fluxes based just on the observations that fall within the specified baseline. This prevents information about the light curves at later times from mistakenly inflating model performance. Since the length of the input data ($n(\vec{X}_{\rm in})$) varies with baseline, we redetermine the padding and strides of the two convolutional layers of the model, while keeping the kernel widths tested in the hyperparameter search the same as described in Table \ref{tab:nnparams}. Other than these changes, the models for the baseline tests have the same architecture and training process as described in Section \ref{sec:methods}. The results we report are for the performance of the best model on the \cite{yu18}, \cite{pande18}, and \cite{mcquillan14} test sets selected from the 144 models trained for each baseline.

Figure \ref{fig:all_baseline} demonstrates the model performance as a function of baseline for all of the stellar properties examined in Section \ref{sec:regression}, except for rotation period, which is specially considered as a property of interest in Section \ref{sec:rotation}. Table \ref{tab:shortbaseline} in the Appendix reports the full list of performance metrics, $r^{2}$, $\Delta$, rms, fractional $\Delta$, and fractional rms, for each baseline model. From Figure \ref{fig:all_baseline} and the full list of metrics in Table \ref{tab:shortbaseline}, we find that all stellar properties are recovered remarkably well using short baseline time series data. First, examining the \cite{yu18} stellar properties, we find that the $\Delta\nu$ recovery only degrades slightly with decreasing baseline, with an $r^{2}$ of 0.95 at 97 days and an $r^{2}$ score of 0.93 at 14 days. The rms increases slightly from 97 days down to 14 days from 0.9 $\mu$Hz to 1.1 $\mu$Hz, while the bias fluctuates marginally. The log $g$ predictions exhibit similar behavior with baseline. The $r^{2}$ degrades slightly from 0.97 at 97 days to 0.94 at 14 days, while the rms increases from 0.06 to 0.09 dex and the bias is nearly the same for each of the models. For the $\nu_{\rm max}$ predictions the 97- and 62-day models perform similarly with an $r^{2}$ score of 0.96, while the 27-day model results in and $r^{2}$ score of 0.91 and the 14-day model results in an $r^{2}$ score of 0.94, with the rms and bias of each of the models similarly following the $r^{2}$ score trends.

Considering the \cite{pande18} stellar sample we find similar trends, with only a marginal decrease or fluctuations in the model performance across the tested baselines. For log $g$ the $r^{2}$ score of the best model is 0.92 for the 62-day baseline data, while the 97-day, 27-day, and 14-day models all result in an $r^{2}$ score of 0.89, with the bias and rms for each of the models being similar. For $T_{\rm eff}$ the $r^{2}$ scores follow a similar pattern, with the 62-day model resulting in the highest score of 0.8, while the 97-day and 27-day models have an $r^{2}$ score of 0.79 and the 14-day model with an $r^{2}$ score of 0.77, and again the bias and rms of the four models is similar. Lastly, for the $M_{*}$ predictions for the \cite{mcquillan14} sample, we see that the $r^{2}$ score decreases steadily from the 97-day to the 14-day model, from $r^{2}$ = 0.4 to $r^{2}$ = 0.37, with the rms slightly increasing with decreasing baseline and the bias fluctuating marginally. 

The recovery of stellar properties using short baseline data suggests is that these stellar properties are still sufficiently encoded in the light curve data at these shorter timescales. This results of  Figure \ref{fig:all_baseline} are promising for the prospects of estimating these stellar properties from light curves from surveys such as TESS and LSST. In the following section we explore a similar prospect for the recovery of stellar rotation period, demonstrating how the recovery of $P_{\rm rot}$ changes with baseline, as well as with the cadence of the observations.

%%%%%%%%%%%%%%%%%%%%%%%%%%%%%%%%%%% ROTATION PERIOD %%%%%%%%%%%%%%%%%%%%%%%%%%%%%%%%%%
\section{Rotation period of main sequence stars}
\label{sec:rotation}
\subsection{Baseline}
\label{sec:baseline}
We now focus on stellar rotation as a key stellar property, particularly for gyrochronology studies, and examine the prospects of deriving $P_{\rm rot}$ from light curves with baselines less than 97 days (Section \ref{sec:baseline}), as well as cadences longer than 29.4 minutes (Section \ref{sec:cadence}). For the full list of stellar properties examined in Section \ref{sec:regression}, similar cadence tests can also be performed. However, to limit the scope of this paper, we omit this examination.

First, we investigate how well rotation periods can be recovered from light curves as a function of the observation baseline. In addition to the 97-day rotation model trained in Section \ref{sec:regression}, we train three additional CNN models based on light curves with baselines of 62, 27, and 14 days. The 27-day model is of particular interest, as most stars that will be observed by the TESS mission will have observations spanning 27 days. For these shorter baseline models, we prepare the \textit{Kepler} Q9 light curves and modify the CNN padding and stride values for the hyperparameter search the same as described in Section \ref{sec:all_short_baseline}. Further demonstrating the necessity of the hyperparameter search, a number of the short baseline $P_{\rm rot}$ models result in $r^{2}$ scores of $\sim$0.5, which is significantly worse than the performance of the best models presented here.

\begin{figure*}[tp!]
\centering
\includegraphics[width=2\columnwidth]{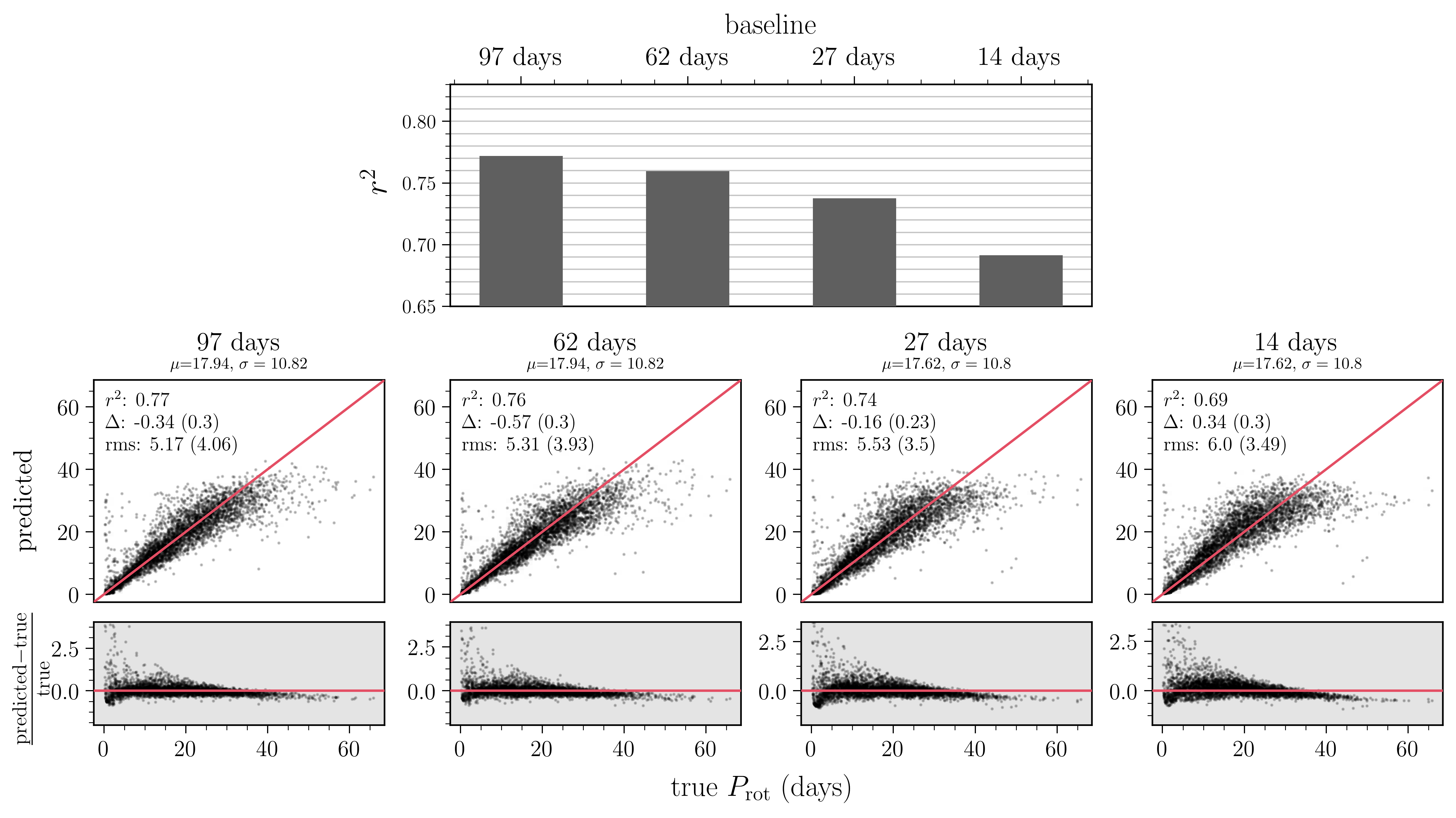}
\caption{Performance of the CNN model based on different light curve baselines for a test set of \cite{mcquillan14} stars. \textit{\textbf{Top panel:}} summary of model performance showing the $r^{2}$ score versus the tested baselines of 97, 62, 27, and 14 days. \textit{\textbf{Bottom panels:}} the predicted versus true rotation period, and the fractional difference between the predicted and true values, for models based on each light curve baseline for the test set of stars. The $r^{2}$, $\Delta$, and rms of the predictions are indicated in each panel, as well as the fractional bias and fractional rms in parenthesis. As discussed in the text, the fractional metrics for the $P_{\rm rot}$ predictions is greatly inflated by the over-prediction of short period stars. If we remove these stars with fractional differences \textgreater~2.5, the fractional bias and fractional rms decrease significantly. \bigskip}
\label{fig:baseline}
\end{figure*}

Figure \ref{fig:baseline} demonstrates how the recovery of stellar rotation period degrades as a function of light curve baseline, with the top panel of the figure summarizing the performance of the models by showing how the $r^{2}$ decreases with decreasing observation lengths. We find that the $r^{2}$ score changes by less than $\Delta r^{2}$ = -0.1, from $r^{2}$ = 0.77 at a baseline of 97-days, to $r^{2}$ = 0.69 at a baseline of 14-days, with the 27-day ``TESS"-baseline model resulting in an $r^{2}$ of 0.74. Examining the model performance in more detail, for each baseline the bottom half of Figure \ref{fig:baseline} shows how both the light curve predicted (top panel) and the fractional difference between the predicted and true rotation period (bottom panel) vary as a function of the true rotation period for the test set of \citep{mcquillan14} stars. For ease of comparison, the 97-day model from Figure \ref{fig:regression_results} is also included. Across the four baselines considered, the bias of the models computed across the entire range of true rotation periods remains $|\Delta|$ \textless~0.6 days without a clear trend with baseline, while the rms of the models increase with decreasing baseline by $\sim$1 day from the $\sim$5 days for the 97-day model to $\sim$6 days for the 14-day model. 

Visually inspecting the bias and rms of the models as a function of the true rotation period, we find that for each baseline, the rms increases as rotation period increases, while the fractional rms decreases marginally by 0.05. In general, shorter rotation periods are recovered with less bias, except for the fraction of fast rotators whose rotation periods are over-predicted, as discussed in Section \ref{sec:cnn_results}. In the fractional difference plots for the models shown in Figure \ref{fig:baseline}, which span -0.5$\sigma$ to 1$\sigma$ along the $y$-axis, 35, 41, 32, and 42 short period stars (comprising just $\sim$1\% of the test set) are excluded from the plots for the 97-day, 62-day, 27-day, and 14-day models respectively. As in Section \ref{sec:cnn_results}, if we remove the short rotation period stars with the highest fractional differences, the fractional bias and fractional rms metrics decrease significantly. For the 62-day model, removing the over-predicted $\sim$9\% of the stars with true rotation periods \textless~6.2 days results in a fractional bias and fractional rms of 0.012 and 0.30 respectively. For the 27-day model, removing the over-predicted $\sim$6\% of the stars with true rotation periods \textless~6.5 days results in a fractional bias and fractional rms of 0.032 and 0.36 respectively. And lastly for the 14-day model, removing the over-predicted $\sim$7\% of the stars with true rotation periods \textless~8 days results in a fractional bias and fractional rms of 0.081 and 0.38 respectively.

Another feature of the model performance we notice is that as the baseline decreases, the bias of the predictions at rotation periods \textgreater~35 days marginally increases. While all of the models exhibit this behavior to an extent, as evident in the bottom panels of Figure \ref{fig:baseline}, the 27-day and 14-day models in particular do not predict rotation periods greater than $\sim$35 days, with slowly rotating stars in the test set having their rotation periods under-predicted. The degradation of the predictions for stars with rotation periods longer than $\sim$35 days could be due to a number of factors. One such factor is that for these more slowly rotating stars, fewer cycles of the rotation period are imprinted in the light curves, and in some cases only a fraction of one full rotation period is present. However, from Figure \ref{fig:baseline}, we see that even for rotation periods longer than the baseline, the model can still recover rotation, although at decreasing precision. Another factor that could be impacting the model's ability to precisely recover rotation periods longer than $\sim$35 days is the distribution of the training data. The mean rotation period of the \citep{mcquillan14} sample is $\sim$18 days with a standard deviation of $\sim$11 days. As seen in Figure \ref{fig:data_density}, there are few stars rotation periods longer than $\sim$35 days, predominantly due to the fact that instrumental systematics in the \textit{Kepler} data become more prominent on monthly timescales. The model could be less effective in predicting long rotation periods also because there are few examples to learn from. Given a more complete rotation period coverage in the training data, the model may be better able to learn the rotation periods of more slowly rotating stars, making unbiased predictions even beyond the baseline of the data. 

\subsection{Cadence}
\label{sec:cadence}
\begin{figure*}[tp!]
\centering
\includegraphics[width=2\columnwidth]{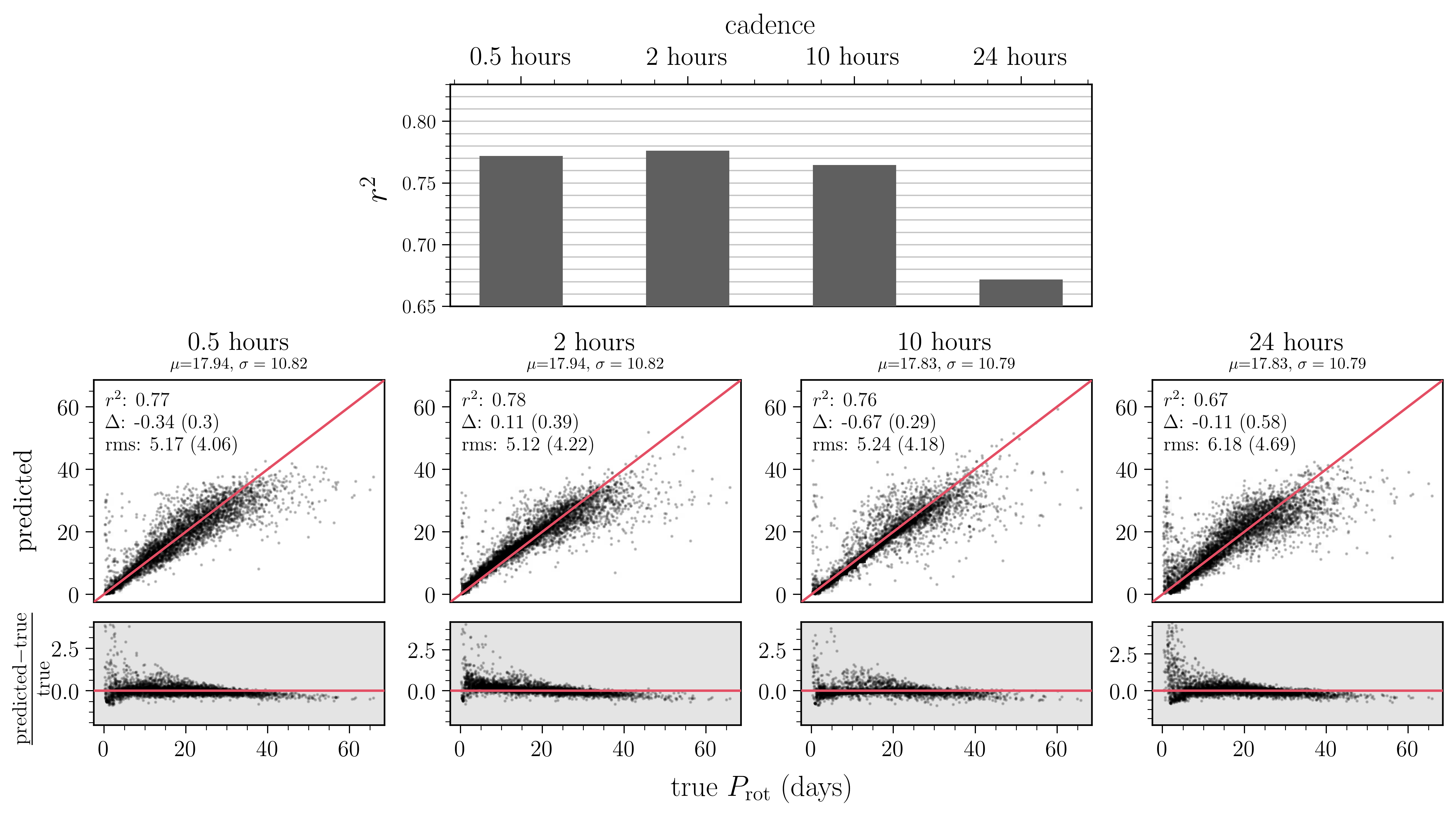}
\caption{Performance of the CNN model based on different light curve cadences for a test set of \cite{mcquillan14} stars. \textit{\textbf{Top panel:}} summary of model performance showing the $r^{2}$ score versus the tested cadences of 0.5 hours, 2 hours, 10 hours, and 24 hours. \textit{\textbf{Bottom panels:}} the predicted versus true rotation period, and the fractional difference between the predicted and true values, for models based on each light curve cadence for the test set of stars. The $r^{2}$, $\Delta$, and rms of the predictions are indicated in each panel, as well as the fractional bias and fractional rms in parenthesis. As discussed in the text, the fractional metrics for the $P_{\rm rot}$ predictions is greatly inflated by the over-prediction of short period stars. If we remove these stars with fractional differences \textgreater~2.5, the fractional bias and fractional rms decrease significantly. \bigskip}
\label{fig:cadence}
\end{figure*}

In addition to baseline, we also investigate how well we can recover rotation periods from light curves as a function of cadence. We train a CNN to predict rotation periods from light curves with an observation every 2 hours, 10 hours, and 24 hours, all with a baseline of 97 days. Being able to measure rotation periods from less frequently sampled photometric time series is also of interest, as upcoming surveys will observe a substantial number of stars at much sparser cadences than those of the \textit{Kepler} 29.4-minute sampling. In particular, the cadence of LSST observations will be irregular, and the minimum separation between subsequent observations is tentatively $\sim$3 days.

To modify the light curve data for the cadence models, we again use the same sample of \cite{mcquillan14} stars described in Section \ref{sec:data_params}, but instead of using the full light curve we only select every 4th, 21st, and 49th flux observation to achieve light curves with cadences of 2 hours, 10 hours, and 24 hours, respectively. As we did for the baseline models, to prevent information leakage, we compute the standard deviation of the light curves based just on the selected $n$th observations for each tested cadence, which is passed to the first fully-connected layer of the model as shown in Figure \ref{fig:nn}. Since the length of the input data also varies with cadence, we redetermine the padding and strides of the two convolutional layers. For the 2-hour and 10-hour models, we test the same kernel widths as listed in Table \ref{tab:nnparams}. However, for the 24-hour model, since there are only 98 flux observations for each light curve, instead of testing kernel widths of $K_{W, 1}$ = 12 and $K_{W, 1}$ = 20, we test two additional smaller kernel widths of $K_{W, 1}$ = 2 and $K_{W, 1}$ = 4. These have a corresponding $K_{W, 2}$ = 3 and $K_{W, 2}$ = 7, respectively. For the 10-hour and 24-hour models, we also modify the size of the fully-connected part of the model. For the 10-hour cadence data with an input size of 228 observations, we change the size of the first fully-connected layer to have $N_{1}$ = 512 hidden neurons, the second fully-connected layer to have $N_{2}$ = 256 hidden neurons, and the last fully-connected layer to have $N_{3}$ = 128 hidden neurons. For the 1-day cadence data with an input size of 98 observations, we change the size of the first fully-connected layer to have $N_{1}$ = 256 hidden neurons, the second fully-connected layer to have $N_{2}$ = 128 hidden neurons, and the last fully-connected layer to have $N_{3}$ = 64 hidden neurons. Other than modifying the kernel widths explored and the reduction of the capacity of the fully-connected part of the model, the cadence models have the same architecture and training process as described in Section \ref{sec:methods}. The results we report here are the performance of the best model on the \cite{mcquillan14} test set selected from the 144 models trained for each cadence. 
Figure \ref{fig:cadence} demonstrates how the recovery of rotation period degrades as a function of the light curve cadence, with the top panel of the figure summarizing the performance of the models by showing how the $r^{2}$ decreases with sparser cadence. As seen in the figure, in terms of the $r^{2}$ score, the 0.5-hour, 2-hour, and 10-hour cadence models have nearly identical model performance with $r^{2}$ values of 0.77, 0.78, and 0.76, respectively. However, for the 24-hour cadence light curves, the performance of the model decreases to an $r^{2}$ of 0.67. Examining the performance of the model on the \cite{mcquillan14} test set in more detail, for each cadence the bottom half of Figure \ref{fig:cadence} shows how both the light curve predicted (top panel) and the fractional difference between the predicted and true rotation period (bottom panel) vary as a function of the true rotation period. For ease of comparison, the 29.4-minute model from Figure \ref{fig:regression_results} is also included. Considering the four cadences tested, the bias of the models computed for the entire range of true rotation periods remains $|\Delta|$ \textless~0.7 days, without a clear trend with cadence. The rms of the 0.5-hour, 2-hour, and 10-hour models are all $\sim$5.2 days, while the rms of the 24-hour model is somewhat larger, at $\sim$6.2 days, and the fractional rms increases marginally by $\sim$0.06 from the 0.5-hour to 24-hour cadence models. If we remove the short rotation period stars with the highest fractional differences, the fractional bias and fractional rms metrics decrease significantly. For the 2-hour model, removing the over-predicted $\sim$7.5\% of the stars with true rotation periods \textless~6.2 days results in a fractional bias and fractional rms of 0.10 and 0.32 respectively. For the 10-hour model, removing the over-predicted $\sim$9.4\% of the stars with true rotation periods \textless~5.6 days results in a fractional bias and fractional rms of -0.02 and 0.27 respectively. And lastly for the 24-hour model, removing the over-predicted $\sim$25\% of the stars with true rotation periods \textless~5.2 days results in a fractional bias and fractional rms of 0.042 and 0.40 respectively.

Inspecting the bias and rms of the models as a function of rotation period, we see that the 0.5-hour, 2-hour, and 10-hour cadence models all perform similarly across the range of rotation periods, with the shorter and longer rotation periods being recovered less precisely than rotation periods spanning $\sim$5-35 days, as with the baseline models discussed in Section \ref{sec:baseline}. In the fractional difference plots shown in Figure \ref{fig:cadence}, 35, 31, and 44 short rotation period stars are excluded by plot limits of the 0.5-hour, 2-hour, and 10-hour cadence models, which comprise just $\sim$1\% of the test set. Of the cadences we test, the 24-hour model is the only model that exhibits a markedly different performance behavior across the range of rotation periods. We take note of two significant differences in the performance of this model compared to the models trained on the higher cadence light curves. The first is that similarly to the shorter baseline models, the model trained on the 24-hour cadence light curves is biased at longer rotation periods, with rotation periods longer than 40 days being under-predicted. For this model we also find that the rotation periods of a larger fraction of more quickly rotating stars, with $P_{\rm rot}$ \textless~5 days, are predicted incorrectly. As evident in fractional difference plot for the 24-hour model in Figure \ref{fig:cadence}, the rotation periods for a more significant number of fast rotating stars are over-predicted, in some cases by up to $\sim$40 days. Additionally, 89 short period stars (comprising $\sim$2\% of the test set) are excluded by the plot limits due to their large over-predictions. The more significant degradation in model performance for the stars with the fastest rotation periods makes sense given that one full rotation cycle for these stars is only sampled a few times when the cadence is as sparse as 24 hours.

%%%%%%%%%%%%%%%%%%%%%%%%%%%%%%%%%%% DISCUSSION %%%%%%%%%%%%%%%%%%%%%%%%%%%%%%%%%%
\section{Discussion}
\label{sec:discussion}
We have systematically explored the recovery of a set of stellar properties from photometric time series data, examining the prediction performance across different baselines and cadences. Our approach, using a 1D CNN model, requires minimal data-processing and no by-hand feature engineering. Using the \textit{Kepler} 97-day, 29.4-minute Q9 data, we first construct a CNN classification model to predict stellar evolutionary state. We then use three catalogues to build training sets to predict the continuous stellar properties of: $\Delta\nu$, $\nu_{\rm max}$, log $g$, $T_{\rm eff}$ and $P_{\rm rot}$. We implement a CNN regression model, optimizing over a grid of possible hyperparameters, to successfully recover these properties to a high fidelity across the parameter space of the training examples. 

Our CNN modeling approach is demonstrative of the information content in the data and how this information is preserved across various baselines and cadences. We expect that our modeling choice is not the primary limitation in our prediction precision. Rather, we expect the information contained in the data, the precision on the input properties and the stellar property range of the training data are the primary drivers of our results. Nevertheless, there are several alternative types of models that can capture the structure of time series data like light curves. Recurrent Neural Networks (RNNs), for example, are a well suited class of neural networks that model temporal data structure, using a recurrence relationship between new outputs of the model and the previous states of the model. Similar to CNNs, RNNs include weight sharing from different parts of the time series throughout the model training process \citep{rnnreview}.

Indeed one downside of CNN models, compared to RNN models, is that the shape of the input data must be similar across the entire dataset. CNNs do not lend themselves to working with unevenly sampled data, and as discussed in Section \ref{sec:time_space}, to overcome this issue we take the simplest approach and replace missing flux time steps in the time series with zeros values. This results in good model performance, however more well-motivated imputation approaches could be tested, including: simple interpolation of the light curves fluxes between points as done for the ACF processing, modelling-based imputation, providing a missing value mask a second input channel to the CNN, or testing the use of alternative modeling approaches like RNNs, however RNN models are typical more difficult to train than CNN models. We leave exploring these options as a task for future work. This paper establishes a baseline of performance expectation, with the adopted model, hyperparameter optimisation choices and other assumptions like the zero-imputation of missing data.

Another alternative choice is to take a generative, rather than a discriminative, approach to the modeling presented here. Generative models take a probabilistic approach to learning the joint distribution of the data ($X$) and label ($Y$), $P(X, Y)$, which is then used to infer $P(Y|X)$. Generative models also include methods that learn a (typically lower dimensional) latent representation of the data itself from which the original data vector can be generated (e.g. variational autoencoders), with the latent space informing $P(Y|X)$. A generative approach lends itself better to understanding the data generation process, which is arguably more aligned with science goals than a discriminative approach. However, as discussed in \cite{fawaz18}, generative models are typically less accurate than discriminative models when it comes to performance on a specific task, and generative models for times series data are not trivial to implement in practice. Future work could certainly include taking a generative approach to the problem. This would perhaps promote understanding of the data generation process, as well as enable the derivation of well-motivated, datapoint-by-datapoint probability distributions for the inferred labels using fast inference methods like variational inference \citep{blei16}, delivering effective errors on the stellar properties inferred for each individual star. 

Lastly, the discriminative end-to-end nature of many deep learning approaches (including our own) makes these models difficult to interpret, often hindering their ability to be useful for understanding the underlying physical theory. There are numerous definitions of what it means for deep learning to be interpretable, as well as numerous proposed methods for making some of these interpretations \citep[e.g.][]{simonyan13, yosinski15, binder16, montavon17}. Specifically, for our case of predicting stellar properties from time series data, it would be insightful to know what features of the light curves contributed most to the prediction of a particular stellar property. For instance, if we could identify the most relevant variation timescales for making a prediction, we could make connections between the quality of the stellar property recovery and our physical understanding of stellar physics. This would allow us to confirm existing theories of how the internal physical processes of stars are imprinted in light curves, as well as have the potential to reveal new connections between stellar properties and stellar variability in the time domain.

%%%%%%%%%%%%%%%%%%%%%%%%%%%%%%%%%%% CONCLUSION %%%%%%%%%%%%%%%%%%%%%%%%%%%%%%%%%%
\section{Conclusions}
\label{sec:conclusion}
We have implemented a 1-dimensional convolutional neural network (CNN) architecture to estimate stellar properties from photometric time series data. Constructing training sets based on the 29.4-minute cadence \textit{Kepler} Q9 data and high-quality stellar property catalogs, we predict evolutionary states, stellar properties ($T_{\rm eff}$ and log $g$), asteroseismic parameters ($\Delta\nu$ and $\nu_{\rm max}$) and rotation periods ($P_{\rm rot}$) for main sequence and red-giant stars. We compare the quality of predictions based on learning directly from the time series data to learning from transformations of the data, including the ACF and the frequency domain. We also examine how the prediction quality varies with the baseline of observations, training models based on 97, 62, 27, and 14 days of data. For rotation period, which is of particular interest for gyrochronology, we further examine how the prediction quality varies with the cadence, training models based on time series data with an observation every 0.5, 2, 10, and 24 hours. The main results of this work are summarized as follows:

\begin{itemize}
  \item Training a CNN model to classify stellar evolutionary state, we are able to distinguish red giant stars from main sequence and sub-giant stars to an accuracy of $\sim$90\%. However, the model is not as successful at distinguishing between main sequence and sub-giant stars, with each of these stellar types having a classification accuracy \textless~60\%. We suspect that this is due to the more subtle physical differences (and how these are manifested in the time domain) between main sequence and sub-giant stars and the limited quality of the training labels, as the border between main sequence and sub-giant stars is sensitive to metallicity. 

  \item Based on one quarter of \textit{Kepler} long-cadence data, our CNN regression model recovers $\nu_{\rm max}$ and $\Delta\nu$ to an rms (fractional rms) precision of $\sim$12 $\mu$Hz (0.2) and $\sim$0.9 $\mu$Hz (0.14), respectively, and log $g$ to an rms precision of $\sim$0.06 dex (0.02), for red giant stars trained with the \cite{yu18} catalog. Using the \cite{pande18} catalog as a training set, we predict $T_{\rm eff}$ with relatively little bias across evolutionary states (with $T_{\rm eff}$= 4500 - 6500 K), to an rms precision of $\sim$300 K (0.05). We also predict log $g$ across the range log $g$ = 2 - 4.5 dex to an rms precision of $\sim$0.22 dex (0.07). This performance is in part limited by the precision of the training labels, with a mean reported uncertainty of $\sim$0.25 dex (compared to 0.01 dex for the red giant log $g$ estimates).

  \item For main sequence stars, based on a single quarter of \textit{Kepler} long-cadence data, our CNN regression model predicts rotation periods unbiased from $\approx$5 to 40 days, with an rms precision of $\sim$5.2 days (4.06). Our model becomes biased in parameter spaces with few training examples in the \cite{mcquillan14} catalog (e.g. $P_{\rm rot}$ $\gtrapprox$ 40 days), and over-predicts $\sim$8\% of short-period stars in the test set with $P_{\rm rot}$ \textless~6.2 days. Removing these short-period stars, the fractional rms of the predictions is 0.3. We also imprecisely predict stellar mass without bias for $M_{*}$ $>$ 0.7 $M_{\odot}$, with an rms precision of $\sim$0.16 $M_{\odot}$ (0.28). The performance is in part limited by the large input uncertainty on mass values.
  
  \item For the stellar properties listed above, we compare the performance of the CNN model based on the 97-day time domain data to fully-connected neural network models based on the ACF and frequency domain representations of the same data. We find that the CNN model trained on the original time series data outperforms the models based on the other two data representations. This implies that more information can be gleaned from deep learning models that work closer to the raw data, as transformations and feature engineering of data often results in lost information. 
  
  \item To inform expectations of what can be delivered from observations made by TESS, LSST, and future missions, we train our CNN model to recover stellar properties from light curves with shorter baselines (62, 27, and 14 days). We find that we can predict stellar properties remarkably well for TESS-like data (27-day baseline), including, for red giant stars, log $g$ to an rms precision of $\sim$0.07 dex (0.03), $\Delta\nu$ to an rms precision of $\sim$1.1 $\mu$Hz (0.18), and $\nu_{\rm max}$ to an rms precision of $\sim$17 $\mu$Hz (0.3). Based on the \cite{pande18} training set, we predict log $g$ to an rms precision of $\sim$0.21 dex (0.06) and $T_{\rm eff}$ to an rms precision of $\sim$300 K (0.05). 
  
  \item We predict rotation periods for main sequence stars up to $\approx$35 days days based on 27-day and even 14-day data, with an rms precision of $\lessapprox$ 6 days (3.5). Removing the $\textless$~10\% of short-period stars that are over-predicted, the fractional rms is $\textless$~0.38. Investigating light curves with longer cadences (2, 10, and 24 hours), we find that even for observations spaced 1 day apart for 97 days, we can predict stellar rotation to an rms precision of $\sim$6.2 days (4.69), unbiased over the range from $P_{\rm rot}$ $\approx$ 5 - 35 days. Removing the $\textless$~25\% of over-predicted short-period stars with $P_{\rm rot}$ \textless~5.2 days, the fractional rms is $\approx$0.40.
 
\end{itemize}

With the results described above, we have established a baseline of performance for the stellar property information that can be extracted from this light curve data alone. Our modeling approach is generalizable to other time domain surveys, as well as other stellar property catalogs. The method presented here is not proposed to replace asteroseismology measurements from high-quality data. Instead, we predict these properties to demonstrate the capability of our approach, as well as the prospect of transferring the relationships establish with high-quality data to lower-quality data, where these measurements are more difficult to make.

Our ability to predict stellar properties is in part subject to the uncertainty on the input properties, and we expect given more precisely derived properties we could improve our predictions in some cases. To better determine some stellar properties, we could also incorporate other information such as photometry from \textit{Gaia}, 2MASS and WISE, as well as stellar parallaxes from \textit{Gaia}. For example, adding photometric information to the model would improve the precision with which we can recover $T_{\rm eff}$.

We make our model code publicly available at \href{https://github.com/kblancato/theia-net}{https://github.com/kblancato/theia-net}, which could be used to produce a rotation period catalogue, as well as other stellar property catalogues, for the TESS mission. Some more immediate improvements to our approach could include expanding the extent, as well as the precision quality, of the stellar property training sets, and incorporating \textit{Gaia} photometric and parallax information in the model. More substantial aspects that could be investigated include adapting the model to permit unevenly sampled time series data, taking a generative modeling approach, incorporating the data errors on both the light curves and stellar properties, and interpreting the model in a physically meaningful way.

Looking forward, in the coming years ongoing and future missions will deliver time domain data for millions of stars. Extracting stellar properties from this data will be a rich pursuit, enabling the exciting potential of Galactic archaeology in the time domain.

%%%%%%%%%%%%%%%%%%%%%%%%%%%%%%%%%%% Acknowledgements %%%%%%%%%%%%%%%%%%%%%%%%%%%%%%%%%%
\bigskip
\section*{Acknowledgements}
The authors would like to thank Kathryn Johnston, David Blei, Gabriella Contardo, Maryum Sayeed, and Adam Wheeler for helpful feedback and discussions. We also thank Travis Berger for providing us his revised \textit{Gaia-Kepler} stellar property catalog. We are grateful to the members of the Flatiron Institute's Scientific Computing Core for their support of the Flatiron's Rusty cluster, which was used to train all of the models for this work.

K.B. is supported by the NSF Graduate Research Fellowship under grant number DGE 16-44869. K.B. thanks the LSSTC Data Science Fellowship Program, her time as a Fellow has benefited this work. M.N. and D.H. are supported by the Alfred P. Sloan Foundation. R.A. acknowledges support from NASA award: 80NSSC20K1006. The research was supported by the Research Corporation for Science Advancement through Scialog award $\#$26080. 

This research was partially conducted during the Exostar19 program at the Kavli Institute for Theoretical Physics at UC Santa Barbara, which was supported in part by the National Science Foundation under Grant No. NSF PHY-1748958.  \\

\textit{Software:} \texttt{PyTorch} \citep{pytorch}, \texttt{scikit-learn} \citep{scikit-learn}, \texttt{Astropy} \citep{astropy}.

%%%%%%%%%%%%%%%%%%%% REFERENCES %%%%%%%%%%%%%%%%%%
\bibliographystyle{apj}
\bibliography{lc_main}

\begin{thebibliography}{}
\expandafter\ifx\csname natexlab\endcsname\relax\def\natexlab#1{#1}\fi

\bibitem[{{Aerts} {et~al.}(2010){Aerts}, {Christensen-Dalsgaard}, \&
  {Kurtz}}]{aerts10}
{Aerts}, C., {Christensen-Dalsgaard}, J., \& {Kurtz}, D.~W. 2010,
  {Asteroseismology}

\bibitem[{{Ag{\"u}eros} {et~al.}(2018){Ag{\"u}eros}, {Bowsher}, {Bochanski},
  {Cargile}, {Covey}, {Douglas}, {Kraus}, {Kundert}, {Law}, {Ahmadi}, \&
  {Arce}}]{agueros18}
{Ag{\"u}eros}, M.~A., {Bowsher}, E.~C., {Bochanski}, J.~J., {et~al.} 2018,
  \apj, 862, 33

\bibitem[{{Angus} {et~al.}(2015){Angus}, {Aigrain}, {Foreman-Mackey}, \&
  {McQuillan}}]{angus15}
{Angus}, R., {Aigrain}, S., {Foreman-Mackey}, D., \& {McQuillan}, A. 2015,
  \mnras, 450, 1787

\bibitem[{{Angus} {et~al.}(2018){Angus}, {Morton}, {Aigrain}, {Foreman-Mackey},
  \& {Rajpaul}}]{angus17}
{Angus}, R., {Morton}, T., {Aigrain}, S., {Foreman-Mackey}, D., \& {Rajpaul},
  V. 2018, \mnras, 474, 2094

\bibitem[{{Baglin} {et~al.}(2006){Baglin}, {Michel}, {Auvergne}, \& {COROT
  Team}}]{baglin06}
{Baglin}, A., {Michel}, E., {Auvergne}, M., \& {COROT Team}. 2006, in ESA
  Special Publication, Vol. 624, Proceedings of SOHO 18/GONG 2006/HELAS I,
  Beyond the spherical Sun, 34

\bibitem[{{Baraffe} {et~al.}(1998){Baraffe}, {Chabrier}, {Allard}, \&
  {Hauschildt}}]{baraffe98}
{Baraffe}, I., {Chabrier}, G., {Allard}, F., \& {Hauschildt}, P.~H. 1998, \aap,
  337, 403

\bibitem[{{Barnes}(2003)}]{barnes03}
{Barnes}, S.~A. 2003, \apj, 586, 464

\bibitem[{{Barnes}(2007)}]{barnes07}
---. 2007, \apj, 669, 1167

\bibitem[{{Bastien} {et~al.}(2013){Bastien}, {Stassun}, {Basri}, \&
  {Pepper}}]{bastien13}
{Bastien}, F.~A., {Stassun}, K.~G., {Basri}, G., \& {Pepper}, J. 2013, \nat,
  500, 427

\bibitem[{{Bastien} {et~al.}(2016){Bastien}, {Stassun}, {Basri}, \&
  {Pepper}}]{bastien16}
---. 2016, \apj, 818, 43

\bibitem[{{Bedding} {et~al.}(2010){Bedding}, {Huber}, {Stello}, {Elsworth},
  {Hekker}, {Kallinger}, {Mathur}, {Mosser}, {Preston}, {Ballot}, {Barban},
  {Broomhall}, {Buzasi}, {Chaplin}, {Garc{\'{\i}}a}, {Gruberbauer}, {Hale}, {De
  Ridder}, {Frandsen}, {Borucki}, {Brown}, {Christensen-Dalsgaard},
  {Gilliland}, {Jenkins}, {Kjeldsen}, {Koch}, {Belkacem}, {Bildsten}, {Bruntt},
  {Campante}, {Deheuvels}, {Derekas}, {Dupret}, {Goupil}, {Hatzes}, {Houdek},
  {Ireland}, {Jiang}, {Karoff}, {Kiss}, {Lebreton}, {Miglio}, {Montalb{\'a}n},
  {Noels}, {Roxburgh}, {Sangaralingam}, {Stevens}, {Suran}, {Tarrant}, \&
  {Weiss}}]{bedding10}
{Bedding}, T.~R., {Huber}, D., {Stello}, D., {et~al.} 2010, \apjl, 713, L176

\bibitem[{{Belkacem} {et~al.}(2011){Belkacem}, {Samadi}, \&
  {Goupil}}]{belkacem11}
{Belkacem}, K., {Samadi}, R., \& {Goupil}, M.~J. 2011, in Journal of Physics
  Conference Series, Vol. 271, GONG-SoHO 24: A New Era of Seismology of the Sun
  and Solar-Like Stars, 012047

\bibitem[{{Berger} {et~al.}(2018){Berger}, {Huber}, {Gaidos}, \& {van
  Saders}}]{berger18}
{Berger}, T.~A., {Huber}, D., {Gaidos}, E., \& {van Saders}, J.~L. 2018, \apj,
  866, 99

\bibitem[{{Binder} {et~al.}(2016){Binder}, {Montavon}, {Bach}, {M{\"u}ller}, \&
  {Samek}}]{binder16}
{Binder}, A., {Montavon}, G., {Bach}, S., {M{\"u}ller}, K.-R., \& {Samek}, W.
  2016, arXiv e-prints, arXiv:1604.00825

\bibitem[{{Blei} {et~al.}(2016){Blei}, {Kucukelbir}, \& {McAuliffe}}]{blei16}
{Blei}, D.~M., {Kucukelbir}, A., \& {McAuliffe}, J.~D. 2016, arXiv e-prints,
  arXiv:1601.00670

\bibitem[{{Borucki} {et~al.}(2008){Borucki}, {Koch}, {Basri}, {Batalha},
  {Brown}, {Caldwell}, {Christensen-Dalsgaard}, {Cochran}, {Dunham}, {Gautier},
  {Geary}, {Gilliland }, {Jenkins}, {Kondo}, {Latham}, {Lissauer}, \&
  {Monet}}]{borucki08}
{Borucki}, W., {Koch}, D., {Basri}, G., {et~al.} 2008, in IAU Symposium, Vol.
  249, Exoplanets: Detection, Formation and Dynamics, ed. Y.-S. {Sun},
  S.~{Ferraz-Mello}, \& J.-L. {Zhou}, 17--24

\bibitem[{{Bouvier} {et~al.}(1997){Bouvier}, {Forestini}, \&
  {Allain}}]{bouvier97}
{Bouvier}, J., {Forestini}, M., \& {Allain}, S. 1997, \aap, 326, 1023

\bibitem[{{Brown} {et~al.}(1991){Brown}, {Gilliland}, {Noyes}, \&
  {Ramsey}}]{brown91}
{Brown}, T.~M., {Gilliland}, R.~L., {Noyes}, R.~W., \& {Ramsey}, L.~W. 1991,
  \apj, 368, 599

\bibitem[{{Browning} {et~al.}(2006){Browning}, {Miesch}, {Brun}, \&
  {Toomre}}]{browning06}
{Browning}, M.~K., {Miesch}, M.~S., {Brun}, A.~S., \& {Toomre}, J. 2006, \apjl,
  648, L157

\bibitem[{{Cardini} \& {Cassatella}(2007)}]{cardini07}
{Cardini}, D., \& {Cassatella}, A. 2007, \apj, 666, 393

\bibitem[{{Claytor} {et~al.}(2019){Claytor}, {van Saders}, {Santos}, {Garcia},
  {Mathur}, {Tayar}, {Pinsonneault}, \& {Shetrone}}]{claytor19}
{Claytor}, Z.~R., {van Saders}, J.~L., {Santos}, A. R.~G., {et~al.} 2019, arXiv
  e-prints, arXiv:1911.04518

\bibitem[{{Cranmer} {et~al.}(2014){Cranmer}, {Bastien}, {Stassun}, \&
  {Saar}}]{cranmer14}
{Cranmer}, S.~R., {Bastien}, F.~A., {Stassun}, K.~G., \& {Saar}, S.~H. 2014,
  \apj, 781, 124

\bibitem[{{Curtis} {et~al.}(2019){Curtis}, {Ag{\"u}eros}, {Mamajek}, {Wright},
  \& {Cummings}}]{curtis19}
{Curtis}, J.~L., {Ag{\"u}eros}, M.~A., {Mamajek}, E.~E., {Wright}, J.~T., \&
  {Cummings}, J.~D. 2019, \aj, 158, 77

\bibitem[{{Davies} {et~al.}(2016){Davies}, {Silva Aguirre}, {Bedding}, {Hand
  berg}, {Lund}, {Chaplin}, {Huber}, {White}, {Benomar}, {Hekker}, {Basu},
  {Campante}, {Christensen-Dalsgaard}, {Elsworth}, {Karoff}, {Kjeldsen},
  {Lundkvist}, {Metcalfe}, \& {Stello}}]{davies16}
{Davies}, G.~R., {Silva Aguirre}, V., {Bedding}, T.~R., {et~al.} 2016, \mnras,
  456, 2183

\bibitem[{{De Ridder} {et~al.}(2009){De Ridder}, {Barban}, {Baudin}, {De
  Ridder}, {Barban}, {Baudin}, {De Ridder}, {Barban}, \& {Baudin}}]{deridder09}
{De Ridder}, J., {Barban}, C., {Baudin}, F., {et~al.} 2009, \nat, 459, 298

\bibitem[{{Decressin} {et~al.}(2009){Decressin}, {Mathis}, {Palacios}, {Siess},
  {Talon}, {Charbonnel}, \& {Zahn}}]{decressin09}
{Decressin}, T., {Mathis}, S., {Palacios}, A., {et~al.} 2009, \aap, 495, 271

\bibitem[{{Dieleman} {et~al.}(2015){Dieleman}, {Willett}, \&
  {Dambre}}]{dieleman15}
{Dieleman}, S., {Willett}, K.~W., \& {Dambre}, J. 2015, \mnras, 450, 1441

\bibitem[{{do Nascimento} {et~al.}(2012){do Nascimento}, {da Costa}, \&
  {Castro}}]{do12}
{do Nascimento}, J.~D., {da Costa}, J.~S., \& {Castro}, M. 2012, \aap, 548, L1

\bibitem[{{Dom{\'\i}nguez S{\'a}nchez} {et~al.}(2018){Dom{\'\i}nguez
  S{\'a}nchez}, {Huertas-Company}, {Bernardi}, {Tuccillo}, \&
  {Fischer}}]{sanchez18}
{Dom{\'\i}nguez S{\'a}nchez}, H., {Huertas-Company}, M., {Bernardi}, M.,
  {Tuccillo}, D., \& {Fischer}, J.~L. 2018, \mnras, 476, 3661

\bibitem[{{Douglas} {et~al.}(2016){Douglas}, {Ag{\"u}eros}, {Covey}, {Cargile},
  {Barclay}, {Cody}, {Howell}, \& {Kopytova}}]{douglas16}
{Douglas}, S.~T., {Ag{\"u}eros}, M.~A., {Covey}, K.~R., {et~al.} 2016, \apj,
  822, 47

\bibitem[{{Douglas} {et~al.}(2019){Douglas}, {Curtis}, {Ag{\"u}eros},
  {Cargile}, {Brewer}, {Meibom}, \& {Jansen}}]{douglas19}
{Douglas}, S.~T., {Curtis}, J.~L., {Ag{\"u}eros}, M.~A., {et~al.} 2019, \apj,
  879, 100

\bibitem[{{Dressing} \& {Charbonneau}(2013)}]{dressing13}
{Dressing}, C.~D., \& {Charbonneau}, D. 2013, \apj, 767, 95

\bibitem[{{Freytag} \& {Steffen}(1997)}]{freytag97}
{Freytag}, B., \& {Steffen}, M. 1997, in Astronomische Gesellschaft Abstract
  Series, Vol.~13, 176

\bibitem[{{Gaia Collaboration} {et~al.}(2016){Gaia Collaboration}, {Prusti},
  {de Bruijne}, {Brown}, {Vallenari}, {Babusiaux}, {Bailer-Jones}, {Bastian},
  {Biermann}, {Evans}, {Eyer}, {Jansen}, {Jordi}, {Klioner}, {Lammers},
  {Lindegren}, {Luri}, {Mignard}, {Milligan}, {Panem}, {Poinsignon},
  {Pourbaix}, {Randich}, {Sarri}, {Sartoretti}, {Siddiqui}, {Soubiran},
  {Valette}, {van Leeuwen}, {Walton}, {Aerts}, {Arenou}, {Cropper}, {Drimmel},
  {H{\o}g}, {Katz}, {Lattanzi}, {O'Mullane}, {Grebel}, {Holland}, {Huc},
  {Passot}, {Bramante}, {Cacciari}, {Casta{\~n}eda}, {Chaoul}, {Cheek}, {De
  Angeli}, {Fabricius}, {Guerra}, {Hern{\'a}ndez}, {Jean-Antoine-Piccolo},
  {Masana}, {Messineo}, {Mowlavi}, {Nienartowicz}, {Ord{\'o}{\~n}ez-Blanco},
  {Panuzzo}, {Portell}, {Richards}, {Riello}, {Seabroke}, {Tanga},
  {Th{\'e}venin}, {Torra}, {Els}, {Gracia-Abril}, {Comoretto},
  {Garcia-Reinaldos}, {Lock}, {Mercier}, {Altmann}, {Andrae}, {Astraatmadja},
  {Bellas-Velidis}, {Benson}, {Berthier}, {Blomme}, {Busso}, {Carry},
  {Cellino}, {Clementini}, {Cowell}, {Creevey}, {Cuypers}, {Davidson}, {De
  Ridder}, {de Torres}, {Delchambre}, {Dell'Oro}, {Ducourant}, {Fr{\'e}mat},
  {Garc{\'\i}a-Torres}, {Gosset}, {Halbwachs}, {Hambly}, {Harrison}, {Hauser},
  {Hestroffer}, {Hodgkin}, {Huckle}, {Hutton}, {Jasniewicz}, {Jordan},
  {Kontizas}, {Korn}, {Lanzafame}, {Manteiga}, {Moitinho}, {Muinonen},
  {Osinde}, {Pancino}, {Pauwels}, {Petit}, {Recio-Blanco}, {Robin}, {Sarro},
  {Siopis}, {Smith}, {Smith}, {Sozzetti}, {Thuillot}, {van Reeven}, {Viala},
  {Abbas}, {Abreu Aramburu}, {Accart}, {Aguado}, {Allan}, {Allasia},
  {Altavilla}, {{\'A}lvarez}, {Alves}, {Anderson}, {Andrei}, {Anglada Varela},
  {Antiche}, {Antoja}, {Ant{\'o}n}, {Arcay}, {Atzei}, {Ayache}, {Bach},
  {Baker}, {Balaguer-N{\'u}{\~n}ez}, {Barache}, {Barata}, {Barbier}, {Barblan},
  {Baroni}, {Barrado y Navascu{\'e}s}, {Barros}, {Barstow}, {Becciani},
  {Bellazzini}, {Bellei}, {Bello Garc{\'\i}a}, {Belokurov}, {Bendjoya},
  {Berihuete}, {Bianchi}, {Bienaym{\'e}}, {Billebaud}, {Blagorodnova},
  {Blanco-Cuaresma}, {Boch}, {Bombrun}, {Borrachero}, {Bouquillon}, {Bourda},
  {Bouy}, {Bragaglia}, {Breddels}, {Brouillet}, {Br{\"u}semeister},
  {Bucciarelli}, {Budnik}, {Burgess}, {Burgon}, {Burlacu}, {Busonero}, {Buzzi},
  {Caffau}, {Cambras}, {Campbell}, {Cancelliere}, {Cantat-Gaudin}, {Carlucci},
  {Carrasco}, {Castellani}, {Charlot}, {Charnas}, {Charvet}, {Chassat},
  {Chiavassa}, {Clotet}, {Cocozza}, {Collins}, {Collins}, {Costigan}, {Crifo},
  {Cross}, {Crosta}, {Crowley}, {Dafonte}, {Damerdji}, {Dapergolas}, {David},
  {David}, {De Cat}, {de Felice}, {de Laverny}, {De Luise}, {De March}, {de
  Martino}, {de Souza}, {Debosscher}, {del Pozo}, {Delbo}, {Delgado},
  {Delgado}, {di Marco}, {Di Matteo}, {Diakite}, {Distefano}, {Dolding}, {Dos
  Anjos}, {Drazinos}, {Dur{\'a}n}, {Dzigan}, {Ecale}, {Edvardsson}, {Enke},
  {Erdmann}, {Escolar}, {Espina}, {Evans}, {Eynard Bontemps}, {Fabre},
  {Fabrizio}, {Faigler}, {Falc{\~a}o}, {Farr{\`a}s Casas}, {Faye}, {Federici},
  {Fedorets}, {Fern{\'a}ndez-Hern{\'a}ndez}, {Fernique}, {Fienga}, {Figueras},
  {Filippi}, {Findeisen}, {Fonti}, {Fouesneau}, {Fraile}, {Fraser}, {Fuchs},
  {Furnell}, {Gai}, {Galleti}, {Galluccio}, {Garabato}, {Garc{\'\i}a-Sedano},
  {Gar{\'e}}, {Garofalo}, {Garralda}, {Gavras}, {Gerssen}, {Geyer}, {Gilmore},
  {Girona}, {Giuffrida}, {Gomes}, {Gonz{\'a}lez-Marcos},
  {Gonz{\'a}lez-N{\'u}{\~n}ez}, {Gonz{\'a}lez-Vidal}, {Granvik}, {Guerrier},
  {Guillout}, {Guiraud}, {G{\'u}rpide}, {Guti{\'e}rrez-S{\'a}nchez}, {Guy},
  {Haigron}, {Hatzidimitriou}, {Haywood}, {Heiter}, {Helmi}, {Hobbs},
  {Hofmann}, {Holl}, {Holland }, {Hunt}, {Hypki}, {Icardi}, {Irwin}, {Jevardat
  de Fombelle}, {Jofr{\'e}}, {Jonker}, {Jorissen}, {Julbe}, {Karampelas},
  {Kochoska}, {Kohley}, {Kolenberg}, {Kontizas}, {Koposov}, {Kordopatis},
  {Koubsky}, {Kowalczyk}, {Krone-Martins}, {Kudryashova}, {Kull}, {Bachchan},
  {Lacoste-Seris}, {Lanza}, {Lavigne}, {Le Poncin-Lafitte}, {Lebreton},
  {Lebzelter}, {Leccia}, {Leclerc}, {Lecoeur-Taibi}, {Lemaitre}, {Lenhardt},
  {Leroux}, {Liao}, {Licata}, {Lindstr{\o}m}, {Lister}, {Livanou}, {Lobel},
  {L{\"o}ffler}, {L{\'o}pez}, {Lopez-Lozano}, {Lorenz}, {Loureiro},
  {MacDonald}, {Magalh{\~a}es Fernandes}, {Managau}, {Mann}, {Mantelet},
  {Marchal}, {Marchant}, {Marconi}, {Marie}, {Marinoni}, {Marrese},
  {Marschalk{\'o}}, {Marshall}, {Mart{\'\i}n-Fleitas}, {Martino}, {Mary},
  {Matijevi{\v{c}}}, {Mazeh}, {McMillan}, {Messina}, {Mestre}, {Michalik},
  {Millar}, {Miranda}, {Molina}, {Molinaro}, {Molinaro}, {Moln{\'a}r},
  {Moniez}, {Montegriffo}, {Monteiro}, {Mor}, {Mora}, {Morbidelli}, {Morel},
  {Morgenthaler}, {Morley}, {Morris}, {Mulone}, {Muraveva}, {Musella},
  {Narbonne}, {Nelemans}, {Nicastro}, {Noval}, {Ord{\'e}novic},
  {Ordieres-Mer{\'e}}, {Osborne}, {Pagani}, {Pagano}, {Pailler}, {Palacin},
  {Palaversa}, {Parsons}, {Paulsen}, {Pecoraro}, {Pedrosa}, {Pentik{\"a}inen},
  {Pereira}, {Pichon}, {Piersimoni}, {Pineau}, {Plachy}, {Plum}, {Poujoulet},
  {Pr{\v{s}}a}, {Pulone}, {Ragaini}, {Rago}, {Rambaux}, {Ramos-Lerate},
  {Ranalli}, {Rauw}, {Read}, {Regibo}, {Renk}, {Reyl{\'e}}, {Ribeiro},
  {Rimoldini}, {Ripepi}, {Riva}, {Rixon}, {Roelens}, {Romero-G{\'o}mez},
  {Rowell}, {Royer}, {Rudolph}, {Ruiz-Dern}, {Sadowski}, {Sagrist{\`a}
  Sell{\'e}s}, {Sahlmann}, {Salgado}, {Salguero}, {Sarasso}, {Savietto},
  {Schnorhk}, {Schultheis}, {Sciacca}, {Segol}, {Segovia}, {Segransan},
  {Serpell}, {Shih}, {Smareglia}, {Smart}, {Smith}, {Solano}, {Solitro},
  {Sordo}, {Soria Nieto}, {Souchay}, {Spagna}, {Spoto}, {Stampa}, {Steele},
  {Steidelm{\"u}ller}, {Stephenson}, {Stoev}, {Suess}, {S{\"u}veges}, {Surdej},
  {Szabados}, {Szegedi-Elek}, {Tapiador}, {Taris}, {Tauran}, {Taylor},
  {Teixeira}, {Terrett}, {Tingley}, {Trager}, {Turon}, {Ulla}, {Utrilla},
  {Valentini}, {van Elteren}, {Van Hemelryck}, {van Leeuwen}, {Varadi},
  {Vecchiato}, {Veljanoski}, {Via}, {Vicente}, {Vogt}, {Voss}, {Votruba},
  {Voutsinas}, {Walmsley}, {Weiler}, {Weingrill}, {Werner}, {Wevers},
  {Whitehead}, {Wyrzykowski}, {Yoldas}, {{\v{Z}}erjal}, {Zucker}, {Zurbach},
  {Zwitter}, {Alecu}, {Allen}, {Allende Prieto}, {Amorim},
  {Anglada-Escud{\'e}}, {Arsenijevic}, {Azaz}, {Balm}, {Beck}, {Bernstein},
  {Bigot}, {Bijaoui}, {Blasco}, {Bonfigli}, {Bono}, {Boudreault}, {Bressan},
  {Brown}, {Brunet}, {Bunclark}, {Buonanno}, {Butkevich}, {Carret}, {Carrion},
  {Chemin}, {Ch{\'e}reau}, {Corcione}, {Darmigny}, {de Boer}, {de Teodoro}, {de
  Zeeuw}, {Delle Luche}, {Domingues}, {Dubath}, {Fodor}, {Fr{\'e}zouls},
  {Fries}, {Fustes}, {Fyfe}, {Gallardo}, {Gallegos}, {Gardiol}, {Gebran},
  {Gomboc}, {G{\'o}mez}, {Grux}, {Gueguen}, {Heyrovsky}, {Hoar}, {Iannicola},
  {Isasi Parache}, {Janotto}, {Joliet}, {Jonckheere}, {Keil}, {Kim},
  {Klagyivik}, {Klar}, {Knude}, {Kochukhov}, {Kolka}, {Kos}, {Kutka}, {Lainey},
  {LeBouquin}, {Liu}, {Loreggia}, {Makarov}, {Marseille}, {Martayan},
  {Martinez-Rubi}, {Massart}, {Meynadier}, {Mignot}, {Munari}, {Nguyen},
  {Nordlander}, {Ocvirk}, {O'Flaherty}, {Olias Sanz}, {Ortiz}, {Osorio},
  {Oszkiewicz}, {Ouzounis}, {Palmer}, {Park}, {Pasquato}, {Peltzer}, {Peralta},
  {P{\'e}turaud}, {Pieniluoma}, {Pigozzi}, {Poels}, {Prat}, {Prod'homme},
  {Raison}, {Rebordao}, {Risquez}, {Rocca-Volmerange}, {Rosen}, {Ruiz-Fuertes},
  {Russo}, {Sembay}, {Serraller Vizcaino}, {Short}, {Siebert}, {Silva},
  {Sinachopoulos}, {Slezak}, {Soffel}, {Sosnowska}, {Strai{\v{z}}ys}, {ter
  Linden}, {Terrell}, {Theil}, {Tiede}, {Troisi}, {Tsalmantza}, {Tur},
  {Vaccari}, {Vachier}, {Valles}, {Van Hamme}, {Veltz}, {Virtanen}, {Wallut},
  {Wichmann}, {Wilkinson}, {Ziaeepour}, \& {Zschocke}}]{gaia16}
{Gaia Collaboration}, {Prusti}, T., {de Bruijne}, J.~H.~J., {et~al.} 2016,
  \aap, 595, A1

\bibitem[{{Gaia Collaboration} {et~al.}(2018){Gaia Collaboration}, {Brown},
  {Vallenari}, {Prusti}, {de Bruijne}, {Babusiaux}, {Bailer-Jones}, {Biermann},
  {Evans}, {Eyer}, {Jansen}, {Jordi}, {Klioner}, {Lammers}, {Lindegren},
  {Luri}, {Mignard}, {Panem}, {Pourbaix}, {Randich}, {Sartoretti}, {Siddiqui},
  {Soubiran}, {van Leeuwen}, {Walton}, {Arenou}, {Bastian}, {Cropper},
  {Drimmel}, {Katz}, {Lattanzi}, {Bakker}, {Cacciari}, {Casta{\~n}eda},
  {Chaoul}, {Cheek}, {De Angeli}, {Fabricius}, {Guerra}, {Holl}, {Masana},
  {Messineo}, {Mowlavi}, {Nienartowicz}, {Panuzzo}, {Portell}, {Riello},
  {Seabroke}, {Tanga}, {Th{\'e}venin}, {Gracia-Abril}, {Comoretto},
  {Garcia-Reinaldos}, {Teyssier}, {Altmann}, {Andrae}, {Audard},
  {Bellas-Velidis}, {Benson}, {Berthier}, {Blomme}, {Burgess}, {Busso},
  {Carry}, {Cellino}, {Clementini}, {Clotet}, {Creevey}, {Davidson}, {De
  Ridder}, {Delchambre}, {Dell'Oro}, {Ducourant},
  {Fern{\'a}ndez-Hern{\'a}ndez}, {Fouesneau}, {Fr{\'e}mat}, {Galluccio},
  {Garc{\'\i}a-Torres}, {Gonz{\'a}lez-N{\'u}{\~n}ez}, {Gonz{\'a}lez-Vidal},
  {Gosset}, {Guy}, {Halbwachs}, {Hambly}, {Harrison}, {Hern{\'a}ndez},
  {Hestroffer}, {Hodgkin}, {Hutton}, {Jasniewicz}, {Jean-Antoine-Piccolo},
  {Jordan}, {Korn}, {Krone-Martins}, {Lanzafame}, {Lebzelter}, {L{\"o}ffler},
  {Manteiga}, {Marrese}, {Mart{\'\i}n-Fleitas}, {Moitinho}, {Mora}, {Muinonen},
  {Osinde}, {Pancino}, {Pauwels}, {Petit}, {Recio-Blanco}, {Richards},
  {Rimoldini}, {Robin}, {Sarro}, {Siopis}, {Smith}, {Sozzetti}, {S{\"u}veges},
  {Torra}, {van Reeven}, {Abbas}, {Abreu Aramburu}, {Accart}, {Aerts},
  {Altavilla}, {{\'A}lvarez}, {Alvarez}, {Alves}, {Anderson}, {Andrei},
  {Anglada Varela}, {Antiche}, {Antoja}, {Arcay}, {Astraatmadja}, {Bach},
  {Baker}, {Balaguer-N{\'u}{\~n}ez}, {Balm}, {Barache}, {Barata}, {Barbato},
  {Barblan}, {Barklem}, {Barrado}, {Barros}, {Barstow}, {Bartholom{\'e}
  Mu{\~n}oz}, {Bassilana}, {Becciani}, {Bellazzini}, {Berihuete}, {Bertone},
  {Bianchi}, {Bienaym{\'e}}, {Blanco-Cuaresma}, {Boch}, {Boeche}, {Bombrun},
  {Borrachero}, {Bossini}, {Bouquillon}, {Bourda}, {Bragaglia}, {Bramante},
  {Breddels}, {Bressan}, {Brouillet}, {Br{\"u}semeister}, {Brugaletta},
  {Bucciarelli}, {Burlacu}, {Busonero}, {Butkevich}, {Buzzi}, {Caffau},
  {Cancelliere}, {Cannizzaro}, {Cantat-Gaudin}, {Carballo}, {Carlucci},
  {Carrasco}, {Casamiquela}, {Castellani}, {Castro-Ginard}, {Charlot},
  {Chemin}, {Chiavassa}, {Cocozza}, {Costigan}, {Cowell}, {Crifo}, {Crosta},
  {Crowley}, {Cuypers}, {Dafonte}, {Damerdji}, {Dapergolas}, {David}, {David},
  {de Laverny}, {De Luise}, {De March}, {de Martino}, {de Souza}, {de Torres},
  {Debosscher}, {del Pozo}, {Delbo}, {Delgado}, {Delgado}, {Di Matteo},
  {Diakite}, {Diener}, {Distefano}, {Dolding}, {Drazinos}, {Dur{\'a}n},
  {Edvardsson}, {Enke}, {Eriksson}, {Esquej}, {Eynard Bontemps}, {Fabre},
  {Fabrizio}, {Faigler}, {Falc{\~a}o}, {Farr{\`a}s Casas}, {Federici},
  {Fedorets}, {Fernique}, {Figueras}, {Filippi}, {Findeisen}, {Fonti},
  {Fraile}, {Fraser}, {Fr{\'e}zouls}, {Gai}, {Galleti}, {Garabato},
  {Garc{\'\i}a-Sedano}, {Garofalo}, {Garralda}, {Gavel}, {Gavras}, {Gerssen},
  {Geyer}, {Giacobbe}, {Gilmore}, {Girona}, {Giuffrida}, {Glass}, {Gomes},
  {Granvik}, {Gueguen}, {Guerrier}, {Guiraud}, {Guti{\'e}rrez-S{\'a}nchez},
  {Haigron}, {Hatzidimitriou}, {Hauser}, {Haywood}, {Heiter}, {Helmi}, {Heu},
  {Hilger}, {Hobbs}, {Hofmann}, {Holland}, {Huckle}, {Hypki}, {Icardi},
  {Jan{\ss}en}, {Jevardat de Fombelle}, {Jonker}, {Juh{\'a}sz}, {Julbe},
  {Karampelas}, {Kewley}, {Klar}, {Kochoska}, {Kohley}, {Kolenberg},
  {Kontizas}, {Kontizas}, {Koposov}, {Kordopatis}, {Kostrzewa-Rutkowska},
  {Koubsky}, {Lambert}, {Lanza}, {Lasne}, {Lavigne}, {Le Fustec}, {Le
  Poncin-Lafitte}, {Lebreton}, {Leccia}, {Leclerc}, {Lecoeur-Taibi},
  {Lenhardt}, {Leroux}, {Liao}, {Licata}, {Lindstr{\o}m}, {Lister}, {Livanou},
  {Lobel}, {L{\'o}pez}, {Managau}, {Mann}, {Mantelet}, {Marchal}, {Marchant},
  {Marconi}, {Marinoni}, {Marschalk{\'o}}, {Marshall}, {Martino}, {Marton},
  {Mary}, {Massari}, {Matijevi{\v{c}}}, {Mazeh}, {McMillan}, {Messina},
  {Michalik}, {Millar}, {Molina}, {Molinaro}, {Moln{\'a}r}, {Montegriffo},
  {Mor}, {Morbidelli}, {Morel}, {Morris}, {Mulone}, {Muraveva}, {Musella},
  {Nelemans}, {Nicastro}, {Noval}, {O'Mullane}, {Ord{\'e}novic},
  {Ord{\'o}{\~n}ez-Blanco}, {Osborne}, {Pagani}, {Pagano}, {Pailler},
  {Palacin}, {Palaversa}, {Panahi}, {Pawlak}, {Piersimoni}, {Pineau}, {Plachy},
  {Plum}, {Poggio}, {Poujoulet}, {Pr{\v{s}}a}, {Pulone}, {Racero}, {Ragaini},
  {Rambaux}, {Ramos-Lerate}, {Regibo}, {Reyl{\'e}}, {Riclet}, {Ripepi}, {Riva},
  {Rivard}, {Rixon}, {Roegiers}, {Roelens}, {Romero-G{\'o}mez}, {Rowell},
  {Royer}, {Ruiz-Dern}, {Sadowski}, {Sagrist{\`a} Sell{\'e}s}, {Sahlmann},
  {Salgado}, {Salguero}, {Sanna}, {Santana-Ros}, {Sarasso}, {Savietto},
  {Schultheis}, {Sciacca}, {Segol}, {Segovia}, {S{\'e}gransan}, {Shih},
  {Siltala}, {Silva}, {Smart}, {Smith}, {Solano}, {Solitro}, {Sordo}, {Soria
  Nieto}, {Souchay}, {Spagna}, {Spoto}, {Stampa}, {Steele},
  {Steidelm{\"u}ller}, {Stephenson}, {Stoev}, {Suess}, {Surdej}, {Szabados},
  {Szegedi-Elek}, {Tapiador}, {Taris}, {Tauran}, {Taylor}, {Teixeira},
  {Terrett}, {Teyssand ier}, {Thuillot}, {Titarenko}, {Torra Clotet}, {Turon},
  {Ulla}, {Utrilla}, {Uzzi}, {Vaillant}, {Valentini}, {Valette}, {van Elteren},
  {Van Hemelryck}, {van Leeuwen}, {Vaschetto}, {Vecchiato}, {Veljanoski},
  {Viala}, {Vicente}, {Vogt}, {von Essen}, {Voss}, {Votruba}, {Voutsinas},
  {Walmsley}, {Weiler}, {Wertz}, {Wevers}, {Wyrzykowski}, {Yoldas},
  {{\v{Z}}erjal}, {Ziaeepour}, {Zorec}, {Zschocke}, {Zucker}, {Zurbach}, \&
  {Zwitter}}]{gaia18}
{Gaia Collaboration}, {Brown}, A.~G.~A., {Vallenari}, A., {et~al.} 2018, \aap,
  616, A1

\bibitem[{{Gallet} \& {Bouvier}(2013)}]{gallet13}
{Gallet}, F., \& {Bouvier}, J. 2013, \aap, 556, A36

\bibitem[{{Garc{\'\i}a} {et~al.}(2010){Garc{\'\i}a}, {Mathur}, {Salabert},
  {Ballot}, {R{\'e}gulo}, {Metcalfe}, \& {Baglin}}]{garcia10}
{Garc{\'\i}a}, R.~A., {Mathur}, S., {Salabert}, D., {et~al.} 2010, Science,
  329, 1032

\bibitem[{{Garc{\'\i}a} {et~al.}(2014){Garc{\'\i}a}, {Ceillier}, {Salabert},
  {Mathur}, {van Saders}, {Pinsonneault}, {Ballot}, {Beck}, {Bloemen},
  {Campante}, {Davies}, {do Nascimento}, {Mathis}, {Metcalfe}, {Nielsen},
  {Su{\'a}rez}, {Chaplin}, {Jim{\'e}nez}, \& {Karoff}}]{garcia14}
{Garc{\'\i}a}, R.~A., {Ceillier}, T., {Salabert}, D., {et~al.} 2014, \aap, 572,
  A34

\bibitem[{{Gilliland} {et~al.}(2010){Gilliland}, {Brown},
  {Christensen-Dalsgaard}, {Kjeldsen}, {Aerts}, {Appourchaux}, {Basu},
  {Bedding}, {Chaplin}, {Cunha}, {De Cat}, {De Ridder}, {Guzik}, {Handler},
  {Kawaler}, {Kiss}, {Kolenberg}, {Kurtz}, {Metcalfe}, {Monteiro}, {Szab{\'o}},
  {Arentoft}, {Balona}, {Debosscher}, {Elsworth}, {Quirion}, {Stello},
  {Su{\'a}rez}, {Borucki}, {Jenkins}, {Koch}, {Kondo}, {Latham}, {Rowe}, \&
  {Steffen}}]{gilliland10}
{Gilliland}, R.~L., {Brown}, T.~M., {Christensen-Dalsgaard}, J., {et~al.} 2010,
  \pasp, 122, 131

\bibitem[{{Goodfellow} {et~al.}(2014){Goodfellow}, {Pouget-Abadie}, {Mirza},
  {Xu}, {Warde-Farley}, {Ozair}, {Courville}, \& {Bengio}}]{gans}
{Goodfellow}, I.~J., {Pouget-Abadie}, J., {Mirza}, M., {et~al.} 2014, arXiv
  e-prints, arXiv:1406.2661

\bibitem[{{He} {et~al.}(2015){He}, {Zhang}, {Ren}, \& {Sun}}]{resnet}
{He}, K., {Zhang}, X., {Ren}, S., \& {Sun}, J. 2015, arXiv e-prints,
  arXiv:1512.03385

\bibitem[{{Hekker} {et~al.}(2009){Hekker}, {Kallinger}, {Baudin}, {De Ridder},
  {Barban}, {Carrier}, {Hatzes}, {Weiss}, \& {Baglin}}]{hekker09}
{Hekker}, S., {Kallinger}, T., {Baudin}, F., {et~al.} 2009, \aap, 506, 465

\bibitem[{{Hinton} {et~al.}(2012){Hinton}, {Srivastava}, {Krizhevsky},
  {Sutskever}, \& {Salakhutdinov}}]{hinton12}
{Hinton}, G.~E., {Srivastava}, N., {Krizhevsky}, A., {Sutskever}, I., \&
  {Salakhutdinov}, R.~R. 2012, arXiv e-prints, arXiv:1207.0580

\bibitem[{{Hon} {et~al.}(2017){Hon}, {Stello}, \& {Yu}}]{hon17}
{Hon}, M., {Stello}, D., \& {Yu}, J. 2017, \mnras, 469, 4578

\bibitem[{{Hon} {et~al.}(2018{\natexlab{a}}){Hon}, {Stello}, \& {Yu}}]{hon18-2}
---. 2018{\natexlab{a}}, \mnras, 476, 3233

\bibitem[{{Hon} {et~al.}(2018{\natexlab{b}}){Hon}, {Stello}, \& {Zinn}}]{hon18}
{Hon}, M., {Stello}, D., \& {Zinn}, J.~C. 2018{\natexlab{b}}, \apj, 859, 64

\bibitem[{Hornik(1991)}]{hornik91}
Hornik, K. 1991, Neural Networks, 4, 251

\bibitem[{Hornik {et~al.}(1990)Hornik, Stinchcombe, \& White}]{hornik90}
Hornik, K., Stinchcombe, M., \& White, H. 1990, Neural Networks, 3, 551

\bibitem[{{Huber} {et~al.}(2009){Huber}, {Stello}, {Bedding}, {Chaplin},
  {Arentoft}, {Quirion}, \& {Kjeldsen}}]{huber09}
{Huber}, D., {Stello}, D., {Bedding}, T.~R., {et~al.} 2009, Communications in
  Asteroseismology, 160, 74

\bibitem[{{Huber} {et~al.}(2011){Huber}, {Bedding}, {Stello}, {Hekker},
  {Mathur}, {Mosser}, {Verner}, {Bonanno}, {Buzasi}, {Campante}, {Elsworth},
  {Hale}, {Kallinger}, {Silva Aguirre}, {Chaplin}, {De Ridder}, {Garc{\'\i}a},
  {Appourchaux}, {Frandsen}, {Houdek}, {Molenda-{\.Z}akowicz}, {Monteiro},
  {Christensen-Dalsgaard}, {Gilliland}, {Kawaler}, {Kjeldsen}, {Broomhall},
  {Corsaro}, {Salabert}, {Sanderfer}, {Seader}, \& {Smith}}]{huber11}
{Huber}, D., {Bedding}, T.~R., {Stello}, D., {et~al.} 2011, \apj, 743, 143

\bibitem[{{Huertas-Company} {et~al.}(2018){Huertas-Company}, {Primack},
  {Dekel}, {Koo}, {Lapiner}, {Ceverino}, {Simons}, {Snyder}, {Bernardi},
  {Chen}, {Dom{\'\i}nguez-S{\'a}nchez}, {Lee}, {Margalef-Bentabol}, \&
  {Tuccillo}}]{huertascompany18}
{Huertas-Company}, M., {Primack}, J.~R., {Dekel}, A., {et~al.} 2018, \apj, 858,
  114

\bibitem[{{Ioffe} \& {Szegedy}(2015)}]{batchnorm}
{Ioffe}, S., \& {Szegedy}, C. 2015, arXiv e-prints, arXiv:1502.03167

\bibitem[{{Irwin} {et~al.}(2009){Irwin}, {Aigrain}, {Bouvier}, {Hebb},
  {Hodgkin}, {Irwin}, \& {Moraux}}]{irwin09}
{Irwin}, J., {Aigrain}, S., {Bouvier}, J., {et~al.} 2009, \mnras, 392, 1456

\bibitem[{{Ismail Fawaz} {et~al.}(2018){Ismail Fawaz}, {Forestier}, {Weber},
  {Idoumghar}, \& {Muller}}]{fawaz18}
{Ismail Fawaz}, H., {Forestier}, G., {Weber}, J., {Idoumghar}, L., \& {Muller},
  P.-A. 2018, arXiv e-prints, arXiv:1809.04356

\bibitem[{{Ivezi{\'c}} {et~al.}(2014){Ivezi{\'c}}, {Connelly}, {VanderPlas}, \&
  {Gray}}]{ivezic}
{Ivezi{\'c}}, {\v Z}., {Connelly}, A.~J., {VanderPlas}, J.~T., \& {Gray}, A.
  2014, {Statistics, Data Mining, and Machine Learning in Astronomy}

\bibitem[{{Jenkins} {et~al.}(2010){Jenkins}, {Caldwell}, {Chandrasekaran},
  {Twicken}, {Bryson}, {Quintana}, {Clarke}, {Li}, {Allen}, {Tenenbaum}, {Wu},
  {Klaus}, {Middour}, {Cote}, {McCauliff}, {Girouard}, {Gunter}, {Wohler},
  {Sommers}, {Hall}, {Uddin}, {Wu}, {Bhavsar}, {Van Cleve}, {Pletcher},
  {Dotson}, {Haas}, {Gilliland}, {Koch}, \& {Borucki}}]{jenkins10}
{Jenkins}, J.~M., {Caldwell}, D.~A., {Chandrasekaran}, H., {et~al.} 2010,
  \apjl, 713, L87

\bibitem[{{Kallinger} {et~al.}(2016){Kallinger}, {Hekker}, {Garcia}, {Huber},
  \& {Matthews}}]{kallinger16}
{Kallinger}, T., {Hekker}, S., {Garcia}, R.~A., {Huber}, D., \& {Matthews},
  J.~M. 2016, Science Advances, 2, 1500654

\bibitem[{{Kallinger} {et~al.}(2010){Kallinger}, {Weiss}, {Barban}, {Baudin},
  {Cameron}, {Carrier}, {De Ridder}, {Goupil}, {Gruberbauer}, {Hatzes},
  {Hekker}, {Samadi}, \& {Deleuil}}]{kallinger10}
{Kallinger}, T., {Weiss}, W.~W., {Barban}, C., {et~al.} 2010, \aap, 509, A77

\bibitem[{{Kawaler}(1988)}]{kawaler88}
{Kawaler}, S.~D. 1988, \apj, 333, 236

\bibitem[{{Kawaler}(1989)}]{kawaler89}
---. 1989, \apjl, 343, L65

\bibitem[{{Kingma} \& {Ba}(2014)}]{adam}
{Kingma}, D.~P., \& {Ba}, J. 2014, ArXiv e-prints, arXiv:1412.6980

\bibitem[{{Kjeldsen} \& {Bedding}(1995)}]{kjeldsen95}
{Kjeldsen}, H., \& {Bedding}, T.~R. 1995, \aap, 293, 87

\bibitem[{{Kjeldsen} \& {Bedding}(2011)}]{kjeldsen11}
---. 2011, \aap, 529, L8

\bibitem[{Krizhevsky {et~al.}(2017)Krizhevsky, Sutskever, \& Hinton}]{alexnet}
Krizhevsky, A., Sutskever, I., \& Hinton, G.~E. 2017, Commun. ACM, 60, 84–90

\bibitem[{{Lex} {et~al.}(2014){Lex}, N., H., R., \& H.}]{upset}
{Lex}, A., N., G., H., S., R., V., \& H., P. 2014, IEEE, 20, 1983

\bibitem[{{Lipton} {et~al.}(2015){Lipton}, {Berkowitz}, \& {Elkan}}]{rnnreview}
{Lipton}, Z.~C., {Berkowitz}, J., \& {Elkan}, C. 2015, arXiv e-prints,
  arXiv:1506.00019

\bibitem[{{Lomb}(1976)}]{lomb76}
{Lomb}, N.~R. 1976, \apss, 39, 447

\bibitem[{{Loshchilov} \& {Hutter}(2017)}]{adamw}
{Loshchilov}, I., \& {Hutter}, F. 2017, arXiv e-prints, arXiv:1711.05101

\bibitem[{{LSST Science Collaboration} {et~al.}(2009){LSST Science
  Collaboration}, {Abell}, {Allison}, {Anderson}, {Andrew}, {Angel}, {Armus},
  {Arnett}, {Asztalos}, {Axelrod}, {Bailey}, {Ballantyne}, {Bankert},
  {Barkhouse}, {Barr}, {Barrientos}, {Barth}, {Bartlett}, {Becker}, {Becla},
  {Beers}, {Bernstein}, {Biswas}, {Blanton}, {Bloom}, {Bochanski}, {Boeshaar},
  {Borne}, {Bradac}, {Brandt}, {Bridge}, {Brown}, {Brunner}, {Bullock},
  {Burgasser}, {Burge}, {Burke}, {Cargile}, {Chand rasekharan}, {Chartas},
  {Chesley}, {Chu}, {Cinabro}, {Claire}, {Claver}, {Clowe}, {Connolly}, {Cook},
  {Cooke}, {Cooray}, {Covey}, {Culliton}, {de Jong}, {de Vries}, {Debattista},
  {Delgado}, {Dell'Antonio}, {Dhital}, {Di Stefano}, {Dickinson}, {Dilday},
  {Djorgovski}, {Dobler}, {Donalek}, {Dubois-Felsmann}, {Durech},
  {Eliasdottir}, {Eracleous}, {Eyer}, {Falco}, {Fan}, {Fassnacht}, {Ferguson},
  {Fernandez}, {Fields}, {Finkbeiner}, {Figueroa}, {Fox}, {Francke}, {Frank},
  {Frieman}, {Fromenteau}, {Furqan}, {Galaz}, {Gal-Yam}, {Garnavich},
  {Gawiser}, {Geary}, {Gee}, {Gibson}, {Gilmore}, {Grace}, {Green}, {Gressler},
  {Grillmair}, {Habib}, {Haggerty}, {Hamuy}, {Harris}, {Hawley}, {Heavens},
  {Hebb}, {Henry}, {Hileman}, {Hilton}, {Hoadley}, {Holberg}, {Holman},
  {Howell}, {Infante}, {Ivezic}, {Jacoby}, {Jain}, {R}, {Jedicke}, {Jee},
  {Garrett Jernigan}, {Jha}, {Johnston}, {Jones}, {Juric}, {Kaasalainen},
  {Styliani}, {Kafka}, {Kahn}, {Kaib}, {Kalirai}, {Kantor}, {Kasliwal},
  {Keeton}, {Kessler}, {Knezevic}, {Kowalski}, {Krabbendam}, {Krughoff},
  {Kulkarni}, {Kuhlman}, {Lacy}, {Lepine}, {Liang}, {Lien}, {Lira}, {Long},
  {Lorenz}, {Lotz}, {Lupton}, {Lutz}, {Macri}, {Mahabal}, {Mandelbaum},
  {Marshall}, {May}, {McGehee}, {Meadows}, {Meert}, {Milani}, {Miller},
  {Miller}, {Mills}, {Minniti}, {Monet}, {Mukadam}, {Nakar}, {Neill}, {Newman},
  {Nikolaev}, {Nordby}, {O'Connor}, {Oguri}, {Oliver}, {Olivier}, {Olsen},
  {Olsen}, {Olszewski}, {Oluseyi}, {Padilla}, {Parker}, {Pepper}, {Peterson},
  {Petry}, {Pinto}, {Pizagno}, {Popescu}, {Prsa}, {Radcka}, {Raddick},
  {Rasmussen}, {Rau}, {Rho}, {Rhoads}, {Richards}, {Ridgway}, {Robertson},
  {Roskar}, {Saha}, {Sarajedini}, {Scannapieco}, {Schalk}, {Schindler},
  {Schmidt}, {Schmidt}, {Schneider}, {Schumacher}, {Scranton}, {Sebag},
  {Seppala}, {Shemmer}, {Simon}, {Sivertz}, {Smith}, {Allyn Smith}, {Smith},
  {Spitz}, {Stanford}, {Stassun}, {Strader}, {Strauss}, {Stubbs}, {Sweeney},
  {Szalay}, {Szkody}, {Takada}, {Thorman}, {Trilling}, {Trimble}, {Tyson}, {Van
  Berg}, {Vand en Berk}, {VanderPlas}, {Verde}, {Vrsnak}, {Walkowicz}, {Wand
  elt}, {Wang}, {Wang}, {Warner}, {Wechsler}, {West}, {Wiecha}, {Williams},
  {Willman}, {Wittman}, {Wolff}, {Wood-Vasey}, {Wozniak}, {Young}, {Zentner},
  \& {Zhan}}]{lsst}
{LSST Science Collaboration}, {Abell}, P.~A., {Allison}, J., {et~al.} 2009,
  arXiv e-prints, arXiv:0912.0201

\bibitem[{{Lund} {et~al.}(2017){Lund}, {Silva Aguirre}, {Davies}, {Chaplin},
  {Christensen-Dalsgaard}, {Houdek}, {White}, {Bedding}, {Ball}, {Huber},
  {Antia}, {Lebreton}, {Latham}, {Handberg}, {Verma}, {Basu}, {Casagrande},
  {Justesen}, {Kjeldsen}, \& {Mosumgaard}}]{lund17}
{Lund}, M.~N., {Silva Aguirre}, V., {Davies}, G.~R., {et~al.} 2017, \apj, 835,
  172

\bibitem[{{Mamajek} \& {Hillenbrand}(2008)}]{mamajek08}
{Mamajek}, E.~E., \& {Hillenbrand}, L.~A. 2008, \apj, 687, 1264

\bibitem[{{Mathis} {et~al.}(2004){Mathis}, {Palacios}, \& {Zahn}}]{mathis04}
{Mathis}, S., {Palacios}, A., \& {Zahn}, J.~P. 2004, \aap, 425, 243

\bibitem[{{Mathur} {et~al.}(2011){Mathur}, {Hekker}, {Trampedach}, {Ballot},
  {Kallinger}, {Buzasi}, {Garc{\'\i}a}, {Huber}, {Jim{\'e}nez}, {Mosser},
  {Bedding}, {Elsworth}, {R{\'e}gulo}, {Stello}, {Chaplin}, {De Ridder},
  {Hale}, {Kinemuchi}, {Kjeldsen}, {Mullally}, \& {Thompson}}]{mathur11}
{Mathur}, S., {Hekker}, S., {Trampedach}, R., {et~al.} 2011, \apj, 741, 119

\bibitem[{{Mathur} {et~al.}(2017){Mathur}, {Huber}, {Batalha}, {Ciardi},
  {Bastien}, {Bieryla}, {Buchhave}, {Cochran}, {Endl}, {Esquerdo}, {Furlan},
  {Howard}, {Howell}, {Isaacson}, {Latham}, {MacQueen}, \& {Silva}}]{mathur17}
{Mathur}, S., {Huber}, D., {Batalha}, N.~M., {et~al.} 2017, \apjs, 229, 30

\bibitem[{{Matt} {et~al.}(2012){Matt}, {MacGregor}, {Pinsonneault}, \&
  {Greene}}]{matt12}
{Matt}, S.~P., {MacGregor}, K.~B., {Pinsonneault}, M.~H., \& {Greene}, T.~P.
  2012, \apjl, 754, L26

\bibitem[{{McQuillan} {et~al.}(2013){McQuillan}, {Aigrain}, \&
  {Mazeh}}]{mcquillan13a}
{McQuillan}, A., {Aigrain}, S., \& {Mazeh}, T. 2013, \mnras, 432, 1203

\bibitem[{{McQuillan} {et~al.}(2014){McQuillan}, {Mazeh}, \&
  {Aigrain}}]{mcquillan14}
{McQuillan}, A., {Mazeh}, T., \& {Aigrain}, S. 2014, \apjs, 211, 24

\bibitem[{{Meibom} {et~al.}(2009){Meibom}, {Mathieu}, \& {Stassun}}]{meibom09}
{Meibom}, S., {Mathieu}, R.~D., \& {Stassun}, K.~G. 2009, \apj, 695, 679

\bibitem[{{Montavon} {et~al.}(2017){Montavon}, {Samek}, \&
  {M{\"u}ller}}]{montavon17}
{Montavon}, G., {Samek}, W., \& {M{\"u}ller}, K.-R. 2017, arXiv e-prints,
  arXiv:1706.07979

\bibitem[{{Mosser} {et~al.}(2009){Mosser}, {Baudin}, {Lanza}, {Hulot},
  {Catala}, {Baglin}, \& {Auvergne}}]{mosser09}
{Mosser}, B., {Baudin}, F., {Lanza}, A.~F., {et~al.} 2009, \aap, 506, 245

\bibitem[{{Mosser} {et~al.}(2010){Mosser}, {Belkacem}, {Goupil}, {Miglio},
  {Morel}, {Barban}, {Baudin}, {Hekker}, {Samadi}, {De Ridder}, {Weiss},
  {Auvergne}, \& {Baglin}}]{mosser10}
{Mosser}, B., {Belkacem}, K., {Goupil}, M.~J., {et~al.} 2010, \aap, 517, A22

\bibitem[{{Ness} {et~al.}(2015){Ness}, {Hogg}, {Rix}, {Ho}, \&
  {Zasowski}}]{ness15}
{Ness}, M., {Hogg}, D.~W., {Rix}, H.~W., {Ho}, A. Y.~Q., \& {Zasowski}, G.
  2015, \apj, 808, 16

\bibitem[{{Ness} {et~al.}(2018){Ness}, {Silva Aguirre}, {Lund}, {Cantiello},
  {Foreman-Mackey}, {Hogg}, \& {Angus}}]{ness18}
{Ness}, M.~K., {Silva Aguirre}, V., {Lund}, M.~N., {et~al.} 2018, \apj, 866, 15

\bibitem[{{Nielsen} {et~al.}(2013){Nielsen}, {Gizon}, {Schunker}, \&
  {Karoff}}]{nielsen13}
{Nielsen}, M.~B., {Gizon}, L., {Schunker}, H., \& {Karoff}, C. 2013, \aap, 557,
  L10

\bibitem[{{Pande} {et~al.}(2018){Pande}, {Bedding}, {Huber}, \&
  {Kjeldsen}}]{pande18}
{Pande}, D., {Bedding}, T.~R., {Huber}, D., \& {Kjeldsen}, H. 2018, \mnras,
  480, 467

\bibitem[{Paszke {et~al.}(2017)Paszke, Gross, Chintala, Chanan, Yang, DeVito,
  Lin, Desmaison, Antiga, \& Lerer}]{pytorch}
Paszke, A., Gross, S., Chintala, S., {et~al.} 2017, in NIPS-W

\bibitem[{Pedregosa {et~al.}(2011)Pedregosa, Varoquaux, Gramfort, Michel,
  Thirion, Grisel, Blondel, Prettenhofer, Weiss, Dubourg, Vanderplas, Passos,
  Cournapeau, Brucher, Perrot, \& Duchesnay}]{scikit-learn}
Pedregosa, F., Varoquaux, G., Gramfort, A., {et~al.} 2011, Journal of Machine
  Learning Research, 12, 2825

\bibitem[{{Price-Whelan} {et~al.}(2018){Price-Whelan}, {Sip{\H{o}}cz},
  {G{\"u}nther}, {Lim}, {Crawford}, {Conseil}, {Shupe}, {Craig}, {Dencheva},
  {Ginsburg}, {VanderPlas}, {Bradley}, {P{\'e}rez-Su{\'a}rez}, {de Val-Borro},
  {Paper Contributors}, {Aldcroft}, {Cruz}, {Robitaille}, {Tollerud},
  {Coordination Committee}, {Ardelean}, {Babej}, {Bach}, {Bachetti}, {Bakanov},
  {Bamford}, {Barentsen}, {Barmby}, {Baumbach}, {Berry}, {Biscani}, {Boquien},
  {Bostroem}, {Bouma}, {Brammer}, {Bray}, {Breytenbach}, {Buddelmeijer},
  {Burke}, {Calderone}, {Cano Rodr{\'\i}guez}, {Cara}, {Cardoso}, {Cheedella},
  {Copin}, {Corrales}, {Crichton}, {D{\textquoteright}Avella}, {Deil},
  {Depagne}, {Dietrich}, {Donath}, {Droettboom}, {Earl}, {Erben}, {Fabbro},
  {Ferreira}, {Finethy}, {Fox}, {Garrison}, {Gibbons}, {Goldstein}, {Gommers},
  {Greco}, {Greenfield}, {Groener}, {Grollier}, {Hagen}, {Hirst}, {Homeier},
  {Horton}, {Hosseinzadeh}, {Hu}, {Hunkeler}, {Ivezi{\'c}}, {Jain}, {Jenness},
  {Kanarek}, {Kendrew}, {Kern}, {Kerzendorf}, {Khvalko}, {King}, {Kirkby},
  {Kulkarni}, {Kumar}, {Lee}, {Lenz}, {Littlefair}, {Ma}, {Macleod},
  {Mastropietro}, {McCully}, {Montagnac}, {Morris}, {Mueller}, {Mumford},
  {Muna}, {Murphy}, {Nelson}, {Nguyen}, {Ninan}, {N{\"o}the}, {Ogaz}, {Oh},
  {Parejko}, {Parley}, {Pascual}, {Patil}, {Patil}, {Plunkett}, {Prochaska},
  {Rastogi}, {Reddy Janga}, {Sabater}, {Sakurikar}, {Seifert}, {Sherbert},
  {Sherwood-Taylor}, {Shih}, {Sick}, {Silbiger}, {Singanamalla}, {Singer},
  {Sladen}, {Sooley}, {Sornarajah}, {Streicher}, {Teuben}, {Thomas},
  {Tremblay}, {Turner}, {Terr{\'o}n}, {van Kerkwijk}, {de la Vega}, {Watkins},
  {Weaver}, {Whitmore}, {Woillez}, {Zabalza}, \& {Contributors}}]{astropy}
{Price-Whelan}, A.~M., {Sip{\H{o}}cz}, B.~M., {G{\"u}nther}, H.~M., {et~al.}
  2018, \aj, 156, 123

\bibitem[{{Reiners} \& {Mohanty}(2012)}]{reiners12}
{Reiners}, A., \& {Mohanty}, S. 2012, \apj, 746, 43

\bibitem[{{Reinhold} \& {Hekker}(2020)}]{reinhold20}
{Reinhold}, T., \& {Hekker}, S. 2020, \aap, 635, A43

\bibitem[{{Reinhold} {et~al.}(2013){Reinhold}, {Reiners}, \&
  {Basri}}]{reinhold13}
{Reinhold}, T., {Reiners}, A., \& {Basri}, G. 2013, \aap, 560, A4

\bibitem[{{Ricker} {et~al.}(2014){Ricker}, {Winn}, {Vanderspek}, {Latham},
  {Bakos}, {Bean}, {Berta-Thompson}, {Brown}, {Buchhave}, {Butler}, {Butler},
  {Chaplin}, {Charbonneau}, {Christensen-Dalsgaard}, {Clampin}, {Deming},
  {Doty}, {De Lee}, {Dressing}, {Dunham}, {Endl}, {Fressin}, {Ge}, {Henning},
  {Holman}, {Howard}, {Ida}, {Jenkins}, {Jernigan}, {Johnson}, {Kaltenegger},
  {Kawai}, {Kjeldsen}, {Laughlin}, {Levine}, {Lin}, {Lissauer}, {MacQueen},
  {Marcy}, {McCullough}, {Morton}, {Narita}, {Paegert}, {Palle}, {Pepe},
  {Pepper}, {Quirrenbach}, {Rinehart}, {Sasselov}, {Sato}, {Seager},
  {Sozzetti}, {Stassun}, {Sullivan}, {Szentgyorgyi}, {Torres}, {Udry}, \&
  {Villasenor}}]{ricker14}
{Ricker}, G.~R., {Winn}, J.~N., {Vanderspek}, R., {et~al.} 2014, in Society of
  Photo-Optical Instrumentation Engineers (SPIE) Conference Series, Vol. 9143,
  \procspie, 914320

\bibitem[{{Ronneberger} {et~al.}(2015){Ronneberger}, {Fischer}, \&
  {Brox}}]{unet}
{Ronneberger}, O., {Fischer}, P., \& {Brox}, T. 2015, arXiv e-prints,
  arXiv:1505.04597

\bibitem[{{Santos} {et~al.}(2019){Santos}, {Garc{\'\i}a}, {Mathur}, {Bugnet},
  {van Saders}, {Metcalfe}, {Simonian}, \& {Pinsonneault}}]{santos19}
{Santos}, A.~R.~G., {Garc{\'\i}a}, R.~A., {Mathur}, S., {et~al.} 2019, \apjs,
  244, 21

\bibitem[{{Scargle}(1982)}]{scargle82}
{Scargle}, J.~D. 1982, \apj, 263, 835

\bibitem[{{Simonyan} {et~al.}(2013){Simonyan}, {Vedaldi}, \&
  {Zisserman}}]{simonyan13}
{Simonyan}, K., {Vedaldi}, A., \& {Zisserman}, A. 2013, arXiv e-prints,
  arXiv:1312.6034

\bibitem[{{Simonyan} \& {Zisserman}(2014)}]{vgg}
{Simonyan}, K., \& {Zisserman}, A. 2014, arXiv e-prints, arXiv:1409.1556

\bibitem[{{Skumanich}(1972)}]{skumanich72}
{Skumanich}, A. 1972, \apj, 171, 565

\bibitem[{{Spada} \& {Lanzafame}(2020)}]{spada20}
{Spada}, F., \& {Lanzafame}, A.~C. 2020, \aap, 636, A76

\bibitem[{{Stello} {et~al.}(2009{\natexlab{a}}){Stello}, {Chaplin}, {Basu},
  {Elsworth}, \& {Bedding}}]{stello09a}
{Stello}, D., {Chaplin}, W.~J., {Basu}, S., {Elsworth}, Y., \& {Bedding}, T.~R.
  2009{\natexlab{a}}, \mnras, 400, L80

\bibitem[{{Stello} {et~al.}(2009{\natexlab{b}}){Stello}, {Chaplin}, {Bruntt},
  {Creevey}, {Garc{\'\i}a-Hern{\'a}ndez}, {Monteiro}, {Moya}, {Quirion},
  {Sousa}, {Su{\'a}rez}, {Appourchaux}, {Arentoft}, {Ballot}, {Bedding},
  {Christensen-Dalsgaard}, {Elsworth}, {Fletcher}, {Garc{\'\i}a}, {Houdek},
  {Jim{\'e}nez-Reyes}, {Kjeldsen}, {New}, {R{\'e}gulo}, {Salabert}, \&
  {Toutain}}]{stello09b}
{Stello}, D., {Chaplin}, W.~J., {Bruntt}, H., {et~al.} 2009{\natexlab{b}},
  \apj, 700, 1589

\bibitem[{{Stello} {et~al.}(2013){Stello}, {Huber}, {Bedding}, {Benomar},
  {Bildsten}, {Elsworth}, {Gilliland}, {Mosser}, {Paxton}, \&
  {White}}]{stello13}
{Stello}, D., {Huber}, D., {Bedding}, T.~R., {et~al.} 2013, in Astronomical
  Society of the Pacific Conference Series, Vol. 479, Progress in Physics of
  the Sun and Stars: A New Era in Helio- and Asteroseismology, ed.
  H.~{Shibahashi} \& A.~E. {Lynas-Gray}, 167

\bibitem[{{Strassmeier}(2009)}]{strassmeier09}
{Strassmeier}, K.~G. 2009, in IAU Symposium, Vol. 259, Cosmic Magnetic Fields:
  From Planets, to Stars and Galaxies, ed. K.~G. {Strassmeier}, A.~G.
  {Kosovichev}, \& J.~E. {Beckman}, 363--368

\bibitem[{{Twicken} {et~al.}(2010){Twicken}, {Chandrasekaran}, {Jenkins},
  {Gunter}, {Girouard}, \& {Klaus}}]{twicken10}
{Twicken}, J.~D., {Chandrasekaran}, H., {Jenkins}, J.~M., {et~al.} 2010, in
  \procspie, Vol. 7740, Software and Cyberinfrastructure for Astronomy, 77401U

\bibitem[{{Ulrich}(1986)}]{ulrich86}
{Ulrich}, R.~K. 1986, \apjl, 306, L37

\bibitem[{{Van Oort} {et~al.}(2019){Van Oort}, {Xu}, {Offner}, \&
  {Gutermuth}}]{vanoort19}
{Van Oort}, C.~M., {Xu}, D., {Offner}, S. S.~R., \& {Gutermuth}, R.~A. 2019,
  \apj, 880, 83

\bibitem[{{van Saders} {et~al.}(2016){van Saders}, {Ceillier}, {Metcalfe},
  {Silva Aguirre}, {Pinsonneault}, {Garc{\'\i}a}, {Mathur}, \&
  {Davies}}]{vansaders16}
{van Saders}, J.~L., {Ceillier}, T., {Metcalfe}, T.~S., {et~al.} 2016, \nat,
  529, 181

\bibitem[{{van Saders} {et~al.}(2019){van Saders}, {Pinsonneault}, \&
  {Barbieri}}]{vansaders19}
{van Saders}, J.~L., {Pinsonneault}, M.~H., \& {Barbieri}, M. 2019, \apj, 872,
  128

\bibitem[{{VanderPlas}(2018)}]{vanderplas18}
{VanderPlas}, J.~T. 2018, \apjs, 236, 16

\bibitem[{{Weber} \& {Davis}(1967)}]{weber67}
{Weber}, E.~J., \& {Davis}, Leverett, J. 1967, \apj, 148, 217

\bibitem[{{Wright} {et~al.}(2011){Wright}, {Drake}, {Mamajek}, \&
  {Henry}}]{wright11}
{Wright}, N.~J., {Drake}, J.~J., {Mamajek}, E.~E., \& {Henry}, G.~W. 2011,
  \apj, 743, 48

\bibitem[{{Xu} {et~al.}(2020){Xu}, {Offner}, {Gutermuth}, \& {Van Oort}}]{xu20}
{Xu}, D., {Offner}, S. S.~R., {Gutermuth}, R., \& {Van Oort}, C. 2020, arXiv
  e-prints, arXiv:2001.04506

\bibitem[{{Yip} {et~al.}(2019){Yip}, {Zhang}, {Wang}, {Zhang}, {Sun},
  {Contardo}, {Villaescusa-Navarro}, {He}, {Genel}, \& {Ho}}]{yip19}
{Yip}, J. H.~T., {Zhang}, X., {Wang}, Y., {et~al.} 2019, arXiv e-prints,
  arXiv:1910.07813

\bibitem[{{Yosinski} {et~al.}(2015){Yosinski}, {Clune}, {Nguyen}, {Fuchs}, \&
  {Lipson}}]{yosinski15}
{Yosinski}, J., {Clune}, J., {Nguyen}, A., {Fuchs}, T., \& {Lipson}, H. 2015,
  arXiv e-prints, arXiv:1506.06579

\bibitem[{{Yu} {et~al.}(2018){Yu}, {Huber}, {Bedding}, {Stello}, {Hon},
  {Murphy}, \& {Khanna}}]{yu18}
{Yu}, J., {Huber}, D., {Bedding}, T.~R., {et~al.} 2018, \apjs, 236, 42

\bibitem[{{Zahn}(1992)}]{zahn92}
{Zahn}, J.~P. 1992, \aap, 265, 115

\bibitem[{{Zhang} {et~al.}(2019){Zhang}, {Wang}, {Zhang}, {Sun}, {He},
  {Contardo}, {Villaescusa-Navarro}, \& {Ho}}]{zhang19}
{Zhang}, X., {Wang}, Y., {Zhang}, W., {et~al.} 2019, arXiv e-prints,
  arXiv:1902.05965

\bibitem[{{Zinn} {et~al.}(2019){Zinn}, {Stello}, {Huber}, \& {Sharma}}]{zinn19}
{Zinn}, J.~C., {Stello}, D., {Huber}, D., \& {Sharma}, S. 2019, arXiv e-prints,
  arXiv:1909.11927

\end{thebibliography}

%%%%%%%%%%%%%%%%% APPENDICES %%%%%%%%%%%%%%%%%%%%%

\appendix

\begin{figure*}[htp!]
\centering
\includegraphics[width=1.\columnwidth]{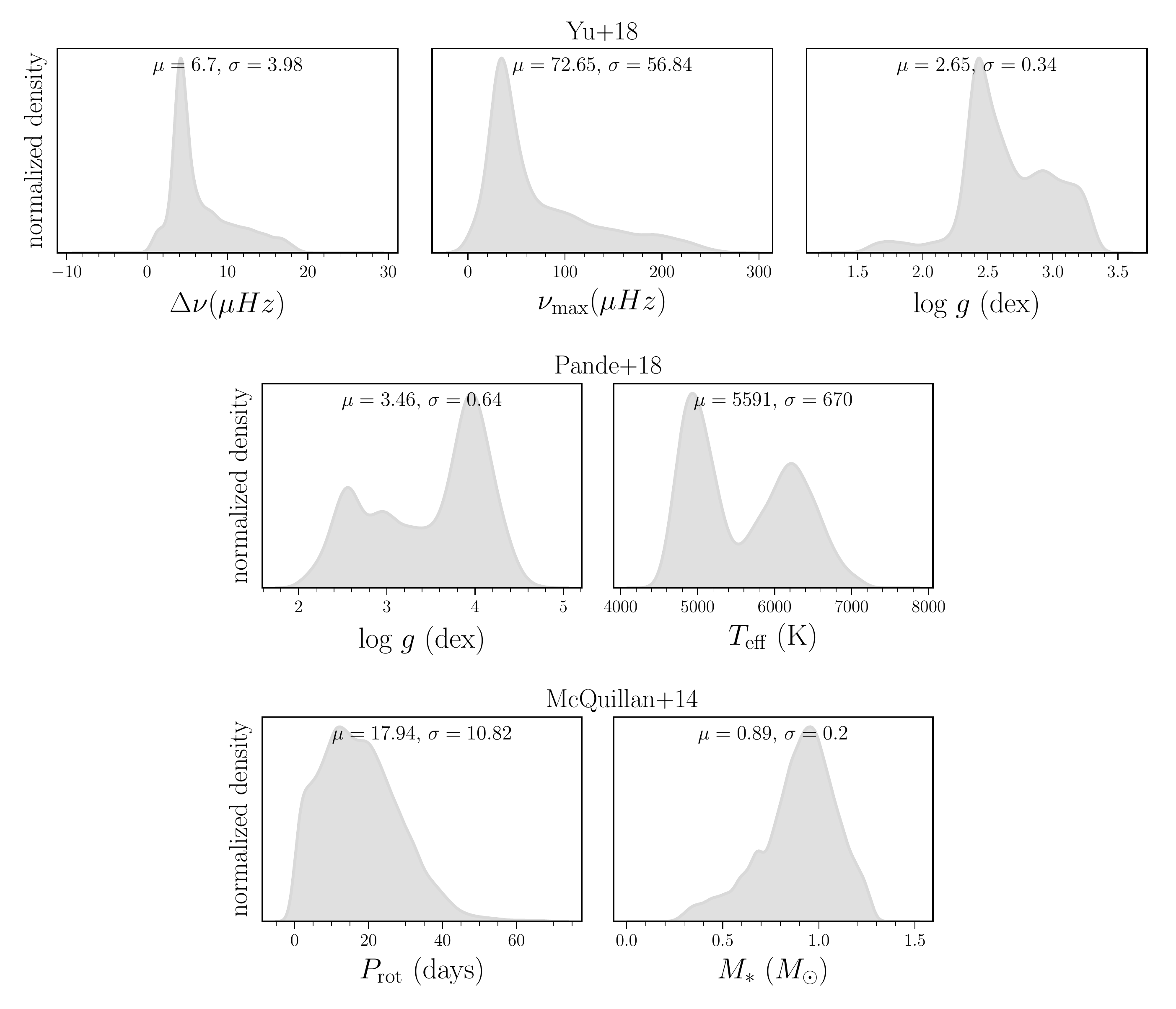}
\caption{The distribution of stellar properties for the \cite{yu18} RGB sample (\textit{\textbf{top row}}: $\Delta\nu$, $\nu_{\rm max}$, and log $g$), the \cite{pande18} RGB and sub-giant sample (\textit{\textbf{middle row}}: log $g$ and $T_{\rm eff}$), and the \cite{mcquillan14} main sequence sample (\textit{\textbf{bottom row}}: $P_{\rm rot}$ and $M_{*}$). \bigskip}
\label{fig:data_density}
\end{figure*}

\begin{sidewaystable}
%\begin{table*}
\begin{centering}
\tabletypesize{\scriptsize}
\caption{Short baseline model performance}
\label{tab:shortbaseline}
\tablewidth{0pt}

\begin{tabular}{lrccccccccccccccc}

& & \multicolumn{3}{c}{97 days} &  & \multicolumn{3}{c}{62 days} &  & \multicolumn{3}{c}{27 days} &  & \multicolumn{3}{c}{14 days} \\

\hline
&  \vline & $r^{2}$ & $\Delta$ & rms & \vline & $r^{2}$ & $\Delta$ & rms & \vline & $r^{2}$ & $\Delta$ & rms  & \vline & $r^{2}$ & $\Delta$ & rms \\ %\cline{1-4} \cline{5-8} \cline{9-12} \cline{13-17}
\hline
\hline

 Yu+18 & $\Delta\nu$ \vline & 0.95 & -0.17(-0.009) & 0.89(0.14) & \vline & 0.94 & -0.02(0.036) & 0.97(0.17) & \vline & 0.93 & -0.3(-0.042) & 1.06(0.18) & \vline & 0.929 & 0.12(0.076) & 1.09(0.23) \\
 
 & $\nu_{\rm max}$ \vline & 0.96 & -3.51(-0.002) & 11.85(0.2) & \vline & 0.96 & -1.14(0.022) & 12.02(0.22) & \vline & 0.91 & -3.21(0.01) & 17.08(0.3) & \vline & 0.94 & 0.22(0.064) & 14.44(0.26) \\

 & log $g$ \vline & 0.97 & 0.008(0.004) & 0.057(0.02) & \vline & 0.97 & 0.006(0.004) & 0.057(0.02) & \vline & 0.96 & 0.006(0.003) & 0.071(0.03) & \vline & 0.94 & -0.0(0.002) & 0.086(0.04) \\

%&  \vline &  &  &  & \vline &  &  &  & \vline &  &  &  & \vline &  &  &  \\
\hline
%&  \vline &  &  &  & \vline &  &  &  & \vline &  &  &  & \vline &  &  &  \\

Pande+18 & log $g$ \vline & 0.89 & 0.07(0.026) & 0.22(0.07) & \vline & 0.92 & 0.01(0.006) & 0.18(0.06) & \vline & 0.89 & -0.03(-0.003) & 0.21(0.06) & \vline & 0.89 & -0.05(-0.011) & 0.21(0.06) \\

& $T_{\rm eff}$ \vline & 0.79 & -66(-0.008) & 309(0.05) & \vline & 0.80 & -55(-0.007) & 297(0.05) & \vline & 0.79 & -30(-0.002) & 306(0.05) & \vline & 0.77 & -35(-0.003) & 320(0.06) \\

%&  \vline &  &  &  & \vline &  &  &  & \vline &  &  &  & \vline &  &  &  \\
\hline
%&  \vline &  &  &  & \vline &  &  &  & \vline &  &  &  & \vline &  &  &  \\

McQuillan+14 & $M_{*}$ \vline & 0.40 & 0.009(0.057) & 0.155(0.28) & \vline & 0.38 & 0.012(0.062) & 0.158(0.29) & \vline & 0.37 & 0.008(0.06) & 0.159(0.3) & \vline & 0.36 & 0.006(0.058) & 0.16(0.3) \\

\hline
\hline

\multicolumn{17}{l}{Model performance as a function of observation baseline (as described in Section \ref{sec:all_short_baseline}). The fractional bias and fractional rms is indicated in parenthesis.} \\

\vspace{-2ex}

%\medskip

\end{tabular}

\end{centering}
\end{sidewaystable}
%\end{table*}

\label{lastpage}
\end{document}